%
%
%
 \let\SELECTOR=R      
%
%
%
\expandafter\ifx\csname phyzzx\endcsname\relax \input phyzzx \fi
%
%
\interdisplaylinepenalty=10000
\overfullrule=0pt
%
%
\newdimen\doublewidth
\doublewidth=12in
\newinsert\LeftPage
\count\LeftPage=0
\dimen\LeftPage=\maxdimen
\def\PageBox{\vbox{\makeheadline \pagebody \makefootline }}
\def\papersize{\hsize=412pt \vsize=585pt \normalspace }
\papersize
\pagebottomfiller=0pt plus 2pt minus 2pt
\skip\footins=20pt plus 8pt minus 6pt
\sectionskip=\bigskipamount
\headskip=\smallskipamount
\if R\SELECTOR
    \mag=833
    \voffset=-20truept
    \hoffset=-35truept
    \output={\ifvoid\LeftPage \insert\LeftPage{\floatingpenalty 20000 \PageBox}
        \else \shipout\hbox to\doublewidth{%
            \box\LeftPage \hfil \PageBox }\fi
        \advancepageno
        \ifnum\outputpenalty>-20000 \else \dosupereject \fi }
    \message{Warning: some DVI drivers cannot handle reduced output!!!}
\else
    \mag=1000
    \voffset=0pt
    \hoffset=0pt
\fi
%
%
\def\tablerule{\vrule height 16pt depth 6pt }
\newif\iffigureexists
\newif\ifepsfloaded
\openin 1 epsf
\ifeof 1 \epsfloadedfalse \else \epsfloadedtrue \fi
\closein 1
\ifepsfloaded \input epsf \fi
\def\checkex#1 {\relax \ifepsfloaded \openin 1 #1
	\ifeof 1 \figureexistsfalse \else \figureexiststrue \fi
	\closein 1
    \else \figureexistsfalse \fi }
\def\diagram#1 {\vcenter{\def\epsfsize##1##2{0.4##1} \epsfbox{#1}}}
%
%

\def\Ch{\mathop{\rm Ch}\nolimits}
 
\def\Re{\mathop{\rm Re}\nolimits}
\def\Im{\mathop{\rm Im}\nolimits}
\def\smallmat#1#2#3{\left( \vcenter{\baselineskip=12pt
	\ialign{\hfil$\scriptstyle{##}$&\kern 8pt\hfil$\scriptstyle{##}$\cr
		#1& #2\cr #2& #3\cr }}\right)}
\def\modulo#1{\enspace\left({\rm mod}\;#1\right)}
%
%
\def\ol{{\scriptscriptstyle\rm 1-loop}}
\def\one{{(1)}}
\def\ib{{\bar\imath}}

\def\Ib{{\bar I}}
\def\Jb{{\bar J}}
%
%
\def\modul{\Phi}
\def\modulb{{\smash{\overline\Phi}\vphantom{\Phi}}}
\def\matter{Q}
\def\matterb{{\smash{\overline\matter}\vphantom{\matter}}}
\def\Sb{{\overline S}}
\def\mod{M}
\def\modb{{\smash{\overline M}\vphantom{M}}}
\def\moo{T}
\def\moob{{\smash{\overline T}\vphantom{T}}}
\def\mot{U}
\def\motb{{\smash{\overline U}\vphantom{U}}}
\def\ts{{\bf 27}}
\def\tsb{{\bf \overline{27}}}
%
%
\def\K{K\"ahler}
\def\Ktot{{\cal K}}
\def\KM{\hat{K}}
\def\Kmod{K}
\def\gstring{g_{\rm string}}
\def\Dtot{\tilde\Delta}
\def\Ddif{\Delta}
\def\Duniv{\Delta^{\rm univ}}
\def\mpl{M_{\rm Pl}}
\def\mstring{M_{\rm string}}
\def\vgs{\delta_{\rm GS}}
%
%
\def\cano{{\bf c}}
\def\cren{{\bf b}}

\def\hoo{h_{(1,1)}}
\def\hot{h_{(1,2)}}
\def\sltwo{SL(2,{\bf Z})}
\def\Alpha{\alpha}
\def\romega{\Omega}
\def\symmat{\Upsilon}
\def\KF{{\cal J}}
\def\clrad{\omega}
%
%
\def\B{{\cal B}}
\def\Rb{\bar{R}}
\def\Fb{\smash{\overline F}\vphantom{F}}
\def\Xb{\smash{\overline X}\vphantom{X}}
\def\psib{\smash{\overline\psi}\vphantom{\psi}}
\def\Nb{\smash{\overline N}\vphantom{N}}
%
%
\def\ldf{\REF}
\def\JJjournal#1#2{\unskip\space{\sfcode`\.=1000\sl #1 \bf #2}\space }
\def\nup#1 {\JJjournal{Nucl. Phys.}{B#1}}
\def\plt#1 {\JJjournal{Phys. Lett.}{#1}}
\def\cmp#1 {\JJjournal{Comm. Math. Phys.}{#1}}
\def\prp#1 {\JJjournal{Phys. Rep.}{#1}}
\def\prl#1 {\JJjournal{Phys. Rev. Lett.}{#1}}
\def\prev#1 {\JJjournal{Phys. Rev.}{#1}}
\def\mplt#1 {\JJjournal{Mod. Phys. Lett.}{#1}}
%
%
%
%
\ldf\GHMR{D.~Gross, J.~Harvey, E.~Martinec and R.~Rohm,
\nup256 (1985) 253; \nup267 (1986) 75.}
\ldf\ginsparg{P.~Ginsparg, \plt  197B (1987) 139.}
%
\ldf\SW{S.~Weinberg, \plt 91B (1980) 51.}
\ldf\VKb{V. Kaplunovsky, \nup307 (1988) 145.}
%
\ldf\LEPData{%
U.~Amaldi, W. de Boer and H.~F\"urstenau, \plt B260 (1991) 447;\brk
J.~Ellis, S.~Kelley and D.V.~Nanopoulos, \plt B260 (1991) 131;\brk
P.~Langacker and M.~Luo, \prev D44 (1991) 817;\brk
P.~Langacker and N.~Polonsky, \prev D47 (1993) 4028.}
%
\ldf\AEKN{I.~Antoniadis, J.~Ellis, S.~Kelley and D.~V.~Nanopoulos,
\plt B272 (1991) 31.}
\ldf\ILR{L.~Ib\'a\~nez, D.~L\"ust and G.~Ross, \plt B272 (1991) 251.}
\ldf\MNS{P.~Mayr, H.-P.~Nilles and S.~Stieberger,\plt B317 (1993) 53.}
%
\ldf\DINDRSW{J.P.~Derendinger, L.E.~Ib\'a\~nez and
  H.P.~Nilles, \plt  155B (1985) 65;\brk
M.~Dine, R.~Rohm, N.~Seiberg and  E.~Witten, \plt  156B (1985) 55.}
\ldf\Tbreak{
A.~Font, L.E.~Ib\'a\~nez, D.~L\"ust and F.~Quevedo,
\plt  B245 (1990) 401;\brk
S.~Ferrara, N.~Magnoli, T.~Taylor and G.~Veneziano,
\plt B245 (1990) 409;\brk
H.~P.~Nilles and M.~Olechowski, \plt B248 (1990) 268;\brk
P.~Bin\'etruy and M. K. Gaillard, \nup358 (1991) 121;\brk
B.~de Carlos, J.~A.~Casas and C.~Mu\~noz, \nup399 (1993) 623.}
%
\ldf\snp{S.H.~Shenker, in the Proceedings of the
1990 Carg\`ese Workshop on {\it Random surfaces and quantum gravity};\brk
A.~Dabholkar, \nup368 (1992) 293;\brk
R.~Brustein and B.~Ovrut,
 \plt B309 (1993) 45, \nup421 (1994) 293;\brk
J.~Lee and P.~Mende, \plt B312 (1993) 433.}
\ldf\DSB{M.~Dine and N.~Seiberg, \plt 162 (1985) 299;\brk
T.~Banks and M.~Dine, \prev D50 (1994) 7454.}
\ldf\KLb{V.~Kaplunovsky and J.~Louis, \nup422 (1994) 57.}
\ldf\Witten{E.~Witten, \plt  155B (1985) 151.}
%
\ldf\DKLb{L.~Dixon, V.~Kaplunovsky and J.~Louis, \nup355 (1991) 649.}
\ldf\ANT{I.~Antoniadis, K.~Narain and T.~Taylor, \plt B267 (1991) 37.}
\ldf\AGN{I.~Antoniadis, E.~Gava and K.~Narain,
          \plt B283 (1992) 209; \nup383 (1992) 93.}
\ldf\MSa{P.~Mayr and S.~Stieberger, \nup407 (1993) 725.}
\ldf\thomas{D.~Bailin, A.~Love, W.~Sabra and S.~Thomas,
Queen Mary preprints QMW-TH-94-18, QMW-TH-94-28.}
\ldf\wb{J.~Wess and J.~Bagger, {\it Supersymmetry and
  Supergravity}, Princeton University Press, 1983.}
%
\ldf\SVa{M.A.~Shifman and A.I.~Vainshtein,\nup277 (1986) 456.}
\ldf\nilles{H.~P.~Nilles, \plt  180B (1986) 240.}
\ldf\DYuri{M.~Dine and Y.~Shirman, \prev D50 (1994) 5389.}
%
\ldf\SVb{M.A.~Shifman and A.I.~Vainshtein, \nup359 (1991)
571.}
\ldf\louisa{J. Louis, in the Proceedings of the
        {\it 2nd International Symposium on Particles, Strings and Cosmology},
        Boston, 1991, ed.~P.~Nath und S.~Reucroft.}
\ldf\DFKZa{J. P. Derendinger, S. Ferrara, C. Kounnas and F. Zwirner,
           \nup372 (1992) 145,\plt B271 (1991) 30.}
\ldf\CLO{G.L.~Cardoso and B. Ovrut, \nup369 (1992) 351; \nup392 (1993) 315.
}
\ldf\ABGG{P. Binetruy, G. Girardi and R. Grimm, \plt B265 (1991) 111;
     P. Adamietz, P. Binetruy, G. Girardi and R. Grimm,\nup 401 (1993) 257.}
\ldf\GT{M.K.~Gaillard and T.~Taylor, \nup381 (1992) 577. }
\ldf\LM{H.S.~Li and K.T.~Mahanthappa, \plt B319 (1993) 152, \prev D49 (1994)
5532.}
\ldf\MSb{P.~Mayr and S.~Stieberger, \nup412 (1994) 502.}
\ldf\IL{L.~Ib\'a\~nez and D.~L\"ust, \nup382 (1992) 305.}
\ldf\CHSW{P.~Candelas, G.~Horowitz,
  A.~Strominger and E.~Witten, \nup258 (1985) 46.}
\ldf\DKLa{L.J.\ Dixon, V.S.\ Kaplunovsky and J.\ Louis,
  \nup329 (1990) 27.}
\ldf\CD{S.~Ferrara and A.~Strominger, in the Proceedings of the
{\it Strings '89} workshop, College Station, 1989;\brk
A.\ Strominger, \cmp 133 (1990) 163;\brk
P.\ Candelas and X.C.\ de la Ossa, \nup355 (1991) 455.}
%
\ldf\BCOV{M.~Bershadsky, S.~Cecotti, H.~Ooguri and C.~Vafa, \nup 405 (1993)279,
          \cmp 165 (1994) 311.}
\ldf\CDGP{P.~ Candelas, X.C.~de la Ossa, P.S.~Green and
            L.~Parkes, \nup359 (1991) 21.}

%
\ldf\JJW{I.~Jack and D.~R.~T.~Jones, \plt258 (1991) 382;
P.~West, \plt258 (1991) 375.}
\ldf\DS{M.~Dine and  N.~Seiberg, \prl 57 (1986) 2625.}
\ldf\IN{L. Ib\'a\~nez and H.P. Nilles, \plt  169B (1986) 354.}
\ldf\KK{E.~Kiritsis and C.~Kounnas, CERN preprint CERN-TH.7472/94.}
%
%
\ldf\konishi{T.E.~Clark, O.~Piquet and K.~Sibold, \nup159 (79) 1;\brk
   K.~Konishi, \plt  135B (1984) 439.}
\ldf\DHVW{L.~Dixon, J.~Harvey, C.~Vafa  and E.~Witten,
  \nup261 (1985) 678; \nup274 (1986) 285.}
\ldf\GPR{For a recent review see,
A.~Giveon, M.~Porrati and E.~Rabinovici, \prp 244 (1994) 77.}
\ldf\kahleref{M.~Cveti\v c, J.~Louis and B.~Ovrut,\plt B206 (1988) 227;\brk
           S.~Ferrara and M.~Porrati, \plt B216 (1989) 289.}
\ldf\CLM{G.L.~Cardoso, D.~L\"ust and T.~Mohaupt,
         Humboldt preprint HUB-IEP-94/50.}
\ldf\lehner{J.~Lehner, {\it Discontinuous groups and automorphic functions},
ed. The American Mathematical Society (1964).}
\ldf\VKa{V. Kaplunovsky, \prl  55 (1985) 1036.}
\ldf\VT{T. R.~Taylor and G.~Veneziano, \plt  212B (1988) 147.}
%
%
\ldf\DG{J.~Distler and B.~Greene, \nup309 (1988) 295.}
\ldf\onepa{D.~Morrison, Duke preprint DUK-M-91-14, (1991);\brk
A.~Klemm and S.~Theisen, \nup389 (1993) 153;\brk
A.\ Font,  \nup391 (1993) 358.}
\ldf\CDFKM{P.~Candelas, A.~Font, S.~Katz and D.~Morrison, \nup429 (1994) 626;
           P.~Candelas, X.C.~de la Ossa, A.~Font, S.~Katz and D.~Morrison,
                 \nup416 (1994) 481.  }
\ldf\KT{A.~Klemm and S.~Theisen,\mplt A9 (1994) 1807;\brk
        S.~Hosono, A.~Klemm, S.~Theisen and S.T.~Yau,\nup433 (1994) 501,
          Harvard preprint HUTMP-93-0801.}
\ldf\BK{P.~Berglund and S.~Katz, \nup420 (1994) 289;\brk
P.~Berglund, S.~Katz and A.~Klemm, IAS preprint IASSNS-HEP-94/106.}
\ldf\MP{D.~Morrison and M.R.~Plesser, IAS preprint IASSNS-HEP-94/82.}
\ldf\dfnt{B.\ de Wit and A.\ Van Proeyen, \nup245 (1984) 89;
B.\ de Wit, P.\ Lauwers and A.\ Van Proeyen, \nup255 (1985) 569;
E.\ Cremmer, C.\ Kounnas, A. \ Van Proeyen, J.P.\ Derendinger,
S.\ Ferrara, B.\ de Wit and L.\ Girardello,  \nup250 (1985) 385.}
\ldf\FKLZ{S.~Ferrara, C.~Kounnas, D.~L\"ust and F.~Zwirner,
\nup365 (1991) 431.}
%
\ldf\LSWrep{For a review see for example,
  W.~Lerche, N.~Schellekens and N.~Warner, \prp 177 (1989) 1.}
\ldf\CV{S.\ Cecotti and C.\ Vafa, \nup367 (1991) 359.}
\ldf\strominger{A.\ Strominger, \cmp133 (1990) 163.}
\ldf\dixon{L. Dixon, talk presented at the M.S.R.I. meeting on
mirror symmetries, Berkeley, May 1991;\brk
M.~Cveti\v c, in the Proceedings of the 26th International Conference on
High Energy Physics, Dallas, 1992.}
\ldf\gepner{D.~Gepner, \plt B199 (1987) 370, \nup296 (1988) 757.}
\ldf\DKV{L.~Dixon, V.~Kaplunovsky and C.~Vafa, \nup294 (1987) 43.}
%
%
\ldf\BDFS{T.~Banks,
     L.~Dixon, D.~Friedan and E.~Martinec, \nup299 (1988) 613.}
\ldf\LSW{W.~Lerche, N.~Schellekens and N.~Warner, \plt B214 (1988) 41;\brk
O.~Lechtenfeld and W.~Lerche, \plt B227 (1989) 373.}
\ldf\DFMS{
  S.~Hamidi and C.~Vafa, \nup279 (1987) 465;\brk
  L.~Dixon, D.~Friedan, E.~Martinec and S.~Shenker,
  \nup282 (1987) 13.}
\ldf\IFQ{A.~Font, L.~Ib\~an\'ez and F.~Quevedo, \plt B217 (1989) 79.}
%

\Pubnum{hep-th/9502077\cr UTTG--24-94\cr LMU--TPW--94--24 }
\date{}
\titlepage
\title{On Gauge Couplings in String Theory
	\foot{Research supported in part by:
	the NSF, under grant PHY--90--09850 (V.~K.);
	the Robert A.~Welch Foundation (V.~K.);
	the Heisenberg Fellowship of the DFG (J.~L.);
	the NATO, under grant CRG~931380 (both authors).}}
\author{Vadim Kaplunovsky
	\foot{Email: \tt vadim@bolvan.ph.utexas.edu}}
\address{Theory Group, Dept.~of Physics, University of Texas\break
	Austin, TX 78712, USA}
\andauthor{Jan Louis
	\foot{Email: \tt jlouis@lswes8.ls-wess.physik.uni-muenchen.de}}
\address{Sektion Physik, Universit\"at M\"unchen\break
	Theresienstrasse 37, D-80333 M\"unchen, Germany}
\abstract
We investigate the field dependence of the gauge couplings
of $N=1$ string vacua from the point of view of the
low energy effective quantum field theory.
We find that field-theoretical considerations severely constrain
the form of the string loop corrections;
in particular, the dilaton dependence of the gauge couplings is
completely universal at the one-loop level.
The moduli dependence of the string threshold corrections is also
constrained, and we illustrate the power of such constraints with a
detailed discussion of the orbifold vacua and the
$(2,2)$ (Calabi-Yau) vacua of the heterotic string.

\endpage
%
%
\chapter{Introduction and Summary}
A unified fundamental theory of all known forces  has  been
one of the prime  goals  of theoretical  high-energy  physics.
Among the  presently known candidates for such a unified theory, string
theories appear to be free of the mathematical inconsistencies at
short distances that plague the fundamentally-local quantum field theories.
This mild ultraviolet behavior results from an infinite tower of superheavy
particles in the string spectrum;
nevertheless, the long-distance limit of  a string theory can be
described by an effective quantum field theory (EQFT) containing only a
finite number of local fields.
Of particular importance is the heterotic superstring\refmark{\GHMR}
whose massless spectrum can comfortably  include
 the $SU(3)\times SU(2)\times U(1)$ gauge fields
of the Standard Model as well as  families of
chiral fermions with quantum numbers of
the quarks and leptons.

Unlike  conventional ``Grand Unified Theories,''
string theory does not combine  all the gauge forces
into a single simple group;
instead, at energies just below the string scale,
the gauge group  has a product structure $G=\prod_a G_a$.
Nevertheless, at the tree level of  string theory,
all the simple factors  $G_a$ have
related gauge couplings,\refmark{\GHMR,\ginsparg}
$$
g_a ^{-2}\ =\ k_a \ \gstring^{-2}\, ,
\eqn\gtree
$$
where $\gstring$ is the universal string coupling parameter and
$k_a$ denotes  the normalization of the gauge group generators
(in string-theoretical terms, $k_a$ is the level of the
Ka\v c-Moody current algebra giving rise to $G_a$).

The universality of the tree-level gauge couplings \gtree\ is spoiled
at the loop level by the low-energy renormalization and  by
finite threshold corrections due to loops of  charged superheavy
particles that decouple from the low-energy EQFT.\refmark{\SW,\VKb}
At the one-loop level of the string theory (and also of the EQFT),
the running effective gauge couplings are given by
$$
g_a ^{-2}(p)\ =\ k_a  \gstring^{-2}\
+\ {\cren_a \over 16\pi^2}\,\log{\mstring^2\over p^2}
+\ {\Ddif_a  \over 16 \pi^2}  \, ,
\eqn\oloopgst
$$
where $p$ is the momentum scale at which the effective couplings are
measured (which is assumed to be much less than the mass of any
superheavy string mode) and
$\cren_a/16\pi^2$ is the coefficient of the
one-loop $\beta$-function of the low-energy EQFT.
Similar to the threshold corrections in ordinary GUTs,\refmark{\SW}
the one-loop string-threshold corrections $\Ddif_a$ can be computed
in terms of charges and masses of the superheavy string modes.\refmark{\VKb}

The physical interest of studying the string-threshold corrections is twofold:
First of all, as in any unified theory,
$\Ddif_a$ are part of the high-energy boundary conditions
for the renormalization group equations for the gauge couplings of the
Standard Model and thus affect their low-energy values;
indeed, current electroweak measurements at LEP and SLC are precise enough to
be sensitive to such  threshold corrections. \refmark{\LEPData}
Thus, in string-based models without additional, intermediate-scale thresholds
in the observed sector,
precision electroweak measurements
impose stringent phenomenological constraints
on the physics at the string scale.
Note that for the string unification, the nominal unification scale
(denoted by $\mstring$ in eq.~\oloopgst) is not a free parameter of the
theory (like $M_{\rm GUT}$ in conventional GUTs) but a computable quantity;
at the one-loop level of accuracy, $\mstring\approx\gstring\times
5\cdot 10^{17}$~GeV.\refmark{\VKb}
Therefore, in string theory, the threshold corrections $\Ddif_a$ have much
stronger phenomenological impact than in GUTs and  deserve serious
investigation.\refmark{\AEKN-\MNS}

The second, and for the present investigation more important aspect
of the threshold corrections results from the extreme sensitivity of
various low-energy non-perturbative effects to the ultraviolet values
of the gauge couplings.
A major problem of the string unification is that to all orders in
perturbation theory, supersymmetric ground states of the heterotic string
are not isolated from each other but come in continuous families of exactly
degenerate vacua parametrized by the vacuum expectation values (VEVs) of
gauge-neutral scalar fields
$\modul^i$  usually called   moduli.
Generally,  all  couplings of the low-energy EQFT
depend on the moduli VEVs and hence remain undetermined until some
non-perturbative effects induce a non-trivial effective potential for the
moduli fields and lift the exact degeneracy of the perturbation theory.
This  effective potential is also essential for the spontaneous  breakdown of
spacetime supersymmetry at or just above the weak scale.
\refmark{\DINDRSW, \Tbreak}

It is of course possible that the non-perturbative effects giving rise
to this effective potential are of an inherently stringy nature and thus
are beyond our present knowledge.\refmark{\snp}
However, there are good reasons to assume
that the leading non-perturbative effects are due to infrared-strong
interactions in a ``hidden'' sector of the low-energy EQFT.
\refmark{\DSB}
The energy scale at which such interactions become strong --- and thus
the overall magnitude of all the field-theoretical non-perturbative effects
--- is controlled by asymptotically free gauge interactions,
and hence the shape of the resulting effective potential for the moduli
is extremely sensitive to the field dependence of the relevant gauge couplings.
\refmark{\DINDRSW, \Tbreak, \KLb}

Generally, moduli dependence of the gauge couplings of a string-based EQFT
can be studied in two very different ways.
One approach is to calculate the gauge couplings directly from the
string-theoretical amplitudes involving the gauge fields
and then analyze their moduli dependence.
At the tree level, this approach yields eq.~\gtree\ where $\gstring^{-2}$
depends solely on the dilaton --- a modulus common to all vacuum
families of the heterotic string.
For the $N=1$ supersymmetric vacua, this dependence can be summarized as
\refmark{\Witten}
$$
\left[ f_a\equiv {1\over g_a^2}-{i\theta_a\over 8\pi^2}\right]^{\rm tree}\
=\ k_a S
\eqn\fdilaton
$$
where $S$ is (the bosonic part of) the dilaton/axion chiral superfield.
At the one-loop level, the gauge couplings are given by eqs.~\oloopgst,
in which the
threshold corrections $\Ddif_a$ do not depend on the dilaton but generally
do depend on all the other moduli $\mod^i$ of the vacuum family.
If the masses of  the superheavy string modes are known as analytic
functions of the moduli, then the moduli dependence of the $\Ddif_a$ can
also be evaluated in analytic form.
Following this approach, L.~Dixon and the present authors\refmark{\DKLb}
have calculated the $\Ddif_a$ of factorizable $(0,2)$ orbifolds
as explicit functions of the untwisted moduli;
the same method was subsequently
extended to other classes of string vacua in refs.~\ANT--\thomas.

The other approach to the moduli dependence of the gauge couplings is based
on constraints due to local supersymmetry of the low-energy EQFT.
In the Wilsonian action of the EQFT, a gauge coupling appears in a chiral
superspace integral $\int\!d^4x d^2\Theta {\cal E}\, f_a(\modul)
\tr_a(W^\alpha W_\alpha)$ and hence has to be a harmonic function, \ie,
the real part of a holomorphic
function $f_a(\modul)$ of the complex moduli fields.\refmark{\wb}
The chirality of this action for the gauge superfields leads to a
powerful no-renormalization theorem:
There are no perturbative corrections to the $f_a$ beyond the one-loop level.
\refmark{\SVa, \nilles, \ANT}
On the other hand, the Wilsonian gauge couplings of an EQFT
do not account for the low-energy loops of the light fields and hence
do not immediately connect to physical quantities such as
scattering amplitudes.
\foot{%
    {}From the renormalization theory's point of view, the Wilsonian couplings
    are couplings of the EQFT from which the high-energy degrees of freedom
    are integrated out but the low-energy quantum operators are left
    as they are.
    On the other hand, the generating functional of the 1PI Feynman graphs
    defines a non-local effective classical action that summarizes
    all the quantum effects, both high-energy and low-energy.
    The distinction between the Wilsonian action and the effective
    classical action
    and between the corresponding couplings is described in detail in
    refs.~\SVa,\KLb,\DYuri.
    }
Instead, one may define more physical, momentum-dependent (running)
{\it effective} gauge couplings, which are free of these problems,
although they have complications of their own:
The effective gauge couplings renormalize at all orders
of the perturbation theory and their moduli dependence is non-harmonic.
\refmark{\DKLb,\SVb-\LM,\KLb}
However, this non-harmonicity is a purely low-energy effect and can be
calculated from the low-energy EQFT without any knowledge of the superheavy
particles;
it is the harmonic terms in the moduli-dependent effective gauge couplings
that are sensitive to the physics at the high-energy threshold.
Such terms can always be interpreted as threshold corrections
to the Wilsonian gauge couplings $f_a(\modul)$,
and because of the no-renormalization theorem,
they can arise at the one-loop level of the perturbation theory or
non-perturbatively, but not at any multi-loop level.

The  effective gauge couplings are physical and hence
invariant under any exact symmetry of the low-energy EQFT.
However, in order to  cancel the potential anomalies
arising from  chiral rotation and rescaling of the charged fermions,
the Wilsonian gauge couplings may be subject to non-trivial
transformation laws, which are determined at the one-loop level
of the EQFT in terms
of its tree-level couplings.\refmark{\louisa-\CLO,\KLb}
These  anomalous transformation laws
act as extremely powerful constraints on the holomorphic functions
$f_a(\modul)$;
indeed, if the moduli space of the EQFT modded out by all the discrete
symmetries were a compact non-singular manifold, the $f_a(\modul)$ would be
completely determined (up to constant terms) by their symmetry
transformations alone.\refmark{\KLb}
More generally, the functional form of the $f_a(\modul)$ is
determined by their transformation properties and their asymptotic
behaviors at the singular points of the moduli space and along its
non-compact directions
(\ie, the large radius limit of a Calabi-Yau manifold).

The purpose of this article is to interrelate the string-theoretical
and the field-theoretical approaches, to establish their mutual
consistency and to  demonstrate the power of the field-theoretical
constraints in the context of string theory.
In the following section (2), we discuss generic properties of
four-dimensional,
$N=1$ supersymmetric vacuum families of the heterotic string;
essentially, we impose the  special properties  of the dilaton superfield
$S$ in an otherwise generic EQFT.
We show that at the one-loop level, the dilaton dependence of the effective
gauge couplings is completely universal: In terms of eq.~\oloopgst,
$$
\gstring^{-2}\ =\ \Re S\ +\ {1\over 16\pi^2}\,\Duniv(\mod,\modb) ,
\eqn\unicorr
$$
{}$\mstring$ is $\gstring\mpl$ times a numerical constant and
the gauge-group-specific threshold corrections $\Ddif_a(\mod,\modb)$
are dilaton-independent; this is exactly what one obtains from
the direct string-loop expansion.\refmark{\MSb}

Furthermore, perturbative consistency between the EQFT and the string
theory requires the ``universal'' threshold correction $\Duniv$ in
eq.~\unicorr\ to have exactly the same moduli dependence as the
Green-Schwarz term in the  K\"ahler function, which is the field-theoretical
description of the mixing between the dilaton and the moduli
at the one-loop level of the string.\refmark{\DFKZa-\ABGG}
Consequently, given the functional form
of the string-theoretical threshold corrections $\Ddif_a(\mod,\modb)$,
field-theoretical techniques can use such data to determine both
the Green-Schwarz term and the exact Wilsonian gauge couplings
of the EQFT.\refmark{\louisa-\GT,\IL}
Indeed, we shall see that the non-harmonic part of the moduli dependence of the
combined threshold corrections
$$
\Dtot_a\ =\ \Ddif_a \ +\ k_a\Duniv
\eqn\Ddef
$$
is completely fixed by the low-energy EQFT in terms of the
tree-level K\"ahler function.
Thus, the discrepancy between the non-harmonic parts of the
string-theoretical $\Ddif_a(\mod,\modb)$ and the field-theoretical
constraints on the $\Dtot_a(\mod,\modb)$ determines $\Duniv$
and hence the Green-Schwarz term
up to a holomorphic ambiguity.
At the same time, the remaining, harmonic part of a $\Dtot_a(\mod,\modb)$
determines the moduli dependence of the one-loop correction to the
Wilsonian gauge coupling $f_a(S,\mod)$,
and because of the no-renormalization theorem
for the Wilsonian gauge couplings,
the resulting $f_a$ are exact to all orders of the perturbation theory.

In section 3 we demonstrate the power of the field-theoretical
constraints on the holomorphic functions $f_a(\mod)$ and show
that their exact form can often be obtained from essentially tree-level
properties of the string vacua.
Specifically, we consider families of factorizable $(0,2)$ orbifolds, which
are invariant under a group of discrete symmetries called {\sl
modular transformations}.
These symmetries are exact to all orders of the string perturbation theory,
but the way they act on the
massless charged fields can be fully determined from the tree-level
K\"ahler function of the EQFT.
In Appendix~B, we present a string-theoretical calculation of the
relevant parameters of this function,
which in turn tells us the exact anomalous transformation rules
for the Wilsonian gauge couplings $f_a$.
Furthermore, we show that the holomorphic functions $f_a(\mod)$ have
no singularities for any finite values of the moduli fields and that
their divergences in the decompactification limit are no worse
than power-like with respect to the radii of the internal six-torus.
Together, these data are sufficient to determine the functions
$f_a(\mod)$ up to moduli-independent constants.

For the factorizable orbifolds we thus have two independent means
of calculating the moduli dependence of the gauge couplings in an
analytic form:
The field-theoretical method outlined above, and also the direct
string-theoretical approach of ref.~\DKLb.
We show that the two calculations yield the same functional
form for the moduli-dependent threshold corrections,
but the numerical coefficients of similar terms are given by apparently
unrelated formul\ae.
Nevertheless, for all the orbifolds we have studied we found
those numerical coefficients to fully agree with each other;
a number of examples  are presented in Appendix~C.

Finally, in section~4 of this article we discuss the  $(2,2)$
vacuum families of the heterotic string;
Calabi-Yau compactifications\refmark{\CHSW}
are the best-known examples of such vacua.
The $(2,2)$ families possess intricate tree-level relations between the
couplings of the charged matter fields and the geometry of the
moduli space.\refmark{\DKLa,\CD}
These relations allow us to derive the anomalous transformation rules
for the Wilsonian gauge couplings under discrete symmetries
from the K\"ahler function and the transformation rules
for the moduli fields, without any additional string-theoretical information.
Furthermore, the gauge group of a generic $(2,2)$ vacuum is $E_6\times E_8$,
and the difference between the gauge couplings for  $E_6$
and  $E_8$ turns out to be related to the topological index
$F_1$ defined in ref.~\BCOV.
This index is computable in geometrical terms for the large-radius
Calabi-Yau threefolds and thus provides  additional  information
about the large-radius  behavior of the gauge couplings.
Consequently, given all the symmetries and all the singularities of
a threefold's  moduli space,
one often has enough constraints for the holomorphic functions
$f_6(\mod)$ and $f_8(\mod)$ to completely determine their form.
As an example of this method,
we calculate the dependence of the Wilsonian gauge couplings
on the only $(1,1)$ modulus of the quintic threefold.\refmark{\CDGP,\BCOV}
Alas, we are unable to compare this field-theoretical result
to a direct string-theoretical calculation
because the moduli dependence of the
superheavy particles' masses is not presently known for the quintic threefold.
Since the same is true for most other currently known string vacua,
the field-theoretical method of analysis appears to be indispensable.

\chapter{Effective Quantum Field Theory\break of Generic String Vacua}
\section{Dilaton Dependence of the Wilsonian Couplings.}
At energies below the Planck scale, all particle interactions can be
described in terms of an Effective Quantum Field Theory (EQFT) for
the light modes of the string.
Local supersymmetry imposes severe constraints on the action of the EQFT;
in particular, all  interactions with at most two derivatives
can be summarized
in terms of the K\"ahler function $\Ktot$, the superpotential $W$
and the field-dependent gauge couplings $f_a$.
It is important to distinguish between the Wilsonian couplings of an
EQFT, which are coefficients of local quantum operators comprising
the action of the  theory, and between the momentum-dependent effective
couplings that parametrize the scattering amplitudes.\refmark{\SVa,\KLb}
For EQFTs quantized and cut-off in a manifestly locally supersymmetric
fashion, the {\sl Wilsonian} superpotential $W$
and the {\sl Wilsonian} gauge couplings $f_a$
are holomorphic functions of the chiral superfields.
Furthermore,  $W$ does not renormalize  perturbatively  while
the renormalization of  $f_a$ is exhausted
at the one-loop level of the perturbation theory.
On the other hand, the effective gauge couplings $g_a^{-2}(p^2)$
\foot{In ref.~\DKLb\ the effective gauge couplings were denoted
    $\{g_a(p^2)\}^{-2}$ in order to distinguish them from
    the Wilsonian couplings
    $(g^W_a)^{-2}=\Re f_a^W$.
    In this article, $g_a$ always denote the effective gauge couplings
    while $f_a$ always denote the Wilsonian couplings.}
renormalize in all orders of the perturbation theory
and their dependence on the moduli scalars is non-holomorphic.

The issue of manifestly locally supersymmetric quantization and regularization
of EQFTs is discussed in detail in ref.~\KLb.
For the purposes of this article, let us  iterate a few points:
First, the supersymmetric cutoff discussed in ref.~\KLb\ is purely
perturbative in nature and cannot be used to {\sl define} a locally
supersymmetric EQFT in a manifestly unitary non-perturbative way;
there is no known supersymmetric analogue of the lattice cutoff for
 ordinary gauge theories.
Therefore, our formalism presumes that all the Wilsonian couplings of
the EQFT cut-off at the string threshold are weak enough to use perturbation
theory; physically, this means that all the interactions
{\sl at the string scale} must be perturbatively weak.
Note that this assumption does not exclude strong interactions at
much lower energies.
However, strong interactions right at the string scale would require
a different field-theoretical formalism --- as well as a
non-perturbative string theory.

Second, manifest local supersymmetry of the regularized EQFT is not enough:
One also needs to maintain full $d=4$, $N=1$ gauge invariance of the theory.
(To be precise, the background gauge invariance should be manifest while the
quantum gauge invariance is protected by the BRST symmetry.)
Such a regularization ought to be possible, but the specific prescription
displayed in ref.~\KLb\ presumes that  only the
gauge and the charged matter superfields are affected by
the gauge transformations
while the background gravitational and moduli superfields remain inert.
In particular, we did not allow for
linear  superfields with Chern-Simons
couplings to the gauge superfields
because of technical difficulties with regularizing such couplings.
Fortunately, linear superfields are always dual to chiral superfields,
so one can avoid these difficulties by using the latter rather then the
former.

Although from the field-theoretical point of view there is no harm
(and much benefit) in putting all the scalar particles into chiral
supermultiplets,
from the string-theoretical point of view, using the chiral superfield
$S$ for the dilaton-axion-dilatino multiplet does it serious injustice.
While for all other light particles the relation between the vertex
operator of the string theory and the corresponding
{\sl unnormalized} quantum field
of the EQFT is completely determined at the tree level and suffers from
no corrections at higher orders,
the dilaton, axion and dilatino vertices have similarly fixed relation to
components of the linear superfield $L$, but their relation to the
components of the chiral superfield $S$ has to be adjusted order-by-order in
perturbation theory.
For this reason, whenever the low-energy limit of the heterotic string
is discussed in terms of the generating function
(sometimes called ``the effective classical Lagrangian''),
the linear superfield $L$ gives a clearer picture of the
dilaton-axion physics than the chiral superfield $S$.\refmark{\DFKZa-\GT}
On the other hand, the analytic properties of the  Wilsonian couplings
are more transparent in  the chiral superfield formalism.
Hence, for the purpose of this article, we prefer
to work with $S$ rather than $L$.

With all these preliminaries in mind, let us consider the K\"ahler function
$\Ktot$ of a string-based EQFT.
$\Ktot$ is a real analytic function of all the chiral superfields which
controls their sigma-model-like interactions
and the geometry of the field space.
Generically, expanding $\Ktot$ in powers of the matter superfields
\foot{In our terminology, the ``matter'' consists of all the scalar superfields
    that are not moduli and do not have Planck-sized VEVs.
    All the charged scalar superfields are matter, including the
    ``hidden matter'' charged under a ``hidden'' gauge symmetry.
    Gauge-singlet superfields that are prevented from acquiring Planck-sized
    VEVs by their Yukawa couplings are also treated as matter.}
$\matter^I$ and $\matterb^\Ib$, we have
$$
\Ktot(\modul,\modulb,\matter,\matterb)\
=\ \kappa^{-2}\, \Kmod(\modul,\modulb)\
+\ Z_{\Ib J}(\modul,\modulb)\ \matterb^\Ib e^{2V} \matter^J\
+\ \cdots \ ,
\eqn\Kexpansion
$$
where  $\modul$ stands for all the chiral moduli
superfields, including both the  moduli $\mod^i$ and the
dilaton-axion $S$; the `$\cdots$' stand for the higher-order terms
in $\matter^I$ which are irrelevant for the present discussion.
Note that in our notations the matter superfields $\matter^I$ have canonical
dimension one while the moduli are dimensionless.
(A $\vev\modul=O(1)$ corresponds to a Planck-sized modulus VEV
in conventional units.)
At the tree level of both the string theory and the EQFT,
$$
\Kmod^{\rm tree}(\modul,\modulb)\ =\ -\log(S+\Sb)\ +\ \KM(\mod,\modb)
\eqn\dilatonk
$$
while the kinetic-energy matrix $Z_{\Ib J}$ for the matter fields depends
on the string moduli $\mod^i$ and $\modb^\ib$ but not on the dilaton.
The specific form of the functions $\KM(\mod,\modb)$
and $Z_{\Ib J}(\mod,\modb)$
depends on the details of the world-sheet SCFT defining a particular family
of string vacua;
their properties
for orbifolds and  Calabi-Yau manifolds are discussed later in this article
(sections 3 and 4).

The dilaton and its superpartners arise in the spacetime sector of the
world-sheet SCFT  rather than in its internal sector.
Consequently, its couplings are model independent  at the tree level
(\cf\ eqs.~\fdilaton\ and~\dilatonk), although the loop corrections destroy
this universality.
Thus, eq.~\dilatonk\ becomes
$$
\Kmod(\modul,\modulb)\ =\ -\log(S+\Sb)\ +\ \KM(\mod,\modb)\
+\ {V^{(1)}(\mod,\modb)\over 8\pi^2(S+\Sb)}\
+\ {V^{(2)}(\mod,\modb)\over 64\pi^4(S+\Sb)^2}\
+\ \cdots ;
\eqn\dilatonkloop
$$
similarly,
$$
Z_{\Ib J}(\modul,\modulb)\ =\ Z_{\Ib J}^{(0)}(\mod,\modb)\
+\ {Z_{\Ib J}^{(1)}(\mod,\modb)\over 8\pi^2(S+\Sb)}\
+\ {Z_{\Ib J}^{(2)}(\mod,\modb)\over 64\pi^4(S+\Sb)^2}\
+\ \cdots .
\eqn\Zloopexpansion
$$
Note that in both formul\ae, the dilaton appears only in
combination $(S+\Sb)$ --- this is required by the continuous
Peccei-Quinn symmetry
$S\to S+i\gamma$, which holds to all orders of the perturbation theory.
Furthermore, all the loop corrections come as power series in
$1/8\pi^2(S+\Sb)$, which serves as the string's coupling parameter.
\foot{%
  Actually, the true string-loop counting parameter is $e^{-2\phi}$
  where $\phi$ is the field defined by the dilaton's vertex operator;
  the precise relation between $e^{-2\phi}$ and $1/(16\pi^2\Re S)$ is itself
  subject to  loop corrections.
  Therefore, the $n$-loop order corrections are generally of the order
  $O(1/(16\pi^2\Re S)^n)$, but they also contain sub-leading terms
  of higher orders in $1/(16\pi^2\Re S)$.
  For example, the one-loop effects not only determine the first-order
  coefficients $V^{(1)}(\mod,\modb)$ and $Z_{\Ib J}^{(1)}(\mod,\modb)$
  in the series \dilatonkloop\ and \Zloopexpansion,
  but they also affect the higher order coefficients
  $V^{(2)}$ and $Z_{\Ib J}^{(2)}$, \etc, \etc\space\space
  The way the loop counting works in terms of $S$ is that the terms of
  the relative order $1/(16\pi^2\Re S)^n$ are completely determined at the
  $n$-loop order --- not solely from the genus-$n$ world sheet,
  but from all the genii from zero to $n$.
  }

Now consider the superpotential $W$, which is a holomorphic function of
the chiral superfields.
Generically, it looks like
$$
W(\matter,\modul)\ =\ \half\mu_{IJ}(\modul)\, \matter^I\matter^J\
+\ \coeff13 Y_{IJK}(\modul)\, \matter^I\matter^J\matter^K\
+\ \cdots ,
\eqn\Wexpansion
$$
where the `$\cdots$' stand for the non-renormalizable  higher-order
terms, but
at the tree level of the heterotic string, the couplings $\mu_{IJ}$,
$Y_{IJK}$, \etc,
depend only on the string moduli $\mod^i$ but not on the dilaton $S$.
In field theory, there is no renormalization of
the Wilsonian superpotential
\foot{%
  The two-loop corrections discussed in ref.~\JJW\ affect the
  {\sl effective} Yukawa couplings of a theory with massless chiral
  superfields but not its {\sl Wilsonian} Yukawa couplings.
  }
and even the finite threshold corrections to $W$
are always completely determined at the tree level;
in string theory, the same result follows from the Peccei-Quinn symmetry
in combination with the holomorphicity.\refmark{\DS}
Indeed, $W$ is a holomorphic function of all the chiral superfields
and thus cannot depend on the dilaton $\Re S$ without at the same time
being dependent on the axion $\Im S$.
On the other hand, the Peccei-Quinn symmetry does not allow for any
non-derivative couplings of the  axion field
and hence to all orders of the perturbation theory, the entire Wilsonian
superpotential \Wexpansion\ does not depend on the dilaton-axion
superfield $S$.
Furthermore, since the loop expansion of the string theory is controlled
by the dilaton, it follows that the string-loop corrections
do not affect the Wilsonian superpotential of the EQFT.

Like the superpotential, the Wilsonian gauge couplings $f_a$ are
holomorphic functions of the chiral superfields.
Therefore, their dependence on the dilaton superfield $S$ is also
severely restricted by the Peccei-Quinn symmetry.
Taking into account the tree-level formul\ae\ \fdilaton\ and the loop-counting
property of the dilaton, it is easy to see that to all orders of the
perturbation theory, one must have\refmark{\nilles, \IN}
$$
f_a(S,\mod)\ =\ k_a\cdot S\ +\ \coeff{1}{16\pi^2}f_a^\one(M)\ ,
\eqn\fdef
$$
where the second term on the right hand side is completely
determined at the one-loop level of the perturbation theory
and suffers from no higher-order corrections.
Again, we see that the string theory upholds the no-renormalization theorems
of the supersymmetric field theory, where the {\sl Wilsonian} gauge couplings
do not suffer from either infinite or finite corrections beyond one loop.
\refmark{\SVa}
We also see that the entire moduli dependence of the $f_a$
is controlled by the one-loop effects;
this result is fundamental for the present investigation.

Note that the stringy no-renormalization theorems for the superpotential
and for the Wilsonian gauge couplings depend on the anomalous Peccei-Quinn
symmetry and thus are purely perturbative in nature.
Non-perturbatively, the continuous Peccei-Quinn symmetry is broken
down to its anomaly-free discrete subgroup, whose invariants include the
holomorphic exponential $\exp(-8\pi^2 S)$ of the dilaton superfield.
Thus, beyond the perturbation theory, one expects
$$
W(\matter,\mod,S)\ =\ W_{\rm tree}(\matter,\mod)\
+\ W_{\rm NP}\bigl(\matter,\mod,e^{-8\pi^2 S}\bigr)
\eqn\Wcont
$$
and similar corrections to the Wilsonian gauge couplings $f_a(S,\mod)$.

\section{The Effective Gauge Couplings and their Dilaton Dependence.}
An effective quantum field theory has
two kinds of gauge couplings (as well as Yukawa couplings, \etc):
The momentum-dependent effective gauge couplings
$g_a(p^2)$, which are directly related to  physical quantities such as
scattering amplitudes, and
the Wilsonian gauge couplings $f_a$, which have no such direct physical
meaning but rather serve as the input parameters of the EQFT.
{}From the renormalization theory's point of view, the Wilsonian couplings
are couplings of the EQFT from which the high-energy degrees of freedom
are integrated out but the low-energy quantum operators are left as they are.
On the other hand, the effective couplings account for all the quantum
effects, both high-energy and low-energy.
In the previous section, we discussed the Wilsonian gauge couplings whose
moduli- and dilaton-dependence is severely constrained by the holomorphicity
and by the no-renormalization theorems.
Let us now turn our attention to the effective gauge couplings~$g_a(p^2)$.

In general, one calculates the effective couplings of an EQFT order-by-order
in the perturbative expansion,
by summing up the appropriate Feynman graphs of the regularized theory.
According to Shifman and Vainshtein,\refmark{\SVa}
the effective gauge couplings
of a rigidly-supersymmetric gauge theory
can be calculated exactly in terms of the Wilsonian gauge couplings and
the effective normalization matrix $Z_{\Ib J}^{\rm eff}(p^2)$ for the
charged matter superfields.
\foot{%
    The effective normalization matrix $Z_{\Ib J}^{\rm eff}(p^2)$
    itself is obtained from the perturbative 1PI 2-point Green's functions
    for the matter superfields $\matterb^\Ib$ and $\matter^J$.
    }
In ref.~\KLb, we extended their formula to the locally supersymmetric
EQFTs, in which
$$
\displaylines{
g_a^{-2}(p^2)\ =\ \Re f_a\ +\ {\cren_a\over 16\pi^2}\,\log{\Lambda^2\over p^2}
\hfill\eqname\loopgf\cr
\hfill{}+\ {\cano_a\over 16\pi^2}\,\Kmod\
+\ {T(G_a)\over 8\pi^2}\,\log g_a^{-2}(p^2)\
-\sum_r{T_a(r)\over 8\pi^2}\,\log\det Z_{(r)}^{\rm eff}(p^2) ,\qquad\cr }
$$
where $r$ runs over the representations of the gauge group $G=\prod_a G_a$,
$$
\eqalign{
T_a(r)\ & =\ \Tr_r( T_{(a)}^2 )\quad{\rm for}\quad T_{(a)}\in G_a ,
	\cropen{1\jot}
T(G_a)\ & =\ T_a({\rm adjoint\ of\ } G_a),\cropen{1\jot}
\cren_a\ &=\sum_r n_r T_a(r)\ -\ 3T(G_a),\cr
\cano_a\ &=\sum_r n_r T_a(r)\ -\  T(G_a).\cr
}\eqn\groupdefs
$$
$n_r$ is the number of the matter multiplets in the representation $r$ and
$Z_{(r)}^{\rm eff}$ is the block of the effective normalization matrix
$Z_{\Ib J}^{\rm eff}$
referring to the ``flavor'' indices of those matter multiplets.
Finally, $\Lambda$ is the nominal UV cutoff scale of the regularized EQFT;
to assure that the functions $f_a(\modul)$ correctly
represent the moduli dependence
of the Wilsonian gauge couplings, $\Lambda$ must be independent of all the
moduli.
To be precise from the supergravitational point of view,
$\Lambda$ is constant in Planck units;
thus, without  loss of generality, we may set $\Lambda=\mpl$.

Eq.~\loopgf\ holds to all orders of the perturbation theory, but actual
evaluation of the effective gauge couplings
in string based EQFTs to the $n$-loop order
requires knowing the K\"ahler function $\Kmod$ and the $Z_{\Ib J}^{\rm eff}$
to the $(n-1)$-loop order.
Hence, beyond the one-loop level, analytic study of the moduli dependence
of the effective gauge couplings is rather difficult, but
at the one-loop level, we can use the tree-level approximations for the
terms on the second line of eq.~\loopgf.
Thus, inserting eqs.~\dilatonk\ and \fdef\ into \loopgf, we learn that
the moduli and dilaton dependence of the gauge couplings for any $N=1$
heterotic string vacuum is given by
$$
\displaylines{
g_a^{-2}(p^2;\modul,\modulb)^\ol\,=\
k_a\,\Re S\ +\ {\cren_a\over 16\pi^2}\left( \log{\mpl^2\over p^2}\,
	-\,\log(S+\Sb)\right)
\hfill\qquad\eqname\oneloopgf\cr
\hfill{}+\ {1\over 16\pi^2}\left[
\Re f_a^\one(M)\ +\ \cano_a \KM(\mod,\modb)\
-\sum_r 2T_a(r)\,\log\det Z_{(r)}^{\rm tree}(\mod,\modb) \right]
\quad\cr }
$$
plus a numerical constant of the order $O(1/16\pi^2)$.
(In this article we study the field dependence
  of the gauge couplings and disregard any constant terms.
  However, such terms are important in determining the unification properties
  of the gauge couplings.)

There are two dilaton-dependent terms on the right hand side of eq.~\oneloopgf:
The $k_a\Re S$ term, which is the tree-level coupling,
and the $\log(S+\Sb)$ term, which arises at the one-loop level.
Remarkably, for any gauge coupling of any
string-based $N=1$ supersymmetric EQFT, the coefficient
of the latter one-loop term is precisely the coefficient $\cren_a/16\pi^2$
of the one-loop beta-function.
Hence, the dilaton dependence of the one-loop corrections to the effective
gauge couplings amounts to a universal change of the couplings' unification
scale:
The natural starting point for the renormalization
of the {\sl effective} gauge couplings is not the Planck scale but rather
$\mpl/\sqrt{S+\Sb}\sim
\gstring\mpl\sim(\alpha')^{-1/2}\sim\mstring$.\refmark{\GHMR\,}
\foot{%
  Notice that the dilaton dependence of this string scale
  is a matter of convention.
  {}From the effective supergravity point of view, the Planck
  scale is field-independent while $\mstring\propto (S+\Sb)^{-1/2}$,
  but in the string theory, it is natural to use the string scale as a
  field-independent unit of mass while $\mpl\propto(S+\Sb)^{+1/2}$.
  }
On the other hand, the Wilsonian gauge couplings~\fdef\ unify at the
Planck scale rather than at the string scale.

Actually, this ``disagreement'' as to whether the string threshold is at
$\mstring$ or at $\mpl$ is similar to what
happens at any threshold of a supersymmetric EQFT
and has nothing specifically stringy about it.
Indeed, at in ordinary GUT or at an intermediate-scale threshold,
the effective gauge
couplings of the unbroken part of the gauge group measure the threshold scale
in terms of the physical masses of the heavy gauge bosons and other charged
particles.\refmark{\SW}
On the other hand, the Wilsonian gauge couplings are sensitive to the
{\sl unnormalized} Higgs VEVs\refmark{\KLb} rather than to the physical masses;
furthermore, it is the unnormalized Higgs VEVs that control the residual,
non-renormalizable ``weak'' interactions due
to the broken part of the gauge group.
Thus, for every threshold we have two distinct threshold scales, which differ
from each other by a factor proportional to the gauge coupling.
This is precisely what happens at the string threshold:
The physical masses of the massive string modes are proportional to
$\mstring$ but the strengths of the non-renormalizable
interactions below the string threshold, including the gravitational coupling
$\kappa$, are proportional to  powers of $\gstring/\mstring\sim 1/\mpl$
rather than simply powers of $1/\mstring$.
Furthermore, whenever some particles can be either light or heavy depending on
some moduli VEVs, in order for the physical masses of those particles to be
of the order $\mstring$, the {\sl unnormalized} VEVs of the Higgs-like moduli
have to be of the order $O(\mpl)$.
Thus,  the effective gauge couplings ``feel'' the string threshold
at $\mstring$, while the Wilsonian couplings register that threshold at
the Planck scale.

Thus far, we have discussed the effective gauge couplings $g_a$ from the
point of view of the string-based EQFT.
However, the same couplings can be calculated directly in the perturbative
string theory.
At the one-loop level, the result can be generally expressed
in terms of eqs.~\oloopgst.
Emphasizing the moduli- and dilaton-dependence of all the terms,
we have\refmark{\VKb}
$$
g_a^{-2}(p^2;S,\Sb,\mod,\modb)\
=\ k_a\gstring^{-2}(S,\Sb,\mod,\modb)\
+\ {\cren_a\over 16\pi^2}\,\log{\mstring^2\over p^2}
+\ {\Ddif_a (\mod,\modb)\over 16\pi^2} ,
\eqn\oneloopgst
$$
where, at the required level of accuracy, $\mstring^2$ is indeed given by
the $\mpl^2/(S+\Sb)$ times a numerical constant.
The dilaton-dependence of the ``universal'' coupling $\gstring$ follows
from the usual loop-counting arguments:
At the tree level, $\gstring^{-2}=\Re S$, while at the one-loop level
we have eq.~\unicorr.
The non-universal (\ie, gauge group dependent)
string-threshold corrections $\Ddif_a $ follow from the
spectrum of the massive modes of the heterotic string:\refmark{\VKb}
$$
  \Ddif_a \ = \int_\Gamma\!{d^2\tau\over\tau_2}\,
  \left( \B_a(\tau,\bar\tau)\,-\,\cren_a  \right) ,
 \eqn\mainresult
$$
where the domain of integration $\Gamma$ is the fundamental domain
for the modulus $\tau$ of the world-sheet torus
and
$$
\B_a\ =\ {2\over \left|\eta(\tau)\right|^4}\,
\sum_{\rm even\,\bf s} (-)^{s_1+s_2}\,
{d Z_\Psi({\bf s},\bar\tau) \over 2\pi i\, d\bar\tau}
\Tr_{s_1}\!\left( \left( T^2_{(a)} - \coeff{k_a}{8\pi\tau_2} \right)
     (-)^{s_2 \Fb} q^{L-{11\over12}} {\bar q}^{\bar L-{3\over8}}
     \right)_{\rm int} .
\eqn\beq
$$
Here $T_{(a)}$ is a generator of the gauge group $G_a$, $q=e^{2\pi i\tau}$,
${\bf s}=(s_1,s_2)$ denotes the NSR boundary conditions for the fermions on the
supersymmetric side of the world sheet, $\Fb$ is their fermion number
and $Z_\Psi$ is the partition function of a free complex Weyl fermion;
the $-{k_a\over 8\pi\tau_2}$ term\refmark{\AGN}
is included for the sake of the modular
invariance of the functions $\tau_2\B_a(\tau,\bar\tau)$.
\foot{%
    The combination $( T^2_{(a)} - \coeff{k_a}{8\pi\tau_2})$ emerges
    when the properly regularized Kac-Moody current-current correlator
    $\vev{J_{(a)}(\zeta)J_{(a)}(0)}$ is integrated over the world
    sheet.\refmark{\AGN}
    The $-{k_a\over 8\pi\tau_2}$ term is obviously universal with respect
    to the gauge couplings and thus can be dropped from eq.~\beq\ while
    its effect is absorbed into $\Duniv$ ---
    this is exactly what was done in ref.~\VKb.
    However, it is not the only universal threshold
    correction to all the gauge couplings,
    so in addition to the $-{k_a\over 8\pi\tau_2}$ term in eq.~\beq,
    we also retain the $\Duniv$ term in eq.~\unicorr\ in order to account
    for the other universal corrections.
    Later in this section, we will show that it is this latter $\Duniv$ term
    which is related to the Green-Schwarz term $V^{(1)}$ in eq.~\dilatonkloop.
    }
The trace in eq.~\beq\ is taken over the internal $c=(22,9)$ sector of the
world-sheet SCFT;
in general, it depends on all the moduli of that internal sector, but not
on the dilaton (which originates in the spacetime sector of the SCFT).
Consequently, the  string-threshold corrections
$\Ddif_a (\mod,\modb)$ depend on the string moduli $\mod^i$ and $\modb^\ib$
but not on the dilaton field $S$.
Thus, we conclude that {\sl in string theory, the dilaton dependence of the
effective gauge couplings $g_a$ is exactly as in the field-theoretical
formula}~\oneloopgf.\refmark{\MSb}

\section{Moduli Dependence of the Gauge Couplings}
Having discussed the dilaton dependence of the gauge couplings
we now turn our attention to their dependence on the moduli $\mod^i$
originating in the internal sector of the world-sheet SCFT.
Let us compare the one-string-loop formula \oneloopgst\ for the effective
gauge couplings $g_a$ with the one-loop EQFT formula \oneloopgf.
We have already seen that the dilaton-dependent parts of the two formul\ae\
agree with each other.
To assure agreement between the $S$-independent but moduli-dependent parts
of eqs.~\oneloopgst\ and \oneloopgf, we now need
$$
\Re f_a^\one(M)\
=\ \Dtot_a (\mod,\modb)\
-\ \cano_a \KM(\mod,\modb)\
+ \sum_r 2T_a(r)\,\log\det Z_{(r)}^{\rm tree}(\mod,\modb)
\eqn\fdet
$$
(up to an $O(1)$ numerical constant),
where $\Dtot_a$ are as in eq.~\Ddef.
The obvious meaning of  eqs.~\fdet\ is that they are formul\ae\
for the one-loop
corrections to the {\sl Wilsonian} gauge couplings $f_a$ in terms of
quantities computable in string theory.
The first term on the right hand side originates at the one-loop
level of the heterotic string theory;
the other two terms subtract the one-loop corrections arising in the low-energy
EQFT and thus are computable in terms of the tree-level properties of
the string.
Despite the one-loop-approximate nature of these right-hand terms,
the left-hand side is protected from any higher-loop corrections.
Thus, {\it as far as the Wilsonian gauge couplings $f_a$ are concerned,
eqs.~\fdet\ are exact}
to all orders of the perturbation theory.

Manifest supersymmetry of the low-energy EQFT's Wilsonian Lagrangian
requires holomorphicity of the functions $f_a^\one(\mod)$.
At the same time, none of the terms on the right hand side of eqs.~\fdet\ is
--- or has any a~priori reason to be ---
a harmonic function of the moduli $\mod^i$ and $\modb^\ib$;
mutual cancellation of their non-harmonic parts is a non-trivial constraint.
Thus, eqs.~\fdet\ impose powerful supersymmetric consistency conditions
on $\Dtot_a$
in terms of the tree-level couplings:
$$
\partial_\mod\partial_{\modb}\,\Dtot_a(\mod,\modb)\
=\ \partial_\mod\partial_{\modb}\left( \cano_a\KM(\mod,\modb)\
-\smash{\sum_r}2\Tr_a(r)\,\log\det Z_{(r)}^{\rm tree}(\mod,\modb)\right) .
\eqn\nonharmon
$$
Eqs.~\nonharmon\ apply to all $(d=4,N=1)$--supersymmetric vacuum families
of the heterotic string;
in the following section, we shall verify that they indeed hold true
for the orbifolds and for the large-radius Calabi-Yau manifolds.

Eqs.~\nonharmon\ follow from the requirement of describing the physics of
energies below the string threshold in terms of a locally supersymmetric EQFT.
There are other constraints that follow from supersymmetry of the S-matrix
and of the Green's functions without any reference to an EQFT.
In particular, we argued in ref.~\DKLb\ that the moduli dependence
of the effective gauge couplings $g_a^{-2}(\modul,\modulb)$ is related to the
effective axionic couplings of the moduli scalars:
The CP-odd part of the Green's function for two gauge bosons of $G_a$ and a
modulus scalar $\modul^i$ is proportional to
$\partial g_a^{-2}/\partial\modul^i$.
This relation must hold true in either field theory or string theory,
and we shall use it momentarily  to relate the universal part
$\Duniv$ of the threshold corrections \Ddef\ to the one-loop
mixing \dilatonkloop\
of the dilaton superfield $S$ with the moduli superfields $\mod^i$.

For generic vacuum families of the heterotic string,
the CP-odd one-string-loop scattering amplitudes were calculated by
Antoniadis, Gava and Narain.\refmark{\AGN}
They found
$$
\eqalign{
{\cal A}^{-}(A^\mu,A^\nu,\mod^i)\ &
=\ {+i\epsilon_{\mu\alpha\nu\beta}p_1^\alpha p_2^\beta\over16\pi^2}\times
\partder{\Ddif_a }{\mod^i}\,,\cr
{\cal A}^{-}(A^\mu,A^\nu,\modb^\ib)\ &
=\ {-i\epsilon_{\mu\alpha\nu\beta}p_1^\alpha p_2^\beta\over16\pi^2}\times
\partder{\Ddif_a }{\modb^\ib}\,,\cr
}\eqn\modaxionic
$$
where $\Ddif_a(\mod,\modb)$ are given by eq.~\mainresult\ in which
$\B_a(\tau,\bar\tau)$ of eq.~\beq\ are replaced with
$$
\B_a^{\rm AGN}(\tau,\bar\tau)\ =\ {-1\over \eta^2(\tau)}\,
\Tr_{\Rb}\left( (-)^{\Fb-{3\over2}}\,\Fb\
\left( T^2_{(a)} - \coeff{k_a}{8\pi\tau_2} \right)
q^{L-{11\over12}}\ {\bar q}^{\bar L-{3\over8}} \right)_{\rm int}  \, ,
\eqn\ramondright
$$
 involving the odd Ramond-Ramond sector rather than the 3 even sectors.
\foot{%
    The right-moving fermion number operator $\Fb$ has half-integer values
    in the Ramond sector, hence the $(-1)^{\Fb-{3\over2}}$ sign factor
    for the Ramond-Ramond boundary conditions.
    However, by abuse of notations, this factor is commonly written as
    simply $(-1)^{\Fb}$.
    }
Actually, for all spacetime-supersymmetric vacua $\B^{\rm AGN}_a=\B_a$, ---
this is one of the Riemann identities between characters of
different NSR sectors of $(0,2)$ superconformal algebras;
this particular identity is proven in Appendix~A.

There is a subtle difference between the string-theoretical and
the field-theoretical axionic couplings of the moduli:
In field theory, the axionic couplings related by the spacetime supersymmetry
to the derivatives of the the effective gauge couplings~\oneloopgst\
are 1PI Green's functions,
but the string-theoretical axionic amplitudes~\modaxionic\
are fully-dressed scattering amplitudes.
Diagrammatically, the relation between these fully-dressed amplitudes
and the 1PI Green's functions is
$$
\checkex dressing.eps \iffigureexists
\vcenter to 1.5in{\kern 4pt\hbox{$A^\mu$}\vfil\hbox{$A^\nu$}\kern 4pt}
\vcenter{\epsfysize 1.5in \epsfbox{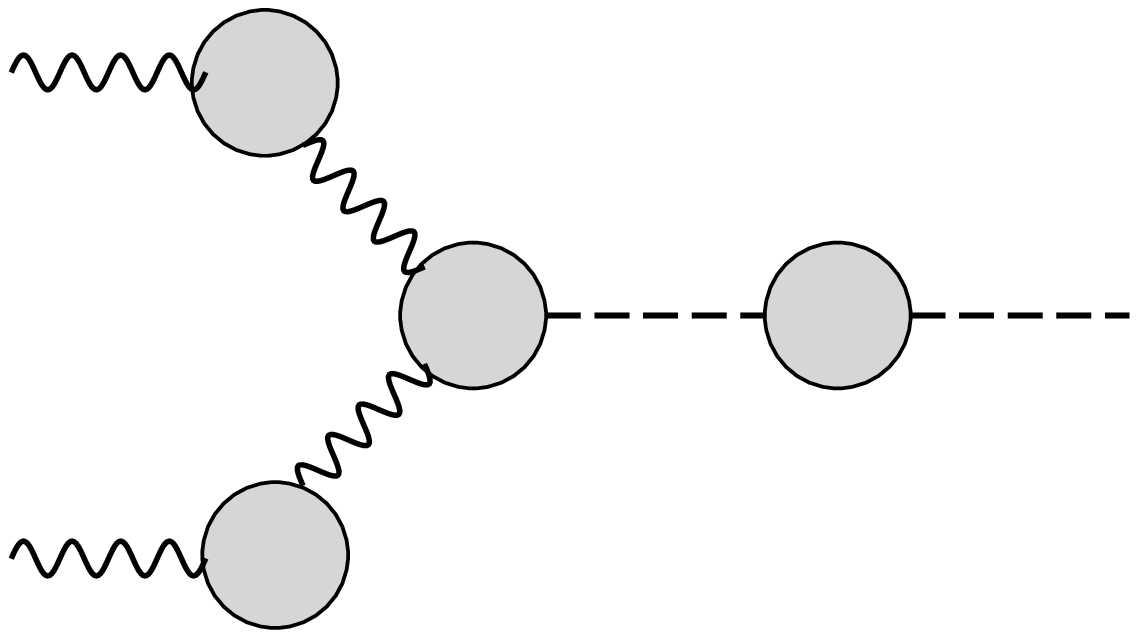}}
\ \mod {\rm\ or\ }\modb ,
\else
\hbox{\bf Missing Diagram},
\fi
\eqn\dressing
$$
which at the one-loop level reduces to
$$
\checkex oneloop1.eps \iffigureexists \checkex oneloop2.eps \fi
\iffigureexists
\vcenter to 1in{\kern -3pt\hbox{$A^\mu$}\vfil\hbox{$A^\nu$}}
\vcenter{\epsfysize 1in \epsfbox{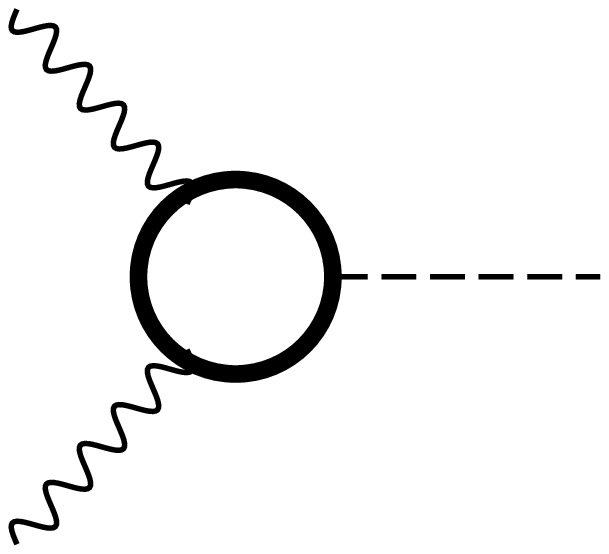}}
\ \mod {\rm\ or\ }\modb
\qquad{\bf +}\qquad
\vcenter to 1in{\kern -3pt\hbox{$A^\mu$}\vfil\hbox{$A^\nu$}}
\vcenter{\epsfysize 1in \epsfbox{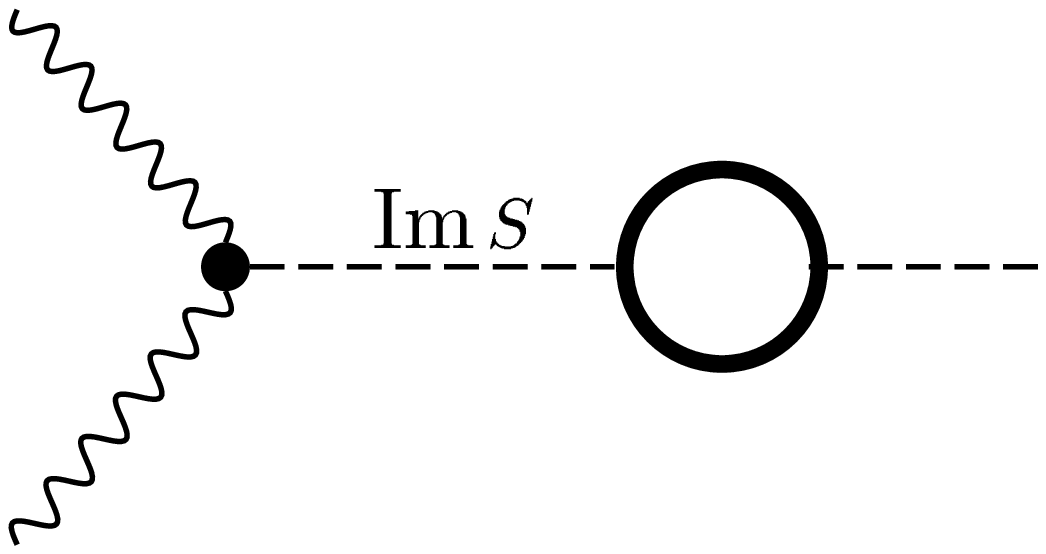}}
\ \mod {\rm\ or\ }\modb .
\else
\hbox{\bf Missing Diagrams}.
\fi
\eqn\oloopdressing
$$
Note that the virtual particle in the second diagram here has to be the
axion $\Im S$ because it is the only scalar with tree-level axionic couplings
to the gauge bosons.
At the one-loop level, the mixing of the axion $\Im S$ with the moduli $\mod^i$
and $\modb^\ib$ is controlled by the ``Green-Schwarz''
term $V^{(1)}(\mod,\modb)$ (\cf~eq.~\dilatonkloop);
thus, in light of eqs.~\oneloopgst\ and \unicorr,
$$
\eqalign{
{\cal A}^{-}(A^\mu,A^\nu,\mod^i)\ &
=\ +i\epsilon_{\mu\alpha\nu\beta}p_1^\alpha p_2^\beta
    \left(\partder{g_a^{-2}}{\mod^i}
	+{k_a\over16\pi^2}\partder{V^{(1)}}{\mod^i}\right)\cr
&=\ {+i\epsilon_{\mu\alpha\nu\beta}p_1^\alpha p_2^\beta\over 16\pi^2}
    \times\partder{}{\mod^i}
	\left( \Ddif_a + k_a\Duniv + k_a V^{(1)}\right) ,\cr
{\cal A}^{-}(A^\mu,A^\nu,\modb^\ib)\ &
=\ -i\epsilon_{\mu\alpha\nu\beta}p_1^\alpha p_2^\beta
    \left(\partder{g_a^{-2}}{\modb^\ib}
	+{k_a\over16\pi^2}\partder{V^{(1)}}{\modb^\ib}\right)\cr
&=\ {-i\epsilon_{\mu\alpha\nu\beta}p_1^\alpha p_2^\beta\over 16\pi^2}
    \times\partder{}{\modb^\ib}
	\left( \Ddif_a + k_a\Duniv + k_a V^{(1)}\right) .\cr
}\eqn\dressedamplitudes
$$
Comparing {\sl these} amplitudes with the string amplitudes \modaxionic,
we now conclude that the spacetime supersymmetry
is indeed consistent, provided
$$
\Duniv(\mod,\modb)\ =\ -V^{(1)}(\mod,\modb)
\eqn\GSDuniv
$$
(up to a moduli-independent constant).
In other words, at the one-loop level, $\gstring^{-2}=\Re S
-(1/8\pi^2)V^{(1)}(\mod,\modb)+\cdots$, in precise agreement with
the definition of the universal string coupling
in the linear multiplet formalism.
\refmark{\DFKZa-\GT,\AGN,\KK}

Ref.~\AGN\ gives an explicit string-theoretical formula for the Green-Schwarz
function $V^{(1)}(\mod,\modb)$, but
the supersymmetric consistency conditions \nonharmon\ allow us to
obtain this function without any additional string-theoretical calculations.
Indeed, eqs.~\nonharmon,  \Ddef\ and \GSDuniv\ together imply
$$
\partial_\mod\partial_{\modb}\,V^{(1)}\
=\ k_a^{-1}\,\partial_\mod\partial_{\modb}\left(
\Ddif_a\ -\ \cano_a\KM\
-\smash{\sum_r}2\Tr_a(r)\,\log\det Z_{(r)}^{\rm tree}\right) ,
\eqn\GSdet
$$
where $\KM$ and $Z_{(r)}^{\rm tree}$ are determined at the tree level of
the string theory while $\Ddif_a$ is computed via
eqs.~\mainresult\ and \beq\ (or \ramondright).
Eq.~\GSdet\ determines the Green-Schwarz function $V^{(1)}(\mod,\modb)$ up to
an arbitrary harmonic function  $H(\mod)$,
$$
V^{(1)}(\mod,\modb)\ \to\ V^{(1)}(\mod,\modb)\ +\ H(\mod)\ +\ H^*(\modb) .
\eqn\GStrans
$$
This remaining indeterminacy is related to a fact that unlike all other fields
of the low-energy EQFT, the chiral dilaton superfield $S$ has no fixed relation
to vertices of the string theory.
Thus, we are free to re-define
$$
S\ \to\ S\ +\ \coeff{1}{8\pi^2}\,H(\mod)
\eqn\diltrans
$$
as long as $H$ is a holomorphic function of the chiral moduli $\mod^i$.
This redefinition naturally affects the analytic form of the \K\ function
\dilatonkloop; at the one-loop level, the effect is precisely~\GStrans.

\chapter{Field Theoretical Constraints for Orbifolds}
Thus far our discussion of the moduli dependent gauge couplings
was completely generic;
we gave a general formula for the moduli and dilaton dependence of the gauge
couplings (eqs.~\oneloopgf)
and outlined how to compute this moduli dependence in string theory
(eqs.~\mainresult, \ramondright\ and \GSdet).
However, for many families of the heterotic string vacua, their special
properties may be used to severely constrain the holomorphic functions
$f_a^\one(\mod)$
and sometimes determine their exact analytic forms  (up to a constant)
without performing any string-loop calculations.
There are several sources of such constraints:
The geometry of the moduli space of a vacuum family controls the locations
of all the singular points (or subspaces) of the functions $f_a^\one(\mod)$
and the types of the respective singularities.
Perturbative consistency of the EQFT imposes limits on the growth
of the $f_a$ along asymptotic directions such as infinite radius.
Finally, vacuum families often have exact discrete symmetries,
which also impose
constraints on the analytic form of the $f_a(\mod)$.

The effective gauge couplings $g_a(p^2)$ are physical quantities and hence
must remain invariant under all the exact symmetries of the theory.
The transformation properties of the Wilsonian gauge couplings $f_a$
are not so obvious because these couplings act as
counterterms cancelling potential
anomalies of the EQFT.\refmark{\louisa-\CLO,\KLb}
Specifically, there are two supersymmetrized Adler-Bell-Jackiw
anomalies at play:
The Konishi anomaly arising when charged chiral matter superfields
are mixed with each other,\refmark{\konishi}
$$
\matter^I\ \to\ \symmat^I_{\,J}(\modul)\,\matter^J\ +\ O(\matter^2/\mpl),
\eqn\mattertrans
$$
and the anomaly of the \K\ transformations
$$
\Kmod\ \to\ \Kmod\ +\ \KF(\modul)\ +\ \KF^*(\modulb),
\qquad W\ \to\ W\times\exp(-\KF(\modul)) ;
\eqn\kahlertrans
$$
cancellation of the combined anomaly requires\refmark{\KLb}
$$
f_a\ \to\ f_a\ -\ {\cano_a\over 8\pi^2} \KF(\modul)\
-\sum_r {T_a(r)\over 4\pi^2}\,\log\det\symmat^{(r)}(\modul) .
\eqn\fLocal
$$

In this section we are going to demonstrate the power of the
field-theoretical constraints on the  Wilsonian gauge couplings.
We shall see that for many orbifolds vacua of the heterotic string
\refmark{\DHVW} such constraints uniquely determine
the functions $f_a^\one(\mod)$;
furthermore, when we restate the results in terms of the moduli dependence
of the physical threshold corrections $\Dtot_a$,
we shall find the latter in complete agreement with the string-theoretical
threshold corrections computed in ref.~\DKLb.
To facilitate this comparison, we focus on exactly the same class of
factorizable abelian $(0,2)$ orbifolds as were discussed in ref.~\DKLb,
although we briefly return to more general orbifolds
at the end of this section.

The best way to describe a generic factorizable orbifold is to build one.
As usual, we begin with a toroidal compactification of the ten-dimensional
heterotic string.
At this stage, we do not allow any Wilson lines, discrete or continuous;
instead, we keep the six internal dimensions completely separate from
the $E_8\times E_8$ degrees of freedom.
Moreover, we split the six internal dimensions into three orthogonal planes
and compactify each plane into a separate two-torus.
The purpose of this restriction is to simplify the moduli space of the theory:
as long as we keep the above constrains, we have two complex moduli $\moo^i$
and $\mot^i$ for each of the three planes ($i = 1,2,3$);
furthermore, there is a separate
$\sltwo$ duality symmetry for each of the six moduli
$\moo^i$ and $\mot^i$.\refmark{\GPR}

At the second stage, we twist the theory by a discrete symmetry group;
this may require freezing of some or all of the $\mot^{i}$ moduli.
We insists that all the group elements avoid mixing the planes
but rotate each plane onto itself;
together with the need to preserve $N=1$ spacetime supersymmetry,
this requirement limits us to the abelian twist groups only.
Finally, we limit the asymmetry of the twists by requiring all rotations
of the three internal planes to be symmetric with respect to the left-moving
and right-moving bosonic operators comprising each plane.
However, beyond the constraints of modular invariance,
we do not ask for any relation
between the twisting of fermionic superpartners $\psi^i$
of the internal coordinates $X^i$
and the way the same twist acts upon the $E_8\times E_8$ degrees of freedom.
Thus, we are not limited to the completely symmetric $(2,2)$ orbifolds
but allow for a rather large class of the $(0,2)$ orbifolds, and
our analysis of the moduli dependent gauge couplings applies
to all the gauge symmetries such an orbifold might have.

In the literature, the term ``modulus'' has an ambiguous meaning in the
orbifold context.
Here we consider  an exact flat
direction of the scalar potential  to be a modulus if and only
if one can vary its expectation value without changing the spectrum of
the light fields that appear in the low-energy EQFT.
In particular, all the moduli must be neutral with respect to all the
low-energy gauge symmetries.
For the purposes of this article, we  keep in the low-energy
EQFT all the gauge symmetries originating in the $E_8\times E_8$
(or $D_{16}$) world-sheet degrees of freedom.
A twisted state of an abelian orbifold is generally charged under at least
one of those gauge symmetries and thus should not be considered a modulus
even if it happens to parametrize an exactly flat direction of the scalar
potential.
Disregarding possible exceptions to this rule,
\foot{If a $(0,2)$ orbifold has sectors where the $X^i$ (and the $\psi^i$)
    are twisted but the $E_8\times E_8$ (or $D_{16}$) degrees of freedom
    remain completely untwisted, then such sectors may give rise to neutral
    massless scalars.
    It is not known if any such neutral twisted scalars have exactly flat
    potentials; if they do, they should be regarded as moduli,
    but we are not going to discuss them  any further in this article.
    }
we limit the present analysis of the moduli-dependent couplings to the
untwisted, \ie, toroidal moduli of factorizable orbifolds.
Furthermore, for the sake of notational simplicity, we  concentrate
on the diagonal moduli $\moo^i$ and $\mot^i$ of the tree two-tori,
although  our analysis could be straightforwardly generalized to include
the off-diagonal toroidal moduli of the $Z_3$, $Z_4$ and $Z'_6$ orbifolds.

On the other hand, treating all the $\moo^i$ and $\mot^i$ as moduli means
that we may only consider the couplings of the gauge symmetries that
originate in the $E_8\times E_8$ (or the $D_{16}$) and thus remain
unbroken for generic values of the $\moo^i$ and $\mot^i$.
For some special values of these moduli, the momenta/windings of the six-torus
give rise to additional massless vector bosons\refmark{\DHVW},
but the couplings associated with such ``accidental'' gauge symmetries do not
belong in the  low-energy EQFT which treats all of the $\moo^i$ and $\mot^i$
as moduli.
Instead, they belong in the EQFT that includes  the ``accidentally'' light
particles and re-interprets some of the $\moo^i$ and $\mot^i$ as charged
Higgs fields rather than moduli.
However, in this article we only concentrate on the more generic gauge
couplings.

Having defined the factorizable orbifolds and their gauge couplings,
let us now consider the symmetry constraints on the $f_a$.
Under the modular symmetries, the toroidal moduli of
an orbifold transform according to\refmark{\GPR}
$$
\mod\ \to\ {aM-ib\over icM+d}\ ,\quad
{\Tenpoint \pmatrix{a&b\cr c&d\cr}}\in\sltwo,
\eqn\modtrans
$$
\ie, $a,b,c,d\in{\bf Z}$ and $ad-bc=1$;
the $\mod$ here is either a $\moo^i$ or a $\mot^i$ and there is a separate
$\sltwo$ matrix for each such modulus.
The tree-level \K\ function for the toroidal moduli of any factorizable
orbifold has the form\refmark{\Witten,\kahleref,\DKLa}
$$
\KM(\mod,\modb)\ =\ -\sum_i \log(\mod^i+\modb^i)
\eqn\orbikahler
$$
where the sum is over all the toroidal moduli;
under the symmetries \modtrans,
this \K\ function transforms according to eq.~\kahlertrans\
with
$$
\KF(\mod)\ =\ \sum_i\log(ic_i\mod^i+d_i).
\eqn\orbiKtrans
$$
Note that although the tree-level \K\ function \orbikahler\ is corrected
by  string loops, the holomorphic function \orbiKtrans\ has to be exact
to all orders of the perturbation theory:
This follows from the fact that the Wilsonian superpotential~$W$
in eq.~\kahlertrans\ is protected from any perturbative renormalization.

The same argument can be used to obtain the exact transformation properties
of the matter fields themselves from the tree-level matter normalization matrix
$Z^{(0)}_{I\Jb}(\mod,\modb)$.
This matrix has to be calculated directly from the string theory;
besides the  spectrum of the light particles, it is the only model-dependent
string-theoretical data we need for our purposes.
The calculation is performed in  Appendix~B
(see also refs.~\DKLa, \IL\ and \AGN); the result is
$$
Z^{(0)}_{I\Jb}(\mod,\modb)\
=\ \delta_{I\Jb}\times\prod_i\left(\mod^i+\modb^i\right)^{-q^i_I} ,
\eqn\metricorbi
$$
where the exponents $q^i_I$ are rational numbers depending on the twist
sector giving rise to a matter field $\matter^I$, on the angle by which the
internal $X^i$ coordinate is rotated by that twist and on the presence
of the $\partial X^i$ or $\partial \Xb^i$ world-sheet operators in the
vertex for the $\matter^I$.
In Appendix~B we give an explicit formula for all the $q^i_I$, but for
the present discussion we only need  the general form of  eq.~\metricorbi.
The transformation of $Z^{(0)}_{I\Jb}$  under
the $\sltwo$  symmetries follow from eqs.~\metricorbi\ and \modtrans\
and in turn determine the transformation rules for the matter fields:
$$
\matter^I\ \to\ \widehat\symmat^I_{\,J}\matter^J
\times \prod_i (ic_i\mod^i+d_i)^{-q^i_I}
\eqn\matterorbi
$$
where $\widehat\symmat^I_{\,J}$ is a moduli-independent unitary matrix.
Again, although the tree-level normalization matrix \metricorbi\
suffers from both  field-theoretical and  string-theoretical higher order
corrections, the transformation rules \matterorbi\ are exact to all orders
of the perturbation theory.

According to eq.~\fLocal, the modular transformation rules \orbiKtrans\ and
\matterorbi\ completely determine the behavior of the Wilsonian gauge couplings
$f_a$ under the same modular transformations.
Since the chiral dilaton superfield $S$ is defined by the string only up to
re-definitions \diltrans, we adopt a convention in which $S$ is completely
inert under all the modular transformations.
Hence, the entire transformation \fLocal\ is due to the Wilsonian threshold
corrections $f^1_a(\mod)$, which thus transform according to
\foot{This also follows from eqs.~\fdet.}
$$
f^1_a\ \to\ f^1_a\ - \sum_i 2\Alpha_a^i\,\log(ic_i\mod^i+d_i)
\eqn\forbitrans
$$
(modulo an imaginary constant), where
$$
\Alpha_a^i\ = \sum_I T_a(\matter^I)\,(1-2q^i_I)\ -\ T(G_a) .
\eqn\Alphadef
$$
Mathematically, eqs~\forbitrans\ resemble the modular transformation rules
for logarithm of the Dedekind's $\eta$ function, which
leads us to conclude that
$$
f_a^1(\mod)\ =\  -\sum_i 4\Alpha_a^i\,\log\eta(i\mod^i)\ + p_a(\mod),
\eqn\feta
$$
where $p_a$ are modular invariant (up to imaginary constants)  holomorphic
functions of the toroidal moduli $\mod^i$.
This form makes manifest the modular invariance of the physical threshold
corrections $\Dtot_a$.
Indeed, substituting eqs.~\orbikahler, \metricorbi\ and \feta\ into \fdet,
we arrive at
$$
\Dtot_a(\mod,\modb)\
= -\sum_i \Alpha_a^i\,\log\left( \left|\eta(i\mod^i)\right|^4\,
    \Re\mod^i\right)\ +\ \Re p_a(\mod) .
\eqn\Deta
$$

We are now going to argue that for factorizable orbifolds
$p_a(\mod)=\rm const$;
as a first step in this direction, let us consider possible
singularities of these functions.
As far as the perturbative string theory is concerned, a toroidal modulus
of a factorizable orbifold is either completely frozen by the twist group
or else can take any finite values in the right half of the complex plane
($\Re\mod>0$).
However, a perfectly regular string vacuum may lead to a singular EQFT if
some particles that are massive for generic values of the moduli become
massless at that particular point (or subspace) of the moduli space.
In factorizable orbifolds, this happens whenever
$\moo^i\equiv\mot^i\modulo\sltwo$,
at which point momenta/windings in the $X^i$ plane give rise to several
massless particles;
however, such ``accidentally massless'' particles are always completely
neutral with respect to any low-energy gauge symmetry originating in
the $E_8\times E_8$ (or $D_{16}$).
Hence, although some of the low-energy EQFT's couplings may become singular
when $\moo^i\equiv\mot^i\modulo\sltwo$,
the gauge couplings we are interested in
do not develop any singularities {\sl at the one-loop level}.
\foot{For a discussion of some of the singular couplings see ref.~\CLM.}
At higher-loop levels, these couplings may also become singular, but the
absence of the one-loop singularities is all we need to conclude that
the {\sl Wilsonian gauge couplings} $f_a$  have no singularities
anywhere in the moduli space.
This argument exemplifies the power of the no-renormalization theorem for the
Wilsonian gauge couplings $f_a$:
They are completely determined at the one-loop level and thus do not care
whether the non-harmonic terms have any higher-loop singularities.

The non-singular behavior of the Wilsonian gauge couplings implies
that the $p_a(\mod)$ in eqs.~\feta\ and \Deta\ are holomorphic functions
without any singularities for $\Re\mod>0$ and they are also modular
invariant (modulo imaginary constants) with respect to separate $\sltwo$
transformations~\modtrans\ for each modulus $\mod^i$.
Mathematically, these two constraints imply that $p_a(\mod)$ are polynomials
(or convergent power series) of $j(i\mod^i)$ where $j$ is the $\sltwo$
invariant function that maps the fundamental domain of the symmetry onto
the complex sphere.\refmark{\lehner}
The $j(i\mod)$ function is finite for any finite $\mod$
in the right half plane,
but it grows exponentially in the $\Re\mod\to\infty$ limit.
Hence, each $p_a$ is either entirely independent of a modulus $\mod^i$
or else it has to grow at least exponentially when $\Re\mod^i$ becomes large.

Now consider the physics of the $\Re\mod^i\to\infty$ limits:
In the $\Re\moo^i\to\infty$ limit, both periods of the two-torus for
the internal complex coordinate $X^i$ become very large.
In the $\Re\mot^i$ limit, one of the periods of the same two-torus becomes
very large while the other period becomes very small;
by duality, the physics of this
limit is the same as if both periods were very large.
Thus, in each of the $\Re\mod^i\to\infty$ limits,
the orbifold decompactifies and the four-dimensional low-energy EQFT
becomes rather singular.
However, we will show momentarily that the singularity of such a limit
is relatively mild; specifically, the gauge couplings do not grow larger than
$O(\Re\mod^i)$.

In order to obtain this bound, let us consider the following double limit of an
orbifold vacuum:  $\Re\moo^1\to\infty$, other $\mod^i$ fixed, $\Re S\to\infty$
while the ratio $\Re\moo^1/\Re S$ is kept finite and small.
Since all the physical couplings of the four-dimensional EQFT are proportional
to the negative powers of the dilaton $\Re S$, they are so small in this limit
that the effective theory below the compactification scale is essentially
classical.
Above the compactification scale $\sqrt{\alpha'/\Re\moo^1}$ of the  complex
coordinate $X^1$, the theory is effectively six-dimensional and its
loop-counting parameter is no longer simply $g_4^2\sim1/\Re S$ but rather
$g_6^2\sim\Re T^1/\Re S$.
This modified loop counting applies not just to the six-dimensional
field theory but to the string theory as well.\refmark{\VKa,\VT}
Thus, as long as $g_6^2$ remains sufficiently small,
we can use the perturbation theory at all energies.
Physically, this implies that the loop corrections to the four-dimensional
couplings should be small compared to their tree-level values.
In particular, we got to have
$$
|\Dtot_a|\ll\Re S\quad\hbox{as long as}\ g_6^2\ll 1,\ \ie,\
Re S\gg \Re\moo^1
\eqn\doublelim
$$
(note that all the other moduli $\mod^i$ are fixed here).
Since the one-loop threshold corrections $\Dtot_a$
do not depend on the dilaton,
the inequality \doublelim\ is nothing but a bound on the large $\Re\moo^1$
limit of the $\Dtot_a$, namely $|\Dtot_a|\le O(\Re\moo^1)$
in the large $\Re\moo^1$ limit.
Naturally, similar bounds
$$
|\Dtot_a|\ \le\ O(\Re\mod)
\eqn\Dbound
$$
apply to all the other $\Re\mod^i\to\infty$ limits.
As an immediate corollary of these bounds, we may finally eliminate
the functions $p_a(\mod)$ in eqs.~\feta\ and \Deta.
Indeed,
$$
-\Alpha_a^i\log\bigl(|\eta(i\mod^i)|^4\Re\mod^i\bigr)\
\approx\ \coeff\pi3 \Alpha_a^i\Re\mod^i\quad
{\rm for\ large}\ \Re\mod^i,
\eqn\Dboundsat
$$
in good agreement with the bounds \Dbound.
On the other hand, the $\Re p_a(M)$ terms in eq.~\Deta\ are either constant
or else they grow exponentially or even faster with $\Re\mod^i\to\infty$.
Having seen that the consistency of the perturbation theory does not allow
for such a rapid growth, we conclude that $p_a$ have to be moduli-independent
constants.

This completes our field-theoretical study of factorizable orbifolds
of the heterotic string.
We have used the following string-theoretical data as input:
The spectrum of the light particles, the {\sl tree-level} couplings
\orbikahler\ and \metricorbi\ and, most importantly, the knowledge
that the $\sltwo$ modular symmetries \modtrans\ are not merely symmetries
of the tree-level couplings but
 exact symmetries of the vacuum states of the heterotic string.
\foot{In general, when using a symmetry to restrict the form of the
    gauge (or other) couplings, it is extremely important to verify
    the symmetry at the string theoretical level
    since the tree-level low-derivative couplings often have apparent
    symmetries that do not survive string loop corrections.
    For example, the tree-level $\sigma$-model-like couplings \orbikahler\
    and \metricorbi\ of factorizable orbifolds are invariant under
    continuous $SL(2,{\bf R})$ symmetries, but only the discrete
    $\sltwo$ symmetries are true symmetries of the string theory.}
Given this string theoretical input, the field-theoretical constraints
then determine the Wilsonian gauge couplings to be exactly
$$
f_a(S,\mod)\ =\ k_a S\ - \sum_i{\Alpha_a^i\over 4\pi^2}\,\log\eta(i\mod^i)\
+\ \rm const
\eqn\ForbiF
$$
while the one-loop threshold corrections to the physical
gauge couplings are precisely
$$
\Dtot_a(\mod,\modb)\
= -\sum_i \Alpha_a^i\,\log\left( \left|\eta(i\mod^i)\right|^4\,
    \Re\mod^i\right)\ +\ \rm const.
\eqn\DorbiF
$$
In the large volume limit, these threshold corrections behave as
$$
\Dtot_a\ \approx\ {\pi\over3}\sum_i\Alpha_a^i\Re\mod^i\
\propto\ ({\rm Radius})^2
\eqn\orbilargerad
$$
(\cf\ ref.~\IN);
in the following section we shall see that for the smooth Calabi-Yau
compactifications the threshold corrections also behave like the square
of the radius in the large radius limit.

Having derived eqs.~\DorbiF\ from the field-theoretical arguments,
let us now compare them to the threshold corrections that follow
from the direct one-string-loop calculations
of ref.~\DKLb.
In string theory, the moduli dependence of the non-universal threshold
corrections $\Ddif_a(\mod,\modb)$ for an
orbifold with a twist group $D$ is determined not by that orbifold itself
but rather by the related orbifolds where the
same internal six-torus is twisted by the little subgroups $D_i$ of the
complex coordinates $X^i$.
That is, $D_i$ is comprised of the members of $D$ that do not rotate the $X^i$;
for example, for the $Z_6$ orbifold whose twist group $D$ is generated
by the rotation $\Theta=(e^{2\pi i/6},e^{2\pi i/3},-1)$, the little
subgroup $D_3$ of the third complex plane is a $Z_3$ generated by the
$\Theta^2=(e^{2\pi i/3},e^{4\pi i/3}, 1)$, the little subgroup $D_2$
of the second plane is a $Z_2$ generated by the $\Theta^3=(-1,+1,-1)$
and the little subgroup of the first plane is trivial.
When the original $D$-orbifold has $N=1$ spacetime supersymmetry,
the $D_i$-orbifolds are $N=2$ supersymmetric;
they are also particularly simple when
the original $D$-orbifold is factorizable,
which is precisely why we have concentrated on the factorizable
orbifolds in ref.~\DKLb.

Translating the main result of ref.~\DKLb\ into the notations of the present
paper, we have for any factorizable $(0,2)$ orbifold
$$
\!\eqalign{
\Ddif_a(\mod, \modb) = {} &
- \sum_{i=1,2,3\atop |D_i|>1}\,{\cren_a^{N=2}(i)\over|D|/|D_i|}\,
\left[ \log\left(\left|\eta(i\moo^i)\right|^4 \Re\moo^i\right)
    +\log\left(\left|\eta(i\mot^i)\right|^4 \Re\mot^i\right)\right]\cr
&+  k_a \sum_{i=1,2,3\atop |D_i|>1}\,{\romega(\moo^i,\mot^i)\over|D|/|D_i|}\
+\ \rm const,\cr
}\!\eqn\DorbiS
$$
where $\cren_a^{N=2}(i)$ are the $\beta$-function coefficients of the $N=2$
supersymmetric $D_i$-orbifold.
The second sum in this formula constitutes an additional universal term,
quite distinct from the $\Duniv$ in eqs.~\unicorr\ and \Ddef;
this term was not computed or even discussed in ref.~\DKLb\ where we
calculated only the differences
between the threshold corrections for the different gauge couplings.

The functional form of the first sum in eq.~\DorbiS\ is exactly the same
as that of the field-theoretical threshold corrections \DorbiF.
Since the latter differ from the string-theoretical threshold corrections
by the Green-Schwarz function $V^{(1)}(\mod,\modb)$ (\cf\ eqs.~\Ddef\
and \GSDuniv), we immediately conclude that
$$
\eqalign{
V^{(1)}(\mod, \modb)\ = {}&
\sum_{i=1,2,3} \vgs^i
  \left[ \log\left(\left|\eta(i\moo^i)\right|^4\, \Re\moo^i\right)\,
    +\,\log\left(\left|\eta(i\mot^i)\right|^4\, \Re\mot^i\right)\right]\cr
&+\sum_{i=1,2,3\atop |D_i|>1}\,
	{\romega(\moo^i,\mot^i,\moob^i,\motb^i)\over|D|/|D_i|}\
  +\ \rm const,\cr
}\eqn\Vorbi
$$
where the $\vgs^i$ are some numerical constants to which we shall return in a
moment.
The non-harmonic part of the first sum in this formula has the same form
as the Green-Schwarz function discussed in refs.~\DFKZa--\ABGG,\AGN.
The harmonic $\log|\eta(i\mod)|^4$ terms in this sum
follow from our convention that the dilaton $S$ (and hence the Green-Schwarz
function $V$) should be inert under the $\sltwo$ modular transformations
\modtrans.
Without this convention, those harmonic terms can be re-defined away
according to eqs.~\GStrans\ and \diltrans.

The second sum in eq.~\Vorbi\ and the role it plays in $N=2$ and $N=1$
orbifolds will be discussed in detail in a forthcoming article.
For the present, we simply state without proof that the function
$\romega(\moo,\mot)$ is non-trivial, that it is
universal for all the factorizable orbifolds and that it is finite but
mildly singular when $\moo^i\equiv\mot^i\modulo\sltwo$.
The simplest way to describe this function is to say that $\romega$ is the
$\left[\sltwo\right]^2$ modular invariant solution of the differential
equations
$$
\left[ (\moo+\moob)^2\partial_\moo\partial_{\moob}-2\right]\romega\
=\ \left[ (\mot+\motb)^2\partial_\mot\partial_{\motb}-2\right]\romega\
=\ \log\left|j(i\moo)-j(i\mot)\right|.
\eqn\romegadef
$$

Turning our attention to the non-universal parts of the threshold corrections
\DorbiF\ and \DorbiS, we see that their functional form is the same but
the coefficients do not seem to be related to each other.
Thus, the field-theoretical and the string-theoretical threshold corrections
agree with each other if and only if
$$
\Alpha_a^i\ =\ {\cren_a^{N=2}(i)\over|D|/|D_i|}\ +\ k_a \vgs^i
\eqn\coeffrel
$$
where the constants $\vgs^i$ are exactly as in eq.~\Vorbi.
In particular, for a plane $X^i$ that is twisted by all the nontrivial
members of the orbifold group $D$, the ``$N=2$'' $D_i$-orbifold is actually
an untwisted six-torus with $N=4$ spacetime supersymmetry
and zero $\cren_a(i)$;
for such a plane we should have $\Alpha_a^i=k_a \vgs^i$ for all the couplings.
Eqs.~\coeffrel\ are just as essential for the perturbative consistency of
the orbifold vacua of the heterotic string as the cancellation of the
ordinary triangle anomalies for all the low-energy gauge symmetries;
unfortunately, they are also just as difficult to prove directly
from the string theory.

In Appendices B and C.1 we prove two general string-theoretical properties
of the field-theoretical ``modular anomaly'' coefficients \Alphadef:
First, whenever a factorizable orbifold has a $(1,2)$ modulus
whose value is not frozen by the orbifold's twist group~$D$,
such modulus always has exactly the same $\Alpha^i_a$
as the $(1,1)$ modulus of the same internal plane.
This agrees with the behavior of the string-theoretical coefficients
$\cren_a^{N=2}(i)/(|D|/|D_i|)$ and also allows us not to distinguish
between the $\moo^i$ and the $\mot^i$ in eqs.~\coeffrel.
Second, whenever a factorizable orbifold has $N=2$ spacetime supersymmetry
because
one of the three internal planes is never rotated by the orbifold's twist
group, the coefficients $\Alpha_a$ for that plane are indeed equal
to the $\beta$-function coefficients $\cren_a$ while for the other two
planes $\Alpha_a^i=\cren_a(i)=0$.
Alas, in the $N=1$ case, we do not have a general string-theoretical proof
of eqs.~\coeffrel\ but only eqs.~\Alphadef\ that give us the coefficients
$\Alpha_a^i$ for any particular orbifold.
In a way of a numerical experiment, we have calculated the $\Alpha_a^i$
for a few scores of orbifolds and found that they all satisfy eqs.~\coeffrel.
A dozen examples of such calculations are presented in Appendices C.2--4.
(See also refs.~\IL\ and \DFKZa.)

We conclude this section with a few words about the orbifolds whose
internal six-tori do not factorize into products of separate two-tori
for each of the three $X^i$.
At the tree level, relaxing the factorizability condition does not affect
in any way the \K\ functions \orbikahler\ and \metricorbi;
thus, we still have modular symmetries that act like \modtrans\ and
\matterorbi\ on the moduli and matter superfields of the theory
and under which the Wilsonian gauge couplings must transform according
to eqs.~\forbitrans.
However, only for the factorizable orbifolds {\sl all} of the $\sltwo$
transformations \modtrans\ are true symmetries of the string theory;
in the non-factorizable case, the group of the true modular symmetries
is only a subgroup of the $[\sltwo]^n$.
Clearly, only the true symmetries constrain the moduli dependence of the
Wilsonian gauge couplings, which therefore need not be exactly as
in eqs.~\ForbiF.

As an example, let us consider the $[SU(3)\times SO(8)]/Z_6$
orbifold  where the period lattice of the internal six-torus
is a deformation of the $SU(3)\times SO(8)$ root lattice and the orbifold group
$D=Z_6$ is generated by the $\Theta=(e^{2\pi i/3},e^{2\pi i/6},-1)$.
This orbifold has the usual three $\moo^i$ moduli for each of the three
eigenplanes of $\Theta$ plus the $\mot^3$ modulus for the third complex plane.
However, the modular group for the third plane is not the full $[\sltwo]^2$
but only its proper subgroup $[\Gamma^0(3)]^2$, \ie, the integer $b$
in eq.~\modtrans\ for either $\moo^3$ or $\mot^3$ must be divisible by 3.
Unlike the full $\sltwo$ group, which has no holomorphic invariants without
either singularities or unacceptable rate of growth in the decompactification
limit, the $\Gamma^0(3)$ group does have several invariants of this kind,
namely
$$
\log\eta\left({i\mod+\lambda\over3}\right)\ -\ \log\eta(i\mod)\qquad
{\rm for}\ \lambda=0,1,2.
\eqn\invariants
$$
Consequently, for this orbifold, the field theoretical constraints
specify the moduli dependence of the gauge couplings only up to an
arbitrary linear combination of the invariants \invariants.
The  coefficients of such invariants apparently can only be determined
by the string theory at the one-loop level.
The techniques for calculating the string-theoretical threshold corrections
for the non-factorizable orbifolds were developed by Mayr and Stieberger
\refmark{\MSa};
their explicit results show that the $\Ddif_a(\mod)$ indeed include
holomorphic invariants such as \invariants.

\chapter{$(2,2)$ Supersymmetric Vacua.}
In the previous section we saw how analytic knowledge of the
moduli dependence of the orbifolds' tree-level couplings can be used
to deduce (or at least severely constrain)
the one-loop corrections to the gauge couplings.
Now we turn our attention to the $(2,2)$-supersymmetric vacua of the heterotic
string, for which we also have some analytic knowledge of the
moduli-dependent tree-level couplings.\refmark{\DKLa,\CD,\CDGP,\DG-\MP}
Calabi-Yau compactifications of the ten-dimensional heterotic string
are the best-known examples of such vacua.
However, the $(2,2)$ vacua can be defined and studied in string-theoretical
terms without any reference to the geometry of the six compact dimensions
and without even assuming that the internal SCFT has any
geometrical interpretation at all.
Let us therefore begin this section with a brief review of the
{\it generic} $(2,2)$ vacua and their known properties.

{}From the world-sheet point of view, a $(2,2)$ vacuum is defined by the
following two features:
First, the internal $c=(22,9)$ SCFT contains an $SO(10)\times E_8$
left-moving Kac-Moody algebra ($k=1$ for both the $SO(10)$ and the $E_8$
factors).
\foot{Alternatively, the internal SCFT of a $(2,2)$ vacuum may contain
    an $SO(26)$ Kac-Moody algebra (also at level $k=1$) instead of an
    $SO(10)\times E_8$.
    In this article, however, we focus on the $SO(10)\times E_8$ case.}
Second, the remaining $c=(9,9)$ part of the SCFT has $N=(2,2)$
world-sheet supersymmetry and both the left- and the right-moving
$N=2$ superalgebras have quantized $U(1)$ charges $F$ and $\Fb$
(these charges are always equal to the respective fermion numbers,
hence the notation).
As usual, the right-moving $N=2$ superalgebra is responsible for
the $N=1$ spacetime supersymmetry;
it is the left-moving $N=2$ superalgebra that leads to
the peculiar features of the $(2,2)$ vacuum families.

The gauge group of any $(2,2)$ vacuum family is always $G=E_6\times E_8$
\foot{For some special vacua within the family (corresponding to special
    points or subspaces of the moduli space), additional
    vector bosons and matter fields might become massless.
    In principle, the low-energy physics of such special vacua (and their
    close neighbors) should be described by a different EQFT that
    accounts for the additional light
    particles and it may also re-interpret some of the moduli as combinations
    of the matter fields.
    Such re-analysis is absolutely essential for studying the ``accidental''
    enlargements of the gauge group and their couplings.
    On the other hand, provided the ``accidental'' gauge symmetries commute
    with the $E_6\times E_8$ and all the accidentally light matter fields
    are neutral under the $E_6\times E_8$ (both conditions are true
    in all the known examples), none of these extra fields have any
    one-loop-level impact on the gauge couplings of
    $E_6$ and  $E_8$.
    Therefore, in this section,
    we limit our attention to the $E_6$ and $E_8$ gauge couplings
    for generic $(2,2)$ vacua.}
and the matter fields $\matter^I$ consist of $\hoo$ $\bf 27$ multiplets
of the $E_6$, $\hot$ $\tsb$ multiplets
\foot{For Calabi-Yau compactifications, $\hoo$ and $\hot$ are the Hodge
    numbers of the manifold, hence the notation.
    For generic $(2,2)$ vacua, $\hoo$ and $\hot$ are simply
    integer parameters.}
and some gauge singlets;
none of the light matter fields is charged under the $E_8$ group.
We limit our discussion to the moduli
 that preserve the $(2,2)$ nature of the vacuum.
Such moduli are in one-to-one
correspondence with the charged matter fields
and thus we distinguish between the $\hoo$ moduli $\moo^i$ related to the
$\ts$ matter fields and the $\hot$ moduli $\mot^i$ related to the $\tsb$'s.
At the tree level, these two types of moduli form separate moduli spaces
and the \K\ function is a sum
$$
\KM\ =\ \KM_1(\moo,\moob)\ +\ \KM_2(\mot,\motb) .
\eqn\ksum
$$
Furthermore, both moduli spaces have special \K\ geometries, so both
$\KM_1$ and $\KM_2$ can be written in terms of holomorphic pre-potentials
${\cal F}_1(\moo)$ and ${\cal F}_2(\mot)$.\refmark{\dfnt,\DKLa,\CD}
Unfortunately, loop corrections do not respect the special \K\ geometry
of the moduli space, so the holomorphicity of the prepotentials does not
lead to any non-renormalization theorems for the \K\ functions.

The one-to-one correspondence between the moduli and the charged matter
fields results in a close relation between their respective metrics:
In matrix notations,
$$
Z^{(\ts)}\ =\ G_1\times\exp\coeff13\bigl(\KM_2-\KM_1\bigr)
\qquad{\rm and}\quad
Z^{(\tsb)}\ =\ G_2\times\exp\coeff13\bigl(\KM_1-\KM_2\bigr)
\eqn\ttrel
$$
where $G_1$ and $G_2$ are the moduli metrics
$$
(G_1)_{\ib j}\
=\ {\partial^2 \KM_1\over\partial \moob^\ib\partial\moo^j}\,
\qquad{\rm and}\quad
(G_2)_{\ib j}\
=\ {\partial^2 \KM_2\over\partial \motb^\ib\partial\mot^j}\,.
\eqno\eq
$$
Eqs.~\ttrel\ and \ksum\ are valid only at the tree level of the heterotic
string, but that is all we need to determine the non-harmonic parts of the
one-loop-level threshold corrections to the gauge couplings.
Indeed, substituting eqs.~\ttrel\ into eq.~\oneloopgf\ and taking
into account the group-theoretical factors,
\foot{In the notations of eq.~\groupdefs,
    $T(E_8)=30$, $T(E_6)=12$, $T_{E_6}(\ts)=T_{E_6}(\tsb)=3$;
    thus $\cano_{E_8}=-30$ and $\cano_{E_6}=3\hoo+3\hot-12$.}
we obtain\refmark{\FKLZ,\AGN}
$$
\eqalign{
\Dtot_{E_8}\ =\ \Re f_8^\one\ &
-\ 30\KM_1\ -\ 30\KM_2,\cr
\Dtot_{E_6}\ =\ \Re f_6^\one\ &
 +\ (5\hoo+\hot-12)\KM_1\ -\ 6\log\det G_1 \cr
&+\ (5\hot+\hoo-12)\KM_2\ -\ 6\log\det G_2\,. \cr
}\eqn\dfeekg
$$
Of particular interest is the difference between these two equations,
$$
\eqalign{
\Dtot_{E_6}\ -\ \Dtot_{E_8}\ =\ \Re f_{6-8}^\one\ &
 +\ 6(3+\hoo-\coeff{1}{12}\chi)\KM_1\ -\ 6\log\det G_1 \cr
&+\ 6(3+\hot+\coeff{1}{12}\chi)\KM_2\ -\ 6\log\det G_2 \cr
}\eqn\sixeightdif
$$
($\chi=2[\hoo-\hot]$ is the Euler number and
$f_{6-8}^\one\equiv f_6^\one-f_8^\one$).
The non-harmonic part on the right hand side here is precisely
12 times the ``holomorphic anomaly'' of the topological index $F_1$
of Bershadsky, Cecotti, Ooguri and Vafa.\refmark{\BCOV}
Furthermore, using classical six-dimensional geometry, they proved that
for the large-radius Calabi-Yau threefolds and their mirror images
one indeed has
$$
\Dtot_{E_6}\ -\ \Dtot_{E_8}\ =\ 12 F_1
\eqn\dedef
$$
and hence $F^{\rm top}$ --- the holomorphic part of the $F_1$ ---
is the same as ${1\over12}f_{6-8}$.
We will show momentarily that eq.~\dedef\ actually
holds for all the $(2,2)$ vacua of the heterotic string,
regardless of whether such a vacuum has a geometrical interpretation
of any kind.

Eqs.~\dfeekg\ are field-theoretical constraints based upon the tree-level
properties of the $(2,2)$ vacua.
However, these vacua also have characteristic features that become important
at the one-loop level of the string theory.
In particular,
the fact that the internal $c=(22,9)$ world-sheet SCFT splits into an
$c=(9,9)$ SCFT plus the $SO(10)\times E_8\ (k=1)$ Ka\v{c}-Moody algebra
allows us to factorize the trace in eq.~\ramondright:
$$
\displaylines{
\B_a\ =\ {-1\over 2 \eta^2(\tau)}
\sum_{\rm even\ \bf s} \Tr_{(s_1,R)}\left( (-1)^{s_2 F}\, (-1)^{\Fb-{3\over2}}
\Fb\,q^{L-{3\over8}}\bar q^{\bar L-{3\over8}}\right)_{(9,9)}
\hfill\qquad\eqname\DEfactor\cr
\hfill{}\times \Tr_{s_1}\left( (-1)^{s_2 F}
\left( T_{(a)}^2-\coeff{1}{8\pi\tau_2}\right)
q^{L-{13\over24}}\right)_{SO(10)\times E_8}\qquad\cr }
$$
where the summation over the NSR boundary conditions now refers to the
left-moving world-sheet fermions.
The second trace in this formula distinguishes between the gauge couplings
of the $E_8$ and of the $E_6$ (for the $E_6$ one uses a generator $T_{(a)}$
in  $SO(10)\subset E_6$ subgroup),
but it is totally insensitive to specific properties of a particular $(2,2)$
vacuum.
All such traces (altogether six,
for the two gauge groups and the three even NSR boundary conditions)
can be easily obtained from the characters of the
$SO(10)\times E_8$ Ka\v{c}-Moody algebra or even from the partition functions
$Z_{E_8}(\tau)$ and $Z_{SO(10)}({\bf s},\tau)=Z_\Psi^5({\bf s},\tau)$
($Z_\Psi$ is the partition function of one complex free fermion
or two real ones);
in terms of the partition functions,
the second line of eq.~\DEfactor\ equals to
$$
Z_{E_8}(\tau)\, Z_\Psi^5({\bf s},\tau)\times
{1\over 2\pi i}\partder{}{\tau} \cases{
    \left(\log Z_\Psi({\bf s},\tau)\,
	+\,\half\log\left(\tau_2|\eta(\tau)|^4\right)\right) &
    for  $E_6$,\cr
    \left(\coeff18\log Z_{E_8}(\tau)\,
	+\,\half\log\left(\tau_2|\eta(\tau)|^4\right)\right) &
    for  $E_8$.\cr
}\eqn\twocharacters
$$

We do not see how one can further simplify  eqs.~\twocharacters,
but we can simplify the difference of the two expressions.
Using the partition function identities
$$
\eqalign{
Z_\Psi(NS,NS)\times Z_\Psi(NS,R)\times Z_\Psi(R,NS)\ & =\ 2,\cr
Z_\Psi^4(NS,NS) - Z_\Psi^4(NS,R) - Z_\Psi^4(R,NS)\ & =\ 0,\cr
Z_\Psi^8(NS,NS) + Z_\Psi^8(NS,R) + Z_\Psi^8(R,NS)\ &
    =\ Z_{E_8},\cr
}\eqn\Zidentities
$$
it is easy to show that for each of the three even NSR boundary conditions,
$$
Z_{E_8}(\tau)\, Z_\Psi^5({\bf s},\tau)\times
    \partder{}{\tau} \left(\log Z_\Psi({\bf s},\tau)\,
	-\,\coeff18\log Z_{E_8}(\tau)\right)\
=\ -24 (-1)^{s_1+s_2}\,\partder{Z_\Psi({\bf s},\tau)}{\tau}\,.
\eqn\BSmap
$$
The left hand side of this equation involves partition functions of altogether
26 real fermions (10 for the $SO(10)$ and 16 for the $E_8$), but only
two real fermions appear on the right hand side.
Such drastic reduction in fermionic degrees of freedom is characteristic
of the so-called {\it bosonic/supersymmetric} map between the heterotic
string and the type~II superstring.\refmark{\LSWrep}
In the context of \twocharacters\ and \DEfactor,
the bosonic/supersymmetric map \BSmap\ immediately gives us
$$
\displaylines{
\B_6\ -\ \B_8\ =\ {12\over\eta^2(\tau)} \sum_{\rm even\ \bf s}
(-1)^{s_1+s_2}\,{\partial Z_\Psi({\bf s},\tau) \over 2\pi i \partial\tau}
\hfill\qquad\eqname\typetwo\cr
\hfill{}\times\Tr_{(s_1,R)}\left( (-1)^{s_2 F}\, (-1)^{\Fb-{3\over2}}
\Fb\,q^{L-{3\over8}}\bar q^{\bar L-{3\over8}}\right)_{(9,9)} .\cr }
$$

Now consider the left-moving $N=2$ world-sheet supersymmetry
of the $c=(9,9)$ SCFT.
The left-moving and the right-moving $N=2$ superalgebras of the $(2,2)$ vacua
are complex conjugates of each other and
satisfy exactly the same $F$ charge quantization condition;
consequently, both superalgebras give rise to the same kind of
Riemann identities between the NSR sectors of the $c=(9,9)$ SCFT.
In Appendix~A, we show that the right-moving $N=2$ superalgebra reduces the sum
over three even NSR sectors in eq.~\beq\ to a single Ramond-Ramond trace
in eq.~\ramondright.
In that proof, only the right-moving degrees of freedom play any role while
the left-moving degrees of freedom simply come along for the ride, but
of course, exactly the same identity would also apply to the left-moving
side of an $(2,2)$ supersymmetric SCFT.
This is precisely the situation we have in eq.~\typetwo, whose left-moving side
looks exactly like the right-moving side of eq.~\beq\ and thus can be reduced
in exactly the same way.
The result is
$$
\B_6\ -\B_8\
=\ 6\,\Tr_{R,R}\left( (-1)^{F-\Fb}\,F\Fb\,
q^{L-{3\over8}}\bar q^{\bar L-{3\over8}}\right)_{(9,9)}
\eqn\indexrel
$$
where the boundary conditions are Ramond-Ramond
on both sides of the $c=(9,9)$ SCFT
and no other world-sheet degrees of freedom are involved.

In the path-integral formulation of $N=(2,2)$ SCFT,
the totally-Ramond characters
(Ramond-Ramond for both sides of the world-sheet)
are given by the zero modes of the conformal fields
\foot{The non-zero modes come in supermultiplets
    --- the totally-Ramond boundary conditions preserve all
    of the world-sheet supersymmetries --- and the bosonic non-zero
    modes cancel against the fermionic ones and vice verse.}
and thus behave as generalized supersymmetry indices of the theory.
The particular  index \indexrel\ and its $d^2\tau$ integral
$$
F_1 \equiv \half \int {d^2\tau\over \tau_2}
\Tr'_{R,R}\left( (-1)^{F-\Fb}\,F\Fb\,
q^{L-{3\over8}}\bar q^{\bar L-{3\over8}}\right)_{(9,9)}
\eqn\Fdef
$$
was first encountered in ref.~\CV\
and later studied in more detail in ref.~\BCOV.
Comparing this $F_1$ with eqs.~\indexrel, \mainresult\ and \Ddef,
we immedaitely see
$12F_1=\Ddif_{E_6}-\Ddif_{E_8}=\Dtot_{E_6}-\Dtot_{E_8}$;
this concludes our proof of eq.~\dedef.

Thus far we discussed the general features of all the $(2,2)$ vacua,
regardless of their geometrical interpretation or lack thereof.
Let us now turn to the vacua which are related to the Calabi-Yau
threefolds and consider how the size of the internal threefold
affects the four-dimensional gauge couplings.
The ``overall radius'' $R$ of the threefold is one of its $(1,1)$ moduli;
according to eq.~\ksum, it does not affect the
$(1,2)$ moduli $\mot^i$, so we may safely disregard the latter
in the following discussion.
In terms of the cohomologically defined $(1,1)$ moduli $\moo^i$,
all the $\Re\moo^i$ are proportional to $R^2$
while the torsions $\Im\moo^i$ are radius-independent.
For each torsion, there is a discrete Peccei-Quinn symmetry;
these symmetries involve neither \K\ transformations nor rescalings
of the charged matter fields and hence should leave the
Wilsonian gauge couplings $\Re f_a$ invariant.
Since the $f_a$ are holomorphic functions of $\moo^i$,
this immediately implies
$$
f_a^\one(T)\ = \sum_i \clrad_{a,i} \moo^i\
+\ {\rm const}\ +\ \delta f_a\left(\exp(-2\pi\moo)\right),
\eqn\dpqsym
$$
where $\clrad_{a,i}$ are some rational proportionality constants
and the last term is exponentially small
in the large radius limit.
Thus, in that limit we have
$$
 \Re f_a^\one\ \to \clrad_a\ R^2 +  \rm const ,
\eqn\fchoice
$$
and the large radius behavior of the $f_a^\one$
depends on whether the constant $\clrad_a$
vanishes or not, and
this cannot be determined by the Peccei-Quinn symmetries alone.

Before we turn to string-theoretical reasons determining $\clrad_a$,
let us consider the field-theoretical non-harmonic contributions to the
threshold corrections $\Dtot_a$.
For Calabi-Yau manifolds that are both large and smooth,
\ie, when {\sl all of the $\Re\moo^i$ are large},
the \K\ function $\KM_1$ can be approximated as\refmark{\strominger,\CD}
$$
\KM_1(\moo,\moob)\
\approx\ -\log\left(d_{ijk}(\Re\moo^i)(\Re\moo^j)(\Re\moo^k)\right),
\eqn\approxiKM
$$
and hence $\exp(\KM_1)$ is proportional to
$R^{-6}$ while the moduli metric matrix $G_1$ scales like
$R^{-4}$.
Substituting these scaling laws into eqs.~\dfeekg, collecting
all the $\log R^2$ terms and using eq.~\fchoice\ gives us\refmark{\dixon}
$$
\Dtot_a\
\approx\ \clrad_a\ R^2 - \cren_a\,\log R^2\
+\ \rm const
\eqn\crenee
$$
for both the $E_6$ and the $E_8$ couplings.
We emphasize that this result depends on all of the $\Re\moo^i$
being large and does not apply to
 degenerate manifolds for which some of the $\moo^i$
are frozen at zero values;
indeed, the orbifold results of section 3 generally disagree
with eq.~\crenee.
\foot{For large, smooth manifolds, the entire moduli metric matrix
    $G_1$ is proportional to $R^{-4}$ while the entire $Z$ matrix for
    the $\ts$ matter fields is proportional to  $R^{-2}$
    (\cf~eq.~\ttrel).
    For  singular $(2,2)$ orbifolds,
    the same is true for the {\sl untwisted} moduli
    and $\ts$ matter fields, but the {\sl twisted} fields have quite
    different scaling properties.
    For example, for the $Z_3$ orbifold, the $Z$ matrix for the twisted
    $\ts$'s is proportional to $R^{-4}$ instead of the usual $R^{-2}$.
    Of course, once the sharp points of an orbifold are blown up (and
    the blow-up radii increase proportionately to the overall radius $R$),
    we do recover the usual $R^{-2}$ scaling properties of the
    $Z^{\ts}_{\rm twisted}$ matrix and thus restore the validity of
    eqs.~\crenee.}

The curious coincidence between the coefficients of the $\log R^2$ term
in eqs.\penalty500\ \crenee\ and the $\beta$-function coefficients
for the respective gauge couplings suggests that perhaps
in eq.~\fchoice, $\clrad_a=0$ and the entire radius-dependence
of the threshold corrections $\Dtot_{E_6}$ and $\Dtot_{E_8}$ amounts
to changing the effective threshold scale from the string scale
$\mstring\sim1/\sqrt{\alpha'}$ to the Kaluza-Klein scale $1/R$.\refmark{\VT}
However, eq.~\dedef\ can be used to show that the
leading term in the large-radius limit of  $\Delta_{E_6}-\Delta_{E_8}$
is proportional to the $R^2$ rather than to the $\log R^2$
and hence $\clrad_6-\clrad_8\neq0$.
Indeed, ref.~\BCOV\ gives  the large-radius limit
of the topological index $F_1$ as
$$
F_1 \rightarrow
{1\over6}\sum_i\,\Re\moo^i\!\int_{\cal M} {\bf k}_i\wedge {\bf C}_2\
=\ {1\over 96\pi^2} \int_{\cal M} \left\|{\cal R}\right\|^2
\eqn\cdef
$$
where ${\bf C}_2$ is the second Chern class of the Calabi-Yau threefold
$\cal M$, $\cal R$ is its Riemannian curvature tensor and the ${\bf k}_i$
form a basis of the cohomology group $H^{(1,1)}$.
The left hand side here is obviously proportional to the $R^2$ while
the right hand side is positive definite;
together, they guarantee that $F_1$ and thus the difference
$\Dtot_{E_6}-\Dtot_{E_8}$ indeed grows like $R^2$ in the large-radius limit.

In terms of the Wilsonian couplings $f_6^\one$ and $f_8^\one$, eq.~\cdef\
tells us that
$$
f_6^\one(\moo)\ -\ f_8^\one(\moo)\
=\ 2\sum_i\,\moo^i\!\int_{\cal M} {\bf k}_i\wedge {\bf C}_2\
+\ {\rm const}\ +\ \delta f_{6-8}\left(\exp(-2\pi\moo)\right) .
\eqn\fdifferenceF
$$
Unfortunately, we do not have a second equation of this kind that would
determine separate $R\to\infty$ limits of the $f_6^\one$
and of the $f_8^\one$.
In general, all we can say is that they have the general form \dpqsym\
and do not grow faster than $R^2$.
However, for some specific Calabi-Yau threefolds,  the
entire analytic form of both  $f_6^\one$ and  $f_8^\one$
can be deduced by  essentially
the same techniques as we used  in section~3.

As an example, consider the  quintic threefold analyzed by
Candelas, de la Ossa, Green and Parkes.\refmark{\CDGP}
It has $\hot=101$ but $\hoo=1$, so its $(1,1)$ moduli
space needs only one complex coordinate;
Candelas \etal\ found it convenient to work with the ``mirror
coordinate'' $\psi$ instead of the flat coordinate $\moo$.
According to eqs.~\ttrel, such coordinate transformation also
entails a linear redefinition of the charged matter fields,
which means that one should also use different sets of Wilsonian
gauge couplings $f_a$ in different coordinate pictures
(\cf~eq.~\fLocal).
We should also account for a possible \K\ transformation between
different coordinate pictures, but fortunately, this transformation
is trivial for the two particular pictures of the quintic discussed here.
Thus,
$$
f^\one_8(\psi)\ =\ f_8^\one(\moo)\qquad{\rm but}\quad
f^\one_6(\psi)\ =\ f_6^\one(\moo)\ -\ 12{d\psi\over d\moo}\,.
\eqn\coordquint
$$

In terms of $\psi$, some of the modular transformations are
monodromies that map $\psi$ onto itself
while others map $\psi\to e^{2\pi i/5}\psi$;
thus all physical quantities must be single-valued functions of
the $\psi^5$.
The \K\ function $\KM_1(\psi,\bar\psi)$ determined in ref.~\CDGP\ is
invariant under all the modular transformations, but
the Yukawa coupling $Y_{\ts}$
of the only $\ts$ multiplet of the theory transforms like
$$
Y_{\ts}(\psi)\ =\ {(2\pi i)^3\over25}\,{\psi^2\over 1-\psi^5}\
\to\ e^{4\pi i/5}\,Y_{\ts}\quad
{\rm when}\quad \psi\to e^{2\pi i/5}\psi ,
\eqn\yukawaquint
$$
which implies that the $\psi$-changing modular transformations are
$R$-symmetries of the charged fields:
$$
\matter_{\ts}\ \to\ e^{-2\pi i/15}\,\matter_{\ts}\,,\quad
\matter_{\tsb}^i\ \to\ e^{+2\pi i/15}\,\matter_{\tsb}^i\quad
{\rm and}\quad W\ \to\ e^{+2\pi i/5}\,W
\eqn\fieldsquint
$$
when $\psi\to e^{2\pi i/5}\psi$.
Assuming as usual that the dilaton superfield $S$ is inert under
all modular transformations, we apply eqs.~\fLocal\
to the transformations \fieldsquint\ and
conclude that both of the
Wilsonian threshold corrections $f_8^\one$ and $f_6^\one$ must
be single-valued (modulo $4\pi i$) functions of $\psi$,
and furthermore,
$$
\left.\eqalign{
f_8^\one(e^{2\pi i/5} \psi)\ &
=\ f_8^\one(\psi) \cr
f_6^\one(e^{2\pi i/5} \psi)\ &
=\ f_6^\one(\psi)\ -\ \coeff{4\pi i}{5}\cr
}\,\right\}\,{\rm modulo}\ 4\pi i .
\eqn\modularquint
$$

The $(2,2)$ vacuum family of the quintic threefold includes the Gepner
$[3]^5$ model\refmark{\gepner}; that particular vacuum
corresponds to $\psi=0$.
The Gepner model has four massless abelian gauge fields
as well as four massless matter superfields that are not present in the
spectra of the generic vacua in the same family;
however, all these ``accidentally'' massless fields are neutral under the
$E_8\times E_6$ gauge group.
Therefore, the physical gauge couplings $g_{E_8}$ and $g_{E_6}$ should
 have no singularities at the ``Gepner point'' $\psi=0$.
At the same point, the metric $G_{\bar\psi\psi}$ of the modulus $\psi$
is non-singular, but the \K\ function has a logarithmic singularity,
$\KM_1(\bar\psi,\psi)=-\log|\psi|^2+\rm finite$.
According to eqs.~\dfeekg, this singularity should be canceled by
appropriately singular terms in the Wilsonian corrections $f_8^\one$
and $f_6^\one$.
Thus, for $\psi\to 0$,
$$
\eqalign{
f_8^\one(\psi)\ &
=\,\enspace -60 \log\psi\ +\ \hbox{finite,}\cr
f_6^\one(\psi)\ &
=\, +188 \log\psi\ +\  \hbox{finite;}\cr
}\eqn\zerolimit
$$
note that the modular transformations of the logarithmic terms here
agrees with eqs.~\modularquint.

Besides the spurious \K\ singularity at the Gepner point
$\psi=0$, the $(1,1)$ moduli space of the quintic has two genuine,
physical singularities:
$\psi\to\infty$ is the large radius limit of the threefold,
and at $\psi^5=1$, the mirror threefold suffers from conifold
degeneration.\refmark{\CDGP}
In the large radius limit  one has
$\moo\approx \log\psi^5$;
thus, in light of eqs.~\dpqsym\ and \coordquint,
the Wilsonian gauge couplings have at most logarithmic divergences
as $\psi\to\infty$.
Taking also into account the modular transformation properties
\modularquint\ of the two $f_a^\one$, their $\psi\to0$ limits
\zerolimit\ and the requirement that there should be no
singularities of any kind except at $\psi=0$, $\psi=\infty$
or $\psi^5=1$
we arrive at
$$
\eqalign{
f_8^\one(\psi)\ &
=\ -12\log\psi^5\ +\ (\clrad_8+12)\log(\psi^5-1)\ +\ p_8(\psi^5-1) ,\cr
f_6^\one(\psi)\ &
=\ +\coeff{188}{5}\log\psi^5\ +\ (\clrad_6-40)\log(\psi^5-1)\
+\ p_6(\psi^5-1) ,\cr
}\eqn\fsingular
$$
where the coefficients $\clrad_8$ and $\clrad_6$
are exactly as in eqs.~\dpqsym.
The functions $p_8$ and $p_6$ here must be single-valued (in terms of
$\psi^5$) and non-singular anywhere except at $\psi^5=1$;
such functions may have poles or essential singularities at that point,
but no logarithmic or other singularities that require branch cuts.

Now consider the physics of the conifold limit $\psi^5\to1$.
In that limit, the \K\ function $\KM_1$ is finite while the metric
$G_{\bar\psi\psi}$ has only a mild logarithmic singularity;
the leading divergences of the threshold corrections $\Dtot_{E_8}$
and $\Dtot_{E_6}$ should therefore come from the Wilsonian terms in
eqs.~\dfeekg.
In light of  eqs.~\fsingular\ one might ask:
 {\sl How can a gauge coupling have a pole or an essential singularity
at $\psi^5=1$?}
Generally, threshold corrections to gauge couplings become singular when
otherwise massive charged fields become massless for some particular values
of the moduli, but such divergences are always logarithmic with respect
to the ``accidentally'' vanishing masses.
Thus, there are only two ways to get a pole or any other singularity that
is stronger than logarithmic:
The first way is for the masses to vanish not like powers of $(\psi^5-1)$ but
exponentially or faster; this is rather implausible in terms of
the known geometry of the conifold limit.
The second way is to have an infinite number of charged fields
that all become massless at the same time,
which means that in string-theory, the conifold limit
would be equivalent to some kind of a decompactification and at $\psi^5=1$
we would effectively have five or more non-compact spacetime dimensions.
In this scenario, the rate at which $f^\one_a$ grow when $\psi^5\to1$
can be limited in essentially the same way as we have limited the large-radius
growth of the orbifolds' $f_a$ in the previous section.
Let us skip the technical details of this argument; the result is that
if the conifold were to decompactify,
the $f^\one_{6,8}$ could be no more singular than $1/(\psi^5-1)^2$.
However, the geometry of the conifold does not seem to support
infinitesimally short
strings wrapping around non-contractible loops and we do not see what other
string modes could lead to an effective decompactification of the conifold
limit.
Let us therefore {\sl conjecture} that there is no decompactification
and only a finite number of particles become massless at $\psi^5=1$.
In this case, the threshold corrections cannot have any poles
or essential singularities at $\psi^5=1$;
in terms of eqs.~\fsingular, this means
$p_8 = p_6 = {\rm const}$.
(Of course, a ``constant'' part of an $f_a^\one(\psi)$ is actually a function
of the 101 moduli $\mot^i$, but the analytic form of {\sl that}
function is beyond the scope of the present discussion.)

The question of the logarithmic singularities of eqs.~\fsingular\
at $\psi^5=1$ is more subtle since such singularities require only
a finite number of otherwise massive charged fields to become massless.
We believe however that even this does not happen for the $E_8$ coupling.
Indeed, consider the heterotic string vertices of hypothetical
massless particles with non-trivial $E_8$ charges.
The $k=1$ Ka\v{c}-Moody algebra responsible for the $E_8$ has no
sources of charge other that the Ka\v{c}-Moody currents $J_{(a)}(z)$
themselves.
Therefore, a heterotic vertex of any massless $E_8$-charged boson
(because of the spacetime supersymmetry, it is enough to consider the bosons)
has to factorize into a product of the form $e^{ip\cdot X}J_{(a)}
\left(\Phi+\coeff{i}{2}(p\cdot\psi)\Psi\right)$
where the operators $\Phi$ and $\Psi$
have conformal dimensions $h=(0,1)$ and $h=(0,\half)$, respectively.
These dimensions mean that
$\Phi(\bar z)$ is a right-moving current while $\Psi(\bar z)$ is
a right-moving free fermion;
the ordinary gauge bosons of the $E_8$ come from such operators
in the spacetime part of the world-sheet SCFT,
$\Phi=\partial X^\mu$ and $\Psi=\psi^\mu$.
However, were there additional operators of this kind in the
internal part of the SCFT, the free right-moving fermion $\Psi$
would have a zero mode in the Ramond sector.
That zero mode would be inseparable from the zero modes of the
four $\psi^\mu$ and thus would allow changing the spacetime chirality
of any fermionic particle without changing the rest of its quantum numbers;
in other words, there would be absolutely no chirality in the particle
spectrum of the four-dimensional theory.
\foot{This argument, adapted from ref.~\DKV,
    applies to any four-dimensional vacuum of the heterotic string,
    spacetime supersymmetric or otherwise.
    For any {\it level $k=1$} Ka\v{c}-Moody algebra, if there are
    any massless scalars in the adjoint representation of the gauge
    group, then the spacetime fermions cannot have any chirality.}

Although the conifold limit of the  quintic threefold corresponds to
a somewhat singular $(2,2)$ vacuum,
we do not believe it is singular enough to eliminate
the non-zero Euler number of the theory and completely
remove the chirality of its spacetime fermions.
Therefore, we find it implausible that any particle that becomes accidentally
massless in the conifold limit can carry an $E_8$ charge.
The $E_8$ gauge coupling thus cannot have even a logarithmic
singularity at the conifold point $\psi^5=1$
and the first eq.~\fsingular\ reduces to simply
$$
f_8^\one(\psi)\ =\ -12\log\psi^5\ +\ \rm const.
\eqn\eightquint
$$
Among other things, this formula gives us the exact large-radius limit
of the $E_8$ coupling: in terms of eq.~\fchoice, $\clrad_8=-12$.

For the gauge coupling of the $E_6$, the situation is somewhat different.
The same argument we have just used for the $E_8$ also rules out any
accidentally massless particles in the adjoint representation of the $E_6$.
Furthermore, the non-trivial chirality of the conifold limit also rules out
any accidental enlargement of the $E_6$ gauge group to an $E_7$ or an $E_8$.
What we cannot rule out, and what we believe might indeed happen is the
accidental masslessness of an $\ts+\tsb$ matter multiplet.
As a result, the $E_6$ coupling would diverge logarithmically, and while
we cannot calculate the coefficient of such divergence without knowing exactly
how many $\ts+\tsb$ multiplets do become massless and the way their masses
depend upon $\psi^5-1$, we can be sure of its sign.
In terms of eq.~\fsingular, we must have $(\clrad_6-40)\le0$.

At this point, we again turn to the results of ref.~\BCOV\
who have calculated the topological integral in eq.~\fdifferenceF\
for the quintic and thus determined the large-radius limit
(and hence the entire analytic form) of the difference
$f_{6-8}(\psi)$.
In our notations, their result amounts to $\clrad_6-\clrad_8=50$,
which indeed agrees with $\clrad_8=-12$ and $\clrad_6\le40$.
Thus we now know the exact analytic form of the Wilsonian $E_6$
coupling,
$$
f_6^\one(\psi)\ =\ +\coeff{188}{5}\log\psi^5\ -\ 2\log(\psi^5-1)\
+\ \rm const.
\eqn\sixquint
$$

We conclude this article by extending the above analysis of the quintic
threefold to the three other threefolds that also have $\hoo=1$ and similar
singularities of the $(1,1)$ moduli space.\refmark{\onepa}
(Analysis of Calabi-Yau threefolds with $\hoo\ge2$ is similar in principle
but technically more difficult because of more complicated singularities.)
Following the  notation of refs.~\onepa,
 the modular-invariant coordinate of the $(1,1)$ moduli
space is $\psi^k$ where $k=6$, 8 or 10, depending on a particular model,
and the singularities are at
$\psi\to\infty$
(the large radius limit) and $\psi^k=1$ (a conifold singularity);
there is also a spurious singularity at the Gepner point $\psi=0$.
In this notation
the quintic threefold --- which follows the same pattern ---
corresponds to  $k=5$.
The asymptotic behavior of the $\KM_1$ and $G_{\bar\psi\psi}$ is
similar for all these models;
it is spelled out in detail in refs.~\onepa.
Given these data --- and the topological integrals~\fdifferenceF\
computed in ref.~\BCOV, --- we obtain
$$
\eqalign{
f_8^\one(\psi)\ &
=\ -60\,\log\psi\ +\ \rm const,\cr
f_6^\one(\psi)\ &
=\ A(k)\,\log\psi\ -\ 2\log(\psi^k-1)\ +\ \rm const,\cr
}\eqn\fKT
$$
where $A(5)=188$ (\cf~eq.~\sixquint), $A(6)=192$, $A(8)=296$ and $A(10)=288$.

Remarkably, all four models have exactly the same logarithmic
divergence of the $E_6$ coupling in the conifold limit $\psi^k\to1$.
The coefficient $(-2)$ of this divergence tells us that some
$\ts+\tsb$ multiplets do become ``accidentally'' light in the conifold limit
and that the product of their masses behaves like
$\left(\psi^k-1\right)^{1/6}$.
It would be interesting to verify this result by a direct string calculation
of the masses.

\par\smallskip
\noindent {\bf Acknowledgements: }
Major parts of this work were performed at SLAC and CERN.
We take this opportunity to thank the members of both  theory
groups  for their warm hospitality, their encouragement and for innumerable
enlightening discussions we enjoyed over the past years.
We would also like to acknowledge many conversations we had with
I.~Antoniadis, P.~Candelas, X.~de la Ossa, L.~Dixon,
E.~Gava,  W.~Lerche, P.~Mayr, K.~Narain,  S.~Stieberger, T.~Taylor,
S.~Theisen,  C.~Vafa and E.~Witten.
J.~L. thanks E.~Witten and the Institute for Advanced Study
for the hospitality during the final stages of this project.
\subpar
The research of V.~K. is supported in part by the NSF,
under grant PHY--90--09850,
and by the Robert A.~Welch Foundation.
The research of J.~L. is supported by the Heisenberg Fellowship
of the DFG.
The collaboration of the two authors is additionally supported by
NATO, under grant CRG~931380.

\APPENDIX{A}{A\break
Riemann Identities for Threshold Corrections.}
In terms of the internal $c=(22,9)$ world-sheet SCFT,
unbroken $N=1$ supersymmetry of the four-dimensional spacetime requires
 extended $N=(0,2)$ world-sheet supersymmetry
(rather than just $N=(0,1)$ required by the heterotic string itself).
The current algebra of this extended supersymmetry contains an abelian current
$J(\bar z)$ whose charge $\oint J$ should be quantized.
Together, $J$ and its quantized charge describe a free chiral boson
($J=\sqrt{3}\bar\partial H$) of radius $\sqrt{3}$,
which  is a universal $c=(0,1)$ part of the internal SCFT of any
spacetime-supersymmetric vacuum.\refmark{\BDFS}
Among other things, this universal part is responsible for the NSR sectors
of the internal SCFT;
the remaining $c=(22,8)$ part --- the part that differs
from vacuum to vacuum --- is the same in all the $4^{\rm genus}$ NSR sectors.
Joining the $c=(22,8)$ and the $c=(0,1)$ SCFTs together involves 3
conjugacy classes and hence only $3^{\rm genus}$ sectors.
This fact leads to linear relations between the partition functions and
characters of the combined world-sheet SCFT for different NSR sectors;
such relations are generally known as {\it Riemann identities}.
\refmark{\LSW, \LSWrep}

The quintessential Riemann identities of the $N=1$, $d=4$ spacetime
supersymmetry are
identities for the characters of the $E_6/D_4$ coset algebra,
which combines the internal $H$ boson with the bosonized fermionic
superpartners of the two transverse space coordinates.
All the other Riemann identities can be derived from these and
at  $\rm genus =1$, there is  one  such identity,
which reads\refmark{\LSW}
$$
\!\sum_{\bf s} (-1)^{s_1+s_2}\Ch^{E_6/D_4}_\rho({\bf s}, \nu_\Psi,\nu_H,\tau)
= 2\Ch^{E_6/D_4}_\rho(RR,\half\nu_\Psi-\coeff{\sqrt{3}}{2}\nu_H,
    \half\nu_H+\coeff{\sqrt{3}}{2}\nu_\psi,\tau) .\!
\eqn\quintriemann
$$
Here the subscript $\rho$ labels the three conjugacy classes of the $E_6$,
$RR$ stands for the Ramond-Ramond sector ${\bf s}=(1,1)$ and
the linear transformation of the $(\nu_\Psi,\nu_H)$ 2-vector on the
right hand side is simply a $\pi/3$ rotation.
The one-loop characters $\Ch^{E_6/D_4}$ can be expressed
in Hamiltonian terms according to
$$
\displaylines{
\Ch^{E_6/D_4}_\rho({\bf s}, \nu_\Psi,\nu_H,\tau)
\hfill\qquad\eqname\characterdef\cr
\hfill{}=\ \Tr_{\rho,s_1}\left( (-1)^{s_2(F_\psi+F_H)}\,q^{L-{1\over12}}\,
    \exp(2\pi i \nu_\Psi F_\Psi + 2\pi i \nu_H F_H/\sqrt{3})\right)_{H+\Psi}\,
.\qquad\cr }
$$
Note the dual role played by the $F_H$ operator here:
$F_H/\sqrt{3}$ is the $J$-charge while $F_H$ itself is the fermion number
due to $H$-related degrees of freedom.
Similarly, $F_\Psi$ is both the fermion number for the two transverse fermions
and also the helicity (or rather the $\Psi$-dependent part of the helicity).

The purpose of this Appendix is to use the Riemann identity~\quintriemann\
(or rather its complex conjugate) to establish the identity between the
right hand sides of eqs.~\beq\ and \ramondright.
Let us therefore start with eq.~\beq\ and
factorize the trace over the internal SCFT into
a trace over the $H$-related part and a trace over the rest:
$$
\displaylines{
\B_a(\tau,\bar\tau)\
=\sum_{\rho=0,\pm1}
\Tr_\rho\left( \left( T_{(a)}^2-\coeff{k_a}{8\pi\tau_2}\right)
    q^{L-{11\over12}}\,\bar q^{\bar L-{1\over 3}}\right)_{(22,8)}
\hfill\qquad\eqname\factorB \cr
\hfill{}\times {2\over |\eta(\tau)|^4}
\sum_{\rm even\ s} (-1)^{s_1+s_2}\,
{\partial Z_\Psi({\bf s},\bar\tau)\over 2\pi i\partial\bar\tau}\,
\Tr_{\rho,s_1}\left( (-1)^{s_2 F}\,\bar q^{\bar L-{1\over24}}\right)_H\,
.\qquad\cr }
$$
Clearly, the expression on the second line here is universal
for all the spacetime-supersymmetric vacua;
it is this expression that we are now going to rewrite
in terms of the characters of the $E_6/D_4$ coset.

The $E_6/D_4$ coset is comprised of the $H$-boson  and of an $SO(2)$
generated by the two transverse fermions; the $Z_\Psi$ in eqs.~\beq\ and
\factorB\ is precisely the partition function of those fermions.
The derivative $\partial Z_\Psi/\partial\bar\tau$ can also be obtained
from the $SO(2)$ characters $\Ch^{SO(2)}({\bf s},\nu_\Psi,\bar\tau)$,
which satisfy differential equations
$$
\left(\partder{}{\bar\tau}-{i\over4\pi}\partder{^2}{\nu^2_\Psi}
    +\partder{\log\eta(\bar\tau)}{\bar\tau}\right)
\Ch^{SO(2)}({\bf s},\nu_\Psi,\bar\tau)\
=\ 0.
\eqn\difeqsotwo
$$
Since for $\nu=0$ the characters are the same as the partition functions,
we can express the $\partial Z_\Psi/\partial\bar\tau$ factors in eq.~\factorB\
in terms of the $SO(2)$ characters and their $\partial^2/\partial\nu^2_\Psi$
derivatives.
Combining the result with the trace over the $H$-boson sector,
we see that the expression on the second line of eq.~\factorB\
equals to
$$
{(1/4\pi^2)\over |\eta(\tau)|^4} \left.
    \left( \partder{^2}{\nu_1^2}\,
	+\,4\pi i\,\partder{\log\eta(\bar\tau)}{\bar\tau}\right)
    \sum_{\rm even\ s} (-1)^{s_1+s_2}\,
    \Ch^{E_6/D_4}_\rho({\bf s},\nu_\Psi,\nu_H,\bar\tau)
    \right|_{\nu_\Psi=\nu_H=0.}
\eqn\characteq
$$
The sum here is over the even sectors $\bf s$ only, but we can extend
it to all the sectors since for $\nu_\Psi=\nu_H=0$, the Ramond-Ramond
character vanishes together with its diagonal second derivatives.
In this manner, we arrive at precisely the character sum on the left
hand side of eq.~\quintriemann.
Now we can use the Riemann identity and relate everything to the
Ramond-Ramond characters, but we still need to apply the differential
operator in \characteq, which gives us
$$
{\sqrt{3}\over4\pi^2|\eta(\tau)|^4}\,
\partder{}{\nu'_\Psi}\,\partder{}{\nu'_H}
\left.\Ch^{E_6/D_4}_\rho(RR,\nu'_\Psi,\nu'_H,\bar\tau)
	\right|_{\nu'_\Psi=\nu'_H=0}\,.
\eqn\ramondchar
$$
Here we have again used the vanishing of the Ramond-Ramond character and
its diagonal second derivatives at $\nu'_\Psi=\nu'_H=0$.

To calculate the remaining derivatives in eq.~\ramondchar,
we factorize the $E_6/D_4$ character into the $SO(2)$ character times
the character of the $H$-boson sector and apply the $\partial/\partial\nu'$
derivatives accordingly.
For the $SO(2)$ character we have
$$
\left.\partder{}{\nu_\Psi}\,\Ch^{SO(2)}(RR,\nu_\Psi,\bar\tau)
	\right|_{\nu_\Psi=0}\
=\ 2\pi i\,\eta^2(\bar\tau) .
\eqno\eq
$$
Thus, the entire complicated expression on the second line of eq.~\factorB\
reduces to
$$
{i\sqrt{3}\over2\pi\eta^2(\tau)}\,\left.\partder{}{\nu_H}\,
    \Ch^H_\rho(RR,\nu_H,\bar\tau)\right|_{\nu_H=0}\
=\ {-1\over\eta^2(\tau)}\,
\Tr_{\rho,R}\left( F\,(-1)^{F+{3\over2}}\,
    \bar q^{\bar L-{1\over24}}\right)_H\,.
\eqn\penult
$$
Finally, we substitute eq.~\penult\ into eq.~\factorB, combine the traces over
the $H$-boson sector and over the model-dependent $c=(22,8)$ sector
and identify
the fermion number $F$ of the $H$-boson sector with the $\bar F$ of the entire
anti-holomorphic side of the internal SCFT.
The result is
$$
\B_a(\tau,\bar\tau)\
=\ {-1\over \eta^2(\tau)}\,
    \Tr_{\Rb}\left( (-)^{\Fb-{3\over2}}\,\Fb\
    \left( T^2_{(a)} - \coeff{k_a}{8\pi\tau_2} \right)
    q^{L-{11\over12}}\ {\bar q}^{\bar L-{3\over8}} \right)_{\rm int}
\eqno\ramondright
$$

\APPENDIX{B}{B\break
    Moduli Dependence of Matter-Field Metrics in Orbifolds.}
In this Appendix, we derive eqs.~\metricorbi\
and calculate the exponents $q_I^i$ in terms of the orbifold
parameters of the respective matter fields $\matter^I$.
Generally, in order to derive the parameters of the low-energy EQFT
from the string theory, one calculates scattering amplitudes using either
the string theory or the EQFT and demands that the two amplitudes for the
same physical process agree with each other in the low-energy limit.
The problem at hand involves the moduli dependence of the tree-level
$Z_{\Ib J}$ matrix for the matter fields, so we are going to calculate
the tree-level four-particle amplitudes
${\cal A}^{(0)}(\modb^\ib,\matterb^\Ib,\matter^J,\mod^j)$ for the scattering of
the moduli scalars $\mod^j$ off the matter scalars $\matter^J$.

In EQFT, tree-level modulus-matter scattering is due to Einsteinian gravity
and also due to sigma-model-like interactions
arising from the moduli dependence of the $Z_{\Ib J}(\mod,\modb)$ matrix.
Thus, following a similar calculation in ref.~\DKLa, we have
$$
\def\onefour{\vcenter to 1in{\kern -3pt
	\hbox{$\modb^\ib$}\vfil\hbox{$\mod^j$}}}
\def\twothree{\vcenter to 1in{\kern -3pt
	\hbox{$\matterb^\Ib$}\vfil\hbox{$\matter^J$}}}

\checkex 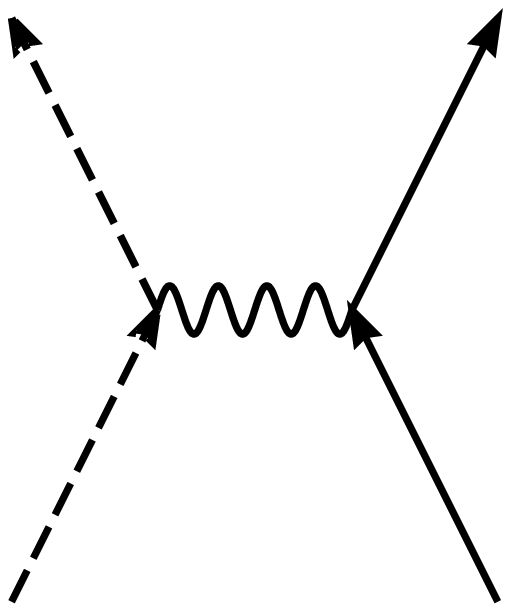
  \iffigureexists \checkex 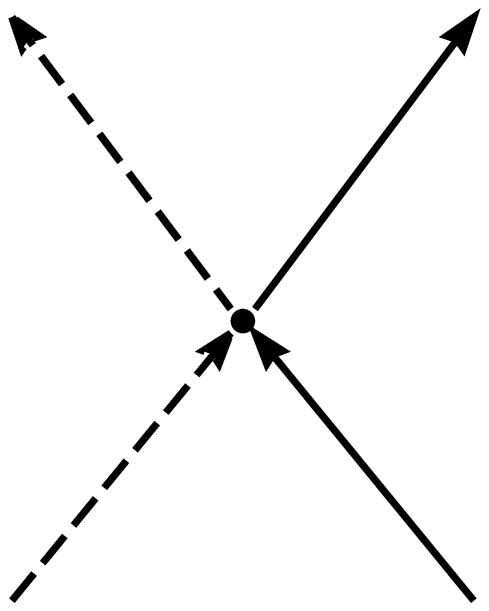
    \iffigureexists \checkex scatter2.eps
    \fi
  \fi
\iffigureexists
  \def\diagline{=\frame{scatter1}&+\frame{scatter2}&+\frame{scatter3}}
\else
  \def\diagline{\multispan3 \bf Missing Diagrams\hfil }
\fi
\displaylines{
{\cal A}^{(0)}_{\rm EQFT}(\modb^\ib,\matterb^\Ib,\matter^J,\mod^j)
  \hfill\eqname\EQFTscattering\crr
\vbox{\ialign{&$\displaystyle{#}$\ \hfil\cr
	\diagline \cr
	=\ \kappa^2{su\over t}\,Z_{\Ib J}G_{\ib j} &
	+\ s\,\partder{^2 Z_{\Ib J}}{\modb^\ib\,\partial\mod_j} &
	-\ s\,\partder{Z_{\Ib K}}{\modb^\ib}\left(Z^{-1}\right)^{K\bar L}
		\partder{Z_{\bar L J}}{\mod^i}\,,\cr
}}\cr
}
$$
where $s=-(k_1+k_2)^2$, $t=-(k_2+k_3)^2$ and $u=-(k_1+k_3)^2$
are Mandelstam's kinematic variables
and $G_{\ib j}=\kappa^{-2}\partial_\ib\partial_j\KM$ is the metric of the
moduli space.

Later in this Appendix, we will show that for the factorizable orbifolds,
the tree-level string-theoretical amplitudes for scattering of the
untwisted moduli off the matter scalars look like
$$
{\cal A}^{(0)}_{\rm string}(\modb^\ib,\matterb^\Ib,\matter^J,\mod^j)\
=\ \kappa^2\delta_{\ib j}\delta_{\Ib J}\left(
	{su\over t}\ +\ sq^j_J\ +\ O(\alpha'k^4)\right).
\eqn\STRscattering
$$
At this point, however, we would like to derive the eq.~\metricorbi\ from
eqs.~\EQFTscattering\ and \STRscattering\ before we proceed to derive
the eq.~\STRscattering\ itself.

As written, the amplitudes \EQFTscattering\ and \STRscattering\ assume
different normalization conventions for the external particles:
The string-theoretical amplitude~\STRscattering\ assumes them to be
canonically normalized while the field-theoretical amplitude~\EQFTscattering\
assumes that the particles are normalized exactly as the fields $\modb^\ib$,
$\matterb^\Ib$, $\matter^J$ and $\mod^i$.
Translating the string amplitude into the field-theoretical normalization
conventions gives
$$
\kappa^2 G_{\ib j}Z_{\Ib J}
\left( {su\over t}\ +\ sq^j_J\ +\ O(\alpha'k^4) \right),
\eqn\STRnormalized
$$
and it this formula that should agree in the low-energy limit with
the field-theoretical amplitude~\EQFTscattering.
By inspection of the two amplitudes \EQFTscattering\ and \STRnormalized,
their low-energy limits are similar and the only not-trivial requirement
of complete agreement has to do with the terms proportional to the
Mandelstam's $s$.
Comparing their coefficients, we arrive at
$$
\eqalign{
\partial_\ib\left( Z^{-1}\partial_j Z\right)^I_{\,J}\ &
\equiv\ \left( Z^{-1}\partial_\ib\partial_j Z
    - Z^{-1}\partial_\ib Z\, Z^{-1}\partial_j Z\right)^I_{\,J}\cr
&=\ \kappa^2 G_{\ib j}\delta^I_J\,q^j_J
    =\ {\delta_{\ib j}\delta^I_J q^j_J \over (\mod^j+\modb^j)^2}\,;\cr
}\eqn\Zdiffeq
$$
the last equality here follows from eq.~\orbikahler.

It remains but to solve the differential equations~\Zdiffeq\ for
the moduli dependence of the $Z_{\Ib J}$ matrix.
It is easy to see that a generic solution of these equations looks like
$$
Z_{\Ib J}(\mod,\modb)\
= \sum_{L} F^*_{\Ib L}(\modb) F_{J L}(\mod)\,
\prod_i\left(\mod^i+\modb^i\right)^{-q^i_L}
\eqn\generalsoln
$$
where $F_{J L}(\mod)$ is an arbitrary non-degenerate
matrix of holomorphic functions of the moduli.
This arbitrariness reflect our freedom to use arbitrary (but holomorphic)
moduli-dependent coordinates for the space of the matter fields $\matter^I$.
Without loss of generality, we may choose those coordinates such that
$F_{JL}(\mod)\equiv\delta_{JL}$; with this choice, eq.~\generalsoln\
reduces to the eq.~\metricorbi.

Let us now turn to the derivation of the string-theoretical tree-level
scattering amplitude~\STRscattering; we use the formalism and many
explicit results of refs.~\DFMS.
In the Hamiltonian formalism for the world-sheet quantities, we have
$$
{\cal A}^{(0)}_{\rm string}(\modb^i,\matterb^I,\matter^J,\mod^j)\
=\ \gstring^2 \int\!\!d^2z\,
\bra{\matter^I} T\left(V_{\modb^i}(w),V_{\mod^i}(z)\right) \ket{\matter^J},
\eqn\stringformula
$$
where $\ket{\matter^J}$ and $\bra{\matter^I}$ are the asymptotic initial
and final states of the matter scalar and
$T$ is the ``time''-ordered product of the vertex operators for the moduli.
For the untwisted moduli of a factorizable orbifold, these operators are
$$
\eqalign{
V_{\moo^j}\ & =\ \exp(ik_\mu X^\mu)\,\partial \Xb^j\,
	\left(\bar\partial X^j + \half k_\mu \psi^\mu \psi^j \right),\cr
V_{\moob^i}\ & =\ \exp(ik_\mu X^\mu)\,\partial X^i\,
	\left(\bar\partial \Xb^i + \half k_\mu \psi^\mu \psib^i \right),\cr
V_{\mot^j}\ & =\ \exp(ik_\mu X^\mu)\,\partial X^j\,
	\left(\bar\partial X^j + \half k_\mu \psi^\mu \psi^j \right),\cr
V_{\motb^i}\ & =\ \exp(ik_\mu X^\mu)\,\partial \Xb^i\,
	\left(\bar\partial \Xb^i + \half k_\mu \psi^\mu \psib^i \right)\cr
}\eqn\modvertices
$$
(in string units $\alpha'=\half$).
For the sake of notational simplicity, we are going to consider
the $(1,1)$ moduli $\moo^i$ first and only then address
the $(1,2)$ moduli $\mot^i$.

We begin by explaining the $\delta_{\ib j}\delta_{\Ib J}$ factor
in eq.~\STRscattering\ (for the $(1,1)$ moduli), which means that
the scattering process $\matter^J+\moo^j\to\matter^I+\moo^i$
is possible only if $i=j$ and $I=J$.
Clearly, $\matter^J$ and $\matter^I$ must belong to
the same twist sector of the orbifold, and if that sector is twisted,
they must originate in the same fixed point (or fixed sub-torus) of the
internal six-torus.
For any given twisted sector of any supersymmetric orbifold,
all the matter fields arising from that sector have
identical structures as far as the right-moving
world-sheet degrees of freedom are concerned.
Similarly, all the matter fields arising from the completely untwisted
sector have the same trivial structure with regard to the left-moving
$\partial X^i$ and $\partial\Xb^i$.
Hence, either the right-moving or the left-moving
creation/annihilation operators contained in the moduli vertex
$V_{\moo^j}$ have to cancel against those in the other vertex $V_{\moob^i}$,
and this is possible only if $i=j$.
On the other hand, if $i=j$, then the operator product
$T\left(V_{\moob^i}(w),V_{\moo^j}(z)\right)$ is completely diagonal
with respect to all the massless matter scalars, so we must have $I=J$
as well.

Next, consider the diagonal matrix elements
$\bra{\matter^I} T\bigl(V_{\moob^i},V_{\moo^i}\bigr) \ket{\matter^I}$.
Like all vertex correlators on the spherical world sheet, these matrix
elements are products of the holomorphic and the
antiholomorphic factors corresponding to the two world-sheet chiralities.
On the left-moving, bosonic side of the heterotic string, we have
$$
\displaylines{
\bra{\matter^I} T\bigl(V_{\moob^i}(w,\bar w),
	V_{\moo^i}(z,\bar z)\bigr) \ket{\matter^I}_{\rm left}
\hfill\eqname\boseside\cr
\noalign{\vskip 5pt plus 10pt}
\hfill{}=\ w^{-s/8} z^{-u/8}(z-w)^{-t/8}\,\bra{\matter^I}
    T\left(\partial X^i(w),\partial\Xb^i(z)\right) \ket{\matter^I} .
\qquad\cr }
$$
Let $2\pi\eta^i_I$ ($0\le\eta^i_I<1$) be the angle by which the $i$th
internal plane is rotated in the twist sector giving rise to the matter
particle $\matter^I$.
Then in that sector, the operators $\partial X^i$ and $\partial\Xb^i$
decompose according to
$$
\eqalign{
\partial\Xb^i(z)\ &
= \sum_{n=-\infty}^{+\infty} \bar\alpha^i_{n+\eta} z^{-1-n-\eta},\cr
\partial X^i(w)\ &
= \sum_{m=-\infty}^{+\infty} \alpha^i_{-m-\eta} w^{-1+m+\eta},\cr
}\eqn\bosedecomp
$$
where $[\bar\alpha^i_{n+\eta},\alpha^i_{-m-\eta}]=(n+\eta)\delta_{nm}$
and $\eta\equiv\eta^i_I$.
Generally, the state $\ket{\matter^I}$ may have $N^i_I\ge0$ quanta
of the $(\bar\alpha_\eta,\alpha_{-\eta})$ oscillator
or $\Nb^i_I\ge0$ quanta of the $(\alpha_{1-\eta},\bar\alpha_{\eta-1})$
oscillator,
\foot{In the untwisted sector and in the twisted sectors with $\eta^i=0$,
    all matter particles have $N^i_I=\Nb^i_I=0$.
    In other twisted sectors, all particles have either $\Nb^i_I=0$ and
    $\eta^i_I N^i_I<1$ or $N^i_I=0$ and $(1-\eta^i_I)\Nb^i_I<1$.
    }
but it must be the ground state of all the other oscillators of the $i$th
plane.
Therefore,
$$
\!\bra{\matter^I} T\left(\partial X^i(w),\partial\Xb^i(z)\right)
\ket{\matter^I}
=w^{\eta-1}z^{-\eta}\left[ {(1-\eta)w+\eta z\over(z-w)^2}
    +{\eta \over z}N^i_I+{(1-\eta)\over w}\Nb_I^i\right]
\eqn\bosematel
$$
(again, $\eta\equiv\eta^i_I$);
substituting this matrix element into eq.~\boseside\ and integrating by parts,
we arrive at
$$
\!\!\eqalign{
\bra{\matter^I} T\bigl(V_{\moob^i}(w,\bar w),
	V_{\moo^i}(z,\bar z)\bigr) &
\ket{\matter^I}_{\rm left}\ =\ {-w^{\eta-1-s/8}} z^{-\eta-u/8}(z-w)^{-1-t/8}\cr
&\times\left[ {u+(1-\eta)t\over 8+t}\
	+\  {\eta t\over 8\eta+u}N^i_I\
	-\ {(1-\eta)t\over 8(1-\eta)+s}\Nb^i_I \right]\cr
}\!\eqn\bosesidedone
$$
plus a total-holomorphic-derivative term that would not contribute to the
integral~\stringformula.

Now consider the right-moving, supersymmetric side of the heterotic string,
where we have
$$
\eqalign{
\bra{\matter^I} T\bigl(V_{\moob^i}(w,\bar w),
	V_{\moo^i}(z,\bar z)\bigr) \ket{\matter^I}_{\rm right}\
    =\ \bar w^{-s/8} &
\bar z^{-u/8} (\bar z-\bar w)^{-t/8}\cr
{}\times\left[ \bra{\matter^I} T\left(\bar\partial\Xb^i(\bar w),
	\bar\partial X^i(\bar z)\right) \ket{\matter^I}\right.\quad\cr
{}+\ {t\over8} \bra{\matter^I}T\left(\psi^\mu(\bar w),
	\psi^\mu(\bar z)\right) \ket{\matter^I} &
\left.\bra{\matter^I}T\left(\psib^i(\bar w),
	\psi^i(\bar z)\right) \ket{\matter^I}\right] .\cr
}\eqn\superside
$$
Clearly, the bosonic matrix elements on the right hand side here
is simply the complex conjugate of eq.~\bosematel\ for the special
case $N^i_I=\Nb^i_I=0$,
$$
\bra{\matter^I} T\left(\bar\partial\Xb^i(w),\bar\partial X^i(z)\right)
\ket{\matter^I}\
=\ \bar w^{\eta-1}\bar z^{-\eta}\,
{(1-\eta)\bar w+\eta \bar z\over(\bar z-\bar w)^2}\,.
\eqn\superXmatel
$$
The fermionic matrix elements in eq.~\superside\ follow from the
decomposition
$$
\eqalign{
\psi^i(z)\ &
= \sum_{r=-\infty}^{+\infty} \psi^i_{r+\eta} z^{-{1\over2}-r-\eta},\cr
\psib^i(w)\ &
= \sum_{p=-\infty}^{+\infty} \psib^i_{-p-\eta} w^{-{1\over2}+p+\eta},\cr
}\eqn\fermidecomp
$$
where $\{\psi^i_{r+\eta},\psib^i_{-p-\eta}\}=\delta_{rp}$
and --- in the sector containing the $\matter^I$ ---
$r$ and $p$ are half-integers and $\eta\equiv\eta^i_I$.
For any of the twisted sectors of the orbifold, all matter states
$\ket{\matter^I}$ are annihilated by all the
$\psi^i_{r+\eta}$ with $r\ge\half$ and all the $\psib^i_{-p-\eta}$
with ${-p}\ge\half$ (because of the GSO projection, this is true
even for the $\psib^i_{{1\over2}-\eta}$ when $\half-\eta^i_I<0$).
In the completely untwisted sector, however, the right-moving sides
of the matter states are formed according to $\psi^\ell_{-{1\over2}}\ket{0}$,
so one should distinguish between the cases $\ell_I=i$ and $\ell_I\neq i$.
Therefore,
$$
\!\bra{\matter^I}T\left(\psib^i(\bar w),\psi^i(\bar z)\right)
\ket{\matter^I}
= {-1\over \bar z-\bar w}\times\cases{
    (\bar w/\bar z)^{\eta^i_I} &
    {\tenrm for all twisted states,}\cr
    1 &
    {\tenrm for untwisted states with} $\ell_I\neq i$,\cr
    (\bar w/\bar z) &
    {\tenrm for untwisted states with} $\ell_I=i$;\cr
    }
\eqn\fermimatel
$$
similarly,
$$
\bra{\matter^I}T\left(\psi^\mu(\bar w),\psi^\mu(\bar z)\right) \ket{\matter^I}\
=\ {-1\over \bar z-\bar w}\,.
\eqn\psimumatel
$$
Substituting the matrix elements \superXmatel, \fermimatel\ and \psimumatel\
into eq.~\superside\ and integrating by parts, we arrive at
$$
\displaylines{
\bra{\matter^I} T\bigl(V_{\moob^i}(w,\bar w),
	V_{\moo^i}(z,\bar z)\bigr) \ket{\matter^I}_{\rm right}
\hfill\eqname\supersidedone\cr
=\ \bar w^{\eta-1-s/8}\bar z^{-\eta-u/8}(\bar z-\bar w)^{-1-t/8}\cr
\hfill{}\times\cases{
	-s^2/8u & for untwisted states with $\ell_I=i$,\cr
	+s/8 & for all other states,\cr
	}\qquad\cr
}
$$
plus a total-antiholomorphic-derivative term.

At this point, all we need to do is to substitute eqs.~\bosesidedone\
and \supersidedone\ into eq.~\stringformula\ and perform the $d^2z$ integral.
Taking the low-energy limit $|s|,|t|,|u|\ll1$ of the resulting expression,
we finally arrive at
$$
{\cal A}^{(0)}_{\rm string}(\moob^i,\matterb^I,\matter^J,\moo^j)\
=\ \coeff18 \gstring^2\,\delta_{ij}\delta_{IJ}\left(
	{su\over t}\ +\ s\,q^i_I\ +\ s\,O(s,t,u)\right).
\eqn\mooscattering
$$
where
$$
q^i_I\ = \cases{
(1-\eta^i_I) +N^i_I -\Nb^i_I &
for all twisted states,\cr
0 &
for untwisted states with $\ell_I\neq i$,\cr
1 &
for untwisted states with $\ell_I=i$.\cr
}\eqn\qformula
$$
Note that eq.~\mooscattering\ is written in string units $\alpha'=\half$;
translating it into the conventional units gives us eq.~\STRscattering\
for the $(1,1)$ moduli.

The above arguments leading to eq.~\mooscattering\ can be applied
almost verbatim to the scattering amplitudes involving the $(1,2)$
moduli $\mot^i$ instead of the $(1,1)$ moduli $\moo^i$.
In light of the vertex operators~\modvertices, the right-moving
degrees of freedom do not distinguish between the two kinds of the
untwisted moduli, so the right-moving matrix element~\supersidedone\
is exactly the same in both cases.
However, on the left-moving (bosonic) side of the heterotic string,
replacing $\moo^i$ with $\mot^i$ and $\moob^i$ with $\motb^i$
requires an interchange between the
$\partial X^i$ and the $\partial\Xb^i$.
Hence, in the left-moving matrix element~\bosesidedone, one should
interchange $N^i_I$ with $\Nb^i_I$ and $\eta^i_I$ with $(1-\eta^i_I)$,
unless $\eta^i_I=0$, in which case there are no modifications at all.
Actually, if a factorizable orbifold does have an unfrozen $(1,2)$ modulus
$\mot^i$, the only allowed values of $\eta^i_I$ are $0$ and $\half$
\foot{If any sector of the orbifold were to rotate the $i$th
    internal plane by any angle other than zero or $\pi$,
    the value of the $\mot^i$ modulus of the internal torus would be
    completely frozen by the orbifolding procedure.
    }
and so the interchange $\eta^i_I\leftrightarrow(1-\eta^i_I)$ is never needed.
Therefore,
$$
{\cal A}^{(0)}_{\rm string}(\motb^i,\matterb^I,\matter^J,\mot^j)\
=\ \coeff18 \gstring^2\,\delta_{ij}\delta_{IJ}\left(
	{su\over t}\ +\ s\,\tilde q^i_I\ +\ s\,O(s,t,u)\right).
\eqn\motscattering
$$
where $\tilde q^i_I$ is exactly as in eq.~\qformula, except for the
interchange $N^i_I\leftrightarrow\Nb^i_I$.

Most matter fields $\matter^I$ have $\tilde q^i_I=q^i_I$ and thus
contribute equally to the ``modular anomaly'' coefficients $\Alpha^i_a$
(\cf~eq.~\Alphadef) for the $(1,2)$ modulus $\mot^i$
and the $(1,1)$ modulus $\moo^i$ of the same internal plane.
The only exception to this rule are the $\matter^I$ for which
$\eta_I^i=\half$ and either $N_I^i=1$ and $\Nb_I^i=0$ or else
$N_I^i=0$ and $\Nb_I^i=1$.
When the orbifold group never twists the $i$th plane by any angle
other than zero or $\pi$, such states always come in pairs:
They have identical gauge and other quantum numbers and the only
difference between the two members of a pair is $N^i_I\leftrightarrow\Nb^i_I$.
Hence, one member of a pair has $q^i_I={3\over2}$ and $\tilde q^i_I=-\half$
while the other has $q^i_I=-\half$ and $\tilde q^i_I={3\over2}$
and the net contribution of such a pair to the coefficients
$\Alpha^i_a$ is zero --- again for either of the two
moduli of the $i$th plane.

Actually, the full story of such pairs of matter field is more
complicated since they allow for non-diagonal scattering amplitudes in
which one member of a pair turns into the other member while the $\mot^i$
modulus turns into the $\moo^i$ or vice verse.
Therefore, we should add some non-diagonal terms to the right-hand side
of eq.~\Zdiffeq, and the resulting moduli-dependent matrix $Z_{\Ib J}$
is not quite as diagonal as in eq.~\metricorbi.
Instead, for the $N^i_I\leftrightarrow\Nb^i_I$ pairs of matter fields
--- and only for such pairs, the rest of the matrix $Z_{\Ib J}(\mod,\modb)$
is exactly as in eq.~\metricorbi, ---
we have inextricably entangled $2\times2$ blocks of rather complicated
moduli dependence.
However, the determinants of such $2\times2$ blocks satisfy exactly the
same differential equations as if the non-diagonal scattering amplitudes
did not exist.
Consequently, while the moduli transformation rules for the
$N^i_I\leftrightarrow\Nb^i_I$ pairs are more complicated than eq.~\matterorbi,
their effect upon the moduli anomalies
of the Wilsonian gauge couplings is exactly as in eq.~\forbitrans,
with the net contribution of each pair to the $\Alpha_a^i$ being exactly
as if the pair had $q^i_I=({3\over2},-\half)$.
Specifically, the net contribution of a $N^i_I\leftrightarrow\Nb^i_I$ pair
of matter fields to the $\Alpha^i_a$ is exactly zero.
This completes the proof that {\sl whenever a factorizable orbifold has an
unfrozen $(1,2)$ modulus $\mot^i$, that modulus has exactly the same
modular anomaly coefficients $\Alpha^i_a$ as the $(1,1)$ modulus
$\moo^i$ of the same internal plane.}

\APPENDIX{C}{C\break Examples of Factorizable Orbifolds}
%
%
\chapternumber=3 \chapterstyle={\Alphabetic}
In this Appendix, we verify eqs.~\coeffrel\ for all factorizable
$N=2$ orbifolds and for several examples of factorizable $N=1$
orbifolds.
To save space, we present only a dozen of the $N=1$ orbifolds here,
but we have actually investigated many more, and for all those
orbifolds we found eqs.~\coeffrel\ holding true for all three
internal planes and all the gauge couplings of the orbifold.

\section{Factorizable $N=2$ Orbifolds.}
A supersymmetric orbifold has unbroken $N=2$ spacetime supersymmetry when
the orbifold group $D$ never rotates one of the internal planes.
In the notations of eq.~\DorbiS,
for a factorizable orbifold of this kind, one has
$D_1=D$ and $\cren_a^{N=2}(1)=\cren_a$ for the untwisted first plane
while for the other two planes, $D_2=D_3=1$
and $\cren_a^{N=2}(2)=\cren_a^{N=2}(3)=0$.
Let us now demonstrate that for all such orbifolds, one also has
$$
\Alpha_a^1\ =\ \cren_a\qquad {\rm and}\quad
\Alpha_a^2\ =\ \Alpha_a^3\ =\ 0.
\eqn\Ntworel
$$
This would not only confirm eqs.~\coeffrel\ for all three planes and all
the gauge couplings but also show that $\vgs^1=\vgs^2=\vgs^3=0$.

The $N=2$ spacetime supersymmetry
has two kinds of massless supermultiplets containing
scalar particles, namely the vector multiplets and the hypermultiplets.
In $N=2$ orbifolds of the heterotic string, scalars belonging to vector
supermultiplets include the dilaton $S$, the two moduli $\moo^1$ and $\mot^1$
of the untwisted plane, and the matter fields $\matter^I$ originating
in completely untwisted string states with $\ell_I=1$;
all other moduli and matter scalars belong to hypermultiplets.
For the matter scalars, this distinction parallels eq.~\qformula\ for $i=1$:
All matter scalars arising from twisted states have $\eta_I^1=0$ and hence
$q_I^1=0$, the untwisted states with $\ell_i\neq1$ also have $q_I^1=0$,
but the untwisted states with $\ell_I=1$ have $q_I^1=1$.
Therefore, eqs.~\Alphadef\ for $i=1$ reduce to
$$
\eqalign{
\Alpha_a^1\ &
= \sum_I^{\rm hyper} T_a(\matter^I)\
	-\sum_I^{\rm vector} T_a(\matter^I)\ -\ T(G_a)\cr
&= \sum_I^{\rm hyper} T_a(\matter^I)\ -\ 2T(G_a)\
	\equiv\ \cren_a\quad ({\rm for}\ N=2).\cr
}\eqn\firstplane
$$

The fact that all the $\matter^I$ belonging to hypermultiplets have $q_i^1=0$
and thus have metrics $Z_{\Ib J}$ that do not depend on $\moo^1$ and $\mot^1$
reflects a universal property of $N=2$ EQFTs:  The metric for
the hypermultiplets does not depend on the vector superfields and vice verse.
Thus, since the moduli $\moo^{2,3}$ and $\mot^{2,3}$ belong to hypermultiplets,
we should also have $q^2_I=q^3_I=0$ for $\matter^I$ in vector
supermultiplets, which is indeed the case according to eq.~\qformula.
On the other hand, the metric matrix for the hypermultiplets may depend on
the $\moo^{2,3}$ and $\mot^{2,3}$, but the resulting field space should
have a quaternionic geometry.
Consequently, for any two matter scalars $\matter^1$ and $\matterb^2$
belonging to the same hypermultiplet, $q_1^i+q_2^i=1$ for $i=2,3$;
because of our present focus on the orbifolds, we prefer to derive this result
from eqs.~\qformula\ instead of going through a more general $N=2$ analysis.

Consider a matter hypermultiplet $(\matter^1,\matterb^2)$ arising from
a twisted sector.
In $N=1$ terms, the two chiral supermultiplets $\matter^1$ and
$\matter^2$ arise from oppositely twisted sectors;
hence, in eqs.~\qformula, one should use $\eta^i_2=1-\eta^i_1$ for $i=2,3$.
Similarly, $N^i_2=\Nb^i_1$ and $\Nb^i_2=N^i_1$;
thus, for the twisted hypermultiplets, eqs.~\qformula\ give $q^i_1+q^i_2=1$
for $i=2,3$.
On the other hand, for the untwisted hypermultiplets, both $\matter^1$ and
$\matter^2$ arise from the same untwisted sector.
However, if $\matter^1$ has $\ell_1=2$ then $\matter^2$ has $\ell_2=3$ and
vice verse, if $\ell_1=3$ then $\ell_2=2$.
Thus, according to eqs.~\qformula, we again have $q^i_1+q^i_2=1$
for $i=2,3$.
Since $\matter^1$ and $\matter^2$ always have exactly opposite gauge
quantum numbers, $T_a(\matter^1)=T_a(\matter^2)$ for all $a$.
Therefore, according to eqs.~\Alphadef,
the net contribution of any matter hypermultiplet to
the modular anomalies $\Alpha_a^{2,3}$ is precisely zero.
Consequently,
$$
\Alpha^2_a\ =\ \Alpha_a^3\
=\sum_I^{\rm vector} T_a(\matter^I)\ -\ T(G_a)\ =\ 0.
\eqn\twistedplanes
$$

Eqs.~\twistedplanes\ do not merely confirm eqs.~\coeffrel\ for the moduli
of the twisted planes, they also verify a stronger string-EQFT
consistency condition required by the unbroken $N=2$ spacetime supersymmetry.
Specifically, in a locally $N=2$ supersymmetric EQFT, the gauge couplings are
not allowed to depend on any hypermultiplets, and any mixing between
the hypermultiplets and the dilaton $S$
(which belongs to a vector supermultiplet) is also forbidden.
Thus, at the one-loop level, we must have $\Alpha_a^2=\Alpha_a^3=0$
and also $\vgs^2=\vgs^3=0$, and both requirements are indeed upheld by
eq.~\twistedplanes.
($\vgs^2=\vgs^3=0$ follows from eq.~\coeffrel\
and the fact that $\cren_a(i=2,3)\equiv0$ for an $N=2$ orbifold.)

On the other hand, the fact that eq.~\coeffrel\ for the untwisted first
plane holds for $\vgs^1=0$ does not have any profound significance from the
$N=2$ point of view.
Both the dilaton $S$ and the moduli $\moo^1$ and $\mot^1$ of the
untwisted plane belong to vector supermultiplets of the $N=2$ supersymmetry,
so there is no reason why they should not mix with each other in
the perturbative string theory.
In fact, they do mix with each other: According to eq.~\Vorbi,
$$
V^{(1)}_{N=2}(\mod,\modb)\ =\ \romega(\moo^1,\mot^1,\moob^1,\motb^1) ,
\eqn\Ntwomixing
$$
where $\romega$ obtains from an explicit calculation of
the entire string-theoretical threshold corrections $\Ddif_a(\mod,\modb)$
rather than the differences $\Ddif_a-\Ddif_{a'}$.
The explicit form of this mixing is given by eq.~\romegadef;
the derivation of this formula and its physical implications will
be presented in a forthcoming article.

\section{$Z_3$ Orbifolds.}
The orbifold group of the $N=1$ supersymmetric $Z_3$ orbifolds\refmark{\DHVW}
is generated by the rotation $\Theta=(e^{2\pi i/3},e^{2\pi i/3},e^{2\pi i/3})$.
There are five inequivalent modular-invariant ways this group may act on the
$E_8\times E_8$ degrees of freedom;
hence, there are five distinct $Z_3$ orbifolds, with unbroken gauge symmetries
being respectively $E_8\times E_8$, $\left(E_6\times SU(3)\right)\times E_8$,
$\left(E_6\times SU(3)\right)\times \left(E_6\times SU(3)\right)$,
$\left(SO(14)\times U(1)\right)\times SU(9)$ and
$\left(SO(14)\times U(1)\right)\times\left( E_7\times U(1)\right)$.

A $Z_3$ orbifold has no $(1,2)$ moduli but nine $(1,1)$ moduli, and
for generic values of these moduli, the orbifold is not quite factorizable
--- the three internal planes are not mutually orthogonal but mix with
each other.
However, since the purpose of this Appendix is to present examples
of factorizable orbifolds, we impose factorizability by fiat, \ie,
we assume that all six of the off-diagonal $(1,1)$ moduli have zero
values and concentrate on the way the gauge couplings depends on the
three diagonal moduli $\moo^{1,2,3}$.
Specifically, we are going to calculate the modular anomaly coefficients
\Alphadef\ and show that
$$
\Alpha_a^i\ =\ k_a \vgs,
\eqn\Zthreecoeffrel
$$
in full agreement with eq.~\coeffrel\ for orbifolds without
$N=2$ supersymmetric sectors.
Because of the obvious symmetry, the coefficient $\vgs$ in eq.~\Zthreecoeffrel\
is always the same for all three internal planes of a $Z_3$ orbifold;
however, the $Z_3$ orbifolds with different gauge groups generally have
different values of~$\vgs$.

Let us start with the left-right symmetric $Z_3$ orbifold whose gauge group is
$G=E_6\times SU(3)\times E_8$.
The matter states for this orbifold are summarized in the following table
$$
\!\vcenter{\Tenpoint
\ialign{%
	\tablerule #&
	\hfil\enspace #\unskip \enspace\hfil &\vrule #&
	\hfil\quad${#}$\quad\hfil &\vrule #&
	\hfil\enspace #\unskip \enspace &\vrule #&
	\hfil\quad${#}$\quad\hfil &\vrule #&
	\hfil\quad${#}$\quad\hfil &\vrule #&
	\hfil\quad${#}$\quad\hfil &\vrule #\cr
\noalign{\hrule}
& sector && E_6\times SU(3)\times E_8 && \# &&
	\ell_I {\rm\ or\ } \eta_I^i && \rm oscillators &&
	{\rm average}\ q_I^i &\cr
\noalign{\hrule}
& untwisted && (\ts,{\bf3},{\bf1}) && 3 &&
	\ell=1,2,3 && \rm none && (\coeff13,\coeff13,\coeff13) &\cr
\noalign{\hrule}
& $\Theta$ && (\ts,{\bf1},{\bf1}) && 27 &&
	\eta=(\coeff13,\coeff13,\coeff13) && \rm none &&
	(\coeff23,\coeff23,\coeff23) &\cr
\noalign{\hrule}
& $\Theta$ && ({\bf1},{\bf\bar3},{\bf1}) && 81 &&
	\eta=(\coeff13,\coeff13,\coeff13) && \sum_i N^i=1,\ \Nb^i=0 &&
	(1,1,1) &\cr
\noalign{\hrule}
}}\!
$$
where the last column gives the average value of the $(q^1,q^2,q^3)$
(calculated according to eq.~\qformula)
for all the states in any given raw.
Substituting this spectrum --- and the values of the $q^i_I$ ---
into eqs.~\Alphadef\ and totalling the sums gives us
$$
\Alpha_{E(6)}^i\ =\ \Alpha_{SU(3)}^i\ =\ \Alpha_{E(8)}^i\ =\ -30.
\eqn\ZthreeLR
$$

Next consider the $Z_3$ orbifold with the completely unbroken
$E_8\times E_8$ gauge group.
This orbifold is somewhat peculiar since it has no untwisted matter
fields at all.
It does have 243 twisted matter fields, but all of them
are singlets under the gauge group and thus do not contribute
the modular anomalies of the gauge couplings.
Therefore,
$$
\Alpha_{E(8)}^i\ =\ \Alpha_{E(8)'}^i\ =\ -30.
\eqn\Zthreeunbr
$$

The next $Z_3$ orbifold has both of the $E_8$ groups twisted
in the same manner as the twisted $E_8$ of the left-right
symmetric orbifold;
its unbroken gauge group is
$G=E_6\times SU(3)\times E_6\times SU(3)$.
The matter states of this orbifold are as follows:
$$
\!\vcenter{\Tenpoint
\ialign{%
	\tablerule #&
	\hfil\enspace #\unskip \enspace\hfil &\vrule #&
	\hfil\quad${#}$\quad\hfil &\vrule #&
	\hfil\enspace #\unskip \enspace &\vrule #&
	\hfil\quad${#}$\quad\hfil &\vrule #&
	\hfil\quad${#}$\quad\hfil &\vrule #&
	\hfil\quad${#}$\quad\hfil &\vrule #\cr
\noalign{\hrule}
& sector && E_6\times SU(3)\times E_6\times SU(3) && \# &&
	\ell_I {\rm\ or\ } \eta_I^i && \rm oscillators &&
	{\rm average}\ q_I^i &\cr
\noalign{\hrule}
& untwisted && (\ts,{\bf3},{\bf1},{\bf1}) && 3 &&
	\ell=1,2,3 && \rm none && (\coeff13,\coeff13,\coeff13) &\cr
\noalign{\hrule}
& untwisted && ({\bf1},{\bf1},\ts,{\bf3}) && 3 &&
	\ell=1,2,3 && \rm none && (\coeff13,\coeff13,\coeff13) &\cr
\noalign{\hrule}
& $\Theta$ && ({\bf1},{\bf\bar3},{\bf1},{\bf\bar3}) && 27 &&
	\eta=(\coeff13,\coeff13,\coeff13) && \rm none &&
	(\coeff23,\coeff23,\coeff23) &\cr
\noalign{\hrule}
}}\!
$$
Therefore,
$$
\Alpha_{E(6)}^i\ =\ \Alpha_{E(6)'}^i\
=\ \Alpha_{SU(3)}^i\ =\ \Alpha_{SU(3)'}^i\
=\ -3.
\eqn\ZthreeLR
$$
Note that for this orbifold eq.~\Zthreecoeffrel\ holds true, but
$\vgs=-3$ rather than $-30$.
We shall see momentarily that the other two $Z_3$ orbifolds in which
both $E_8$ groups are broken also have $\vgs\neq-30$.

Indeed, the $Z_3$ orbifold with the $G=SO(14)\times U(1)\times SU(9)$
gauge group has the following matter fields:
$$
\!\vcenter{\Tenpoint
\ialign{%
	\tablerule #&
	\hfil\enspace #\unskip \enspace\hfil &\vrule #&
	\hfil\quad${#}$\quad\hfil &\vrule #&
	\hfil\enspace #\unskip \enspace &\vrule #&
	\hfil\quad${#}$\quad\hfil &\vrule #&
	\hfil\quad${#}$\quad\hfil &\vrule #&
	\hfil\quad${#}$\quad\hfil &\vrule #\cr
\noalign{\hrule}
& sector && SO(14)\times U(1)\times SU(9) && \# &&
	\ell_I {\rm\ or\ } \eta_I^i && \rm oscillators &&
	{\rm average}\ q_I^i &\cr
\noalign{\hrule}
& untwisted && ({\bf14},-1,{\bf1}) && 3 &&
	\ell=1,2,3 && \rm none && (\coeff13,\coeff13,\coeff13) &\cr
\noalign{\hrule}
& untwisted && ({\bf64},{+\half},{\bf1}) && 3 &&
	\ell=1,2,3 && \rm none && (\coeff13,\coeff13,\coeff13) &\cr
\noalign{\hrule}
& untwisted && ({\bf1},0,{\bf84}) && 3 &&
	\ell=1,2,3 && \rm none && (\coeff13,\coeff13,\coeff13) &\cr
\noalign{\hrule}
& $\Theta$ && ({\bf1},{+\coeff23},{\bf\bar9}) && 27 &&
	\eta=(\coeff13,\coeff13,\coeff13) && \rm none &&
	(\coeff23,\coeff23,\coeff23) &\cr
\noalign{\hrule}
}}\!
$$
where the $U(1)$ charges are normalized according to $k_{U(1)}=2$.
Substituting this table of matter fields into eq.~\Alphadef,
we obtain
$$
\Alpha^i_{SO(14)}\ =\ \Alpha^i_{SU(9)}\ =\ \half\Alpha^i_{U(1)}\
=\ -3.
\eqn\ZthreeSUnine
$$
Similarly, for the remaining $Z_3$ orbifold with the
$G=SO(14)\times U(1)\times E_7\times U(1)$ gauge group, the matter fields are
$$
\!\vcenter{\Tenpoint
\ialign{%
	\tablerule #&
	\hfil\enspace #\unskip \enspace\hfil &\vrule #&
	\hfil $\;#\;$\hfil &\vrule #&
	\hfil\enspace #\unskip \enspace &\vrule #&
	\hfil $\;#\;$\hfil &\vrule #&
	\hfil $\;#\;$\hfil &\vrule #&
	\hfil $\;#\;$\hfil &\vrule #\cr
\noalign{\hrule}
& sector && SO(14)\times U(1)\times E_7\times U(1) && \# &&
	\ell_I {\rm\ or\ } \eta_I^i && \rm oscillators &&
	{\rm average}\ q_I^i &\cr
\noalign{\hrule}
& untwisted && ({\bf14},{+1},{\bf1},0) && 3 &&
	\ell=1,2,3 && \rm none && (\coeff13,\coeff13,\coeff13) &\cr
\noalign{\hrule}
& untwisted && ({\bf64},{+\half},{\bf1},0) && 3 &&
	\ell=1,2,3 && \rm none && (\coeff13,\coeff13,\coeff13) &\cr
\noalign{\hrule}
& untwisted && ({\bf1},0,{\bf56},{+\half}) && 3 &&
	\ell=1,2,3 && \rm none && (\coeff13,\coeff13,\coeff13) &\cr
\noalign{\hrule}
& untwisted && ({\bf1},0,{\bf1},{-1}) && 3 &&
	\ell=1,2,3 && \rm none && (\coeff13,\coeff13,\coeff13) &\cr
\noalign{\hrule}
& $\Theta$ && ({\bf14},{-\coeff13},{\bf1},{+\coeff13}) && 27 &&
	\eta=(\coeff13,\coeff13,\coeff13) && \rm none &&
	(\coeff23,\coeff23,\coeff23) &\cr
\noalign{\hrule}
& $\Theta$ && ({\bf1},{+\coeff23},{\bf1},{-\coeff23}) && 27 &&
	\eta=(\coeff13,\coeff13,\coeff13) && \rm none &&
	(\coeff23,\coeff23,\coeff23) &\cr
\noalign{\hrule}
& $\Theta$ && ({\bf1},{+\coeff23},{\bf1},{+\coeff13}) && 81 &&
	\eta=(\coeff13,\coeff13,\coeff13) && \sum_i N^i=1,\ \Nb^i=0 &&
	(1,1,1) &\cr
\noalign{\hrule}
}}\!
$$
and hence
$$
\Alpha^i_{SO(14)}\ =\ \Alpha^i_{E(7)}\ =\ -12,\qquad
\Alpha^i_{U(1)}\ =\ -12 k_{U(1)}
\eqn\ZthreeEseven
$$
where $k_{U(1)}={2\,0\choose 0\,1}$ is the normalization matrix for the two
abelian gauge charges of the model.
Again, eqs.~\Zthreecoeffrel\ are satisfied
but for $\vgs=-12$ rather than $-30$.

\section{$Z_2\times Z_2$ Orbifolds.}
The orbifold group of the $Z_2\times Z_2$ orbifolds\refmark{\DHVW,\IFQ}
is generated by two
rotations, $\Theta_1=(-1,-1,+1)$ and $\Theta_2=(-1,+1,-1)$;
consequently, there are six untwisted moduli,
$\moo^{1,2,3}$ and $\mot^{1,2,3}$.
Again, there are five inequivalent modular-invariant embeddings of the
orbifold group
into the $E_8\times E_8$ Kac-Moody algebra and hence five distinct
$Z_2\times Z_2$ orbifolds, whose unbroken gauge symmetries are respectively
$\left( E_6\times U(1)^2\right)\times E_8$,
$\left( E_6\times U(1)^2\right)\times SO(16)$,
$\left( E_6\times U(1)^2\right)\times \left( SO(8)\times SO(8)\right)$,
\penalty-1000\space
$\left( SU(8)\times U(1)\right)\times \left( E_7\times SU(2)\right)$ and
$\left( SU(8)\times U(1)\right)\times
 \left( SO(12)\times SU(2)\times SU(2)\right)$.

All three internal planes of a $Z_2\times Z_2$ orbifolds have non-trivial
little groups $D_i=Z_2$ making non-trivial $N=2$ orbifolds.
Depending on a particular $Z_2\times Z_2$ orbifold and on a particular plane,
one may get either of the two $Z_2$ orbifolds:
The first has $G=E_7\times SU(2)\times E_8$ and the hypermultiplet spectrum
consisting of two copies of $({\bf 56},{\bf 2},{\bf 1})$, sixteen copies of
$({\bf 56},{\bf 1},{\bf 1})$ and sixty four copies of
$({\bf 1},{\bf 2},{\bf 1})$; consequently,
$$
\cren^{N=2}_{E(7)}\ =\ \cren^{N=2}_{SU(2)}\ =\ +84,\qquad
\cren^{N=2}_{E(8)}\ =\ -60.
\eqn\firstZtwo
$$
The second $Z_2$ orbifold has $G=E_7\times SU(2)\times SO(16)$ and
the hypermultiplet spectrum consisting of two copies of
$({\bf 56},{\bf 2},{\bf 1})$, two copies of $({\bf 1},{\bf 1},{\bf 128})$
and sixteen copies of $({\bf1},{\bf2},{\bf16})$; consequently
$$
\cren^{N=2}_{E(7)}\ =\ -12,\qquad
\cren^{N=2}_{SU(2)}\ =\ +180,\qquad
\cren^{N=2}_{SO(16)}\ =\ +36.
\eqn\secondZtwo
$$

For the left-right symmetric $Z_2\times Z_2$ orbifold,
each of the three little groups $D_i$ produces the
$N=2$ orbifold with $G^{N=2}=E_7\times SU(2)\times E_8$
thus giving us eqs.~\firstZtwo\ for the $\cren^{N=2}_a(i)$.
The $N=1$ orbifold itself has $G=E_6\times U(1)^2\times E_8$
and the following massless matter fields $\matter^I$:
$$
\!\vcenter{\Tenpoint
\ialign{%
	\tablerule #&
	\hfil\enspace #\unskip \enspace\hfil &\vrule #&
	\hfil\quad${#}$\quad\hfil &\vrule #&
	\hfil\enspace #\unskip \enspace &\vrule #&
	\hfil\enspace${#}$\enspace\hfil &\vrule #&
	\hfil\enspace${#}$\enspace\hfil &\vrule #&
	\hfil\enspace${#}$\enspace\hfil &\vrule #\cr
\noalign{\hrule}
& sector && E_6\times U(1)^2\times E_8 && \# &&
	\ell_I {\rm\ or\ } \eta_I^i && \rm osc. &&
	{\rm average}\ q_I^i &\cr
\noalign{\hrule}
& untwisted &&
	(\ts,+\half,+\half,{\bf1}) + ({\bf1},-\half,+\coeff32,{\bf1})
	+ {\rm c.c}&&
	1 && \ell=1 && 0 && (1,0,0) &\cr
\noalign{\hrule}
& untwisted &&
	(\ts,-\half,+\half,{\bf1}) + ({\bf1},+\half,+\coeff32,{\bf1})
	+ {\rm c.c}&&
	1 && \ell=2 && 0 && (0,1,0) &\cr
\noalign{\hrule}
& untwisted &&
	(\ts,0,-1,{\bf1}) + ({\bf1},+1,0,{\bf1}) + {\rm c.c}&&
	1 && \ell=3 && 0 && (0,0,1) &\cr
\noalign{\hrule}
& $\Theta_1$ &&
	(\ts,0,+\half,{\bf1}) + ({\bf1},0,+\coeff32,{\bf1}) &&
	16 && \eta=(\half,\half,0) && 0 && (\half,\half,0) &\cr
\noalign{\hrule}
& $\Theta_1$ &&
	({\bf1},\pm\half,0,{\bf1}) &&
	32 && \eta=(\half,\half,0) && 1 && (\half,\half,0) &\cr
\noalign{\hrule}
& $\Theta_2$ &&
	(\ts,+\coeff14,-\coeff14,{\bf1})
	+ ({\bf1},+\coeff34,-\coeff34,{\bf1}) &&
	16 && \eta=(\half,0,\half) && 0 && (\half,0,\half) &\cr
\noalign{\hrule}
& $\Theta_2$ &&
	({\bf1},\pm\coeff14,\pm\coeff34,{\bf1}) &&
	32 && \eta=(\half,0,\half) && 1 && (\half,0,\half) &\cr
\noalign{\hrule}
& $\Theta_1\Theta_2$ &&
	(\ts,-\coeff14,-\coeff14,{\bf1})
	+ ({\bf1},-\coeff34,-\coeff34,{\bf1}) &&
	16 && \eta=(0,\half,\half) && 0 && (0,\half,\half) &\cr
\noalign{\hrule}
& $\Theta_1\Theta_2$ &&
	({\bf1},\mp\coeff14,\pm\coeff34,{\bf1}) &&
	32 && \eta=(0,\half,\half) && 1 && (0,\half,\half) &\cr
\noalign{\hrule}
}}\!
$$
(the abelian gauge charges are normalized according to
$k_{U(1)}={1\,0\choose 0\,3}$).
Substituting the above spectrum and the values of $q_I^i$ into
eqs.~\Alphadef, we arrive at
$$
\Alpha^i_{E(8)}\ =\ -30,\qquad
\Alpha^i_{E(6)}\ =\ +42,\qquad
\Alpha^i_{U(1)}\ =\ +42 k_{U(1)},
\eqn\ZtwoZtwoLR
$$
which clearly agrees with eqs.~\firstZtwo\ and \coeffrel\ are for $\vgs^i=0$.

Next consider the $Z_2\times Z_2$ orbifold
in which each of the three twists $\Theta_1$, $\Theta_2$ and $\Theta_1\Theta_2$
would break the $E_8\times E_8$ down to
a $G^{N=2}=E_7\times SU(2)\times SO(16)$;
the combined effect of these twists leaves the $N=1$ orbifold with the unbroken
gauge symmetry $G=E_6\times U(1)^2\times SO(8)\times SO(8)$.
Naturally, for this orbifold, $\cren^{N=2}_a(i)$ are given by eqs.~\secondZtwo,
although a proper interpretation of that result requires paying careful
attention to the way $G$ is embedded into $G^{N=2}(i)$, which is
different for different~$i$.
Thus,
$$
\displaylines{
\cren^{N=2}_{E(6)}(i)\ =\ -12,\quad
\cren^{N=2}_{SO(8)}(i)\ =\ \cren^{N=2}_{SO(8)'}(i)\ =\ +36,
\qquad\hfill\eqname\threeembeddings\crr
\cren^{N=2}_{U(1)}(1)\, = \smallmat{36}{-144}{396},\quad
\cren^{N=2}_{U(1)}(2)\, = \smallmat{36}{+144}{396},\quad
\cren^{N=2}_{U(1)}(3)\, = \smallmat{180}{0}{-36},\hfill\cr }
$$
where on the first line $i=1,2,3$ and on the second line
we use the same basis for the two abelian charges as in the
previous $Z_2\times Z_2$ example.
The matter field spectrum of the present orbifold is
$$
\!\vcenter{\Tenpoint
\ialign{%
	\tablerule #&
	\hfil\enspace #\unskip \enspace\hfil &\vrule #&
	\hfil\enspace${#}$\enspace\hfil &\vrule #&
	\hfil\enspace #\unskip \enspace &\vrule #&
	\hfil\enspace${#}$\enspace\hfil &\vrule #&
	\hfil\enspace${#}$\enspace\hfil &\vrule #&
	\hfil\enspace${#}$\enspace\hfil &\vrule #\cr
\noalign{\hrule}
& sector && E_6\times U(1)^2\times SO(8)\times SO(8) && \# &&
	\ell_I {\rm\ or\ } \eta_I^i && \rm osc. &&
	{\rm avg.}\ q_I^i &\cr
\noalign{\hrule}
& untwisted &&
	(\ts,+\half,+\half,{\bf1},{\bf1})
	+ ({\bf1},-\half,+\coeff32,{\bf1},{\bf1}) + {\rm c.c}&&
	1 && \ell=1 && 0 && (1,0,0) &\cr
\noalign{\hrule}
& untwisted &&
	({\bf1},0,0,{\bf8},{\bf8})&&
	1 && \ell=1 && 0 && (1,0,0) &\cr
\noalign{\hrule}
& untwisted &&
	(\ts,-\half,+\half,{\bf1},{\bf1})
	+ ({\bf1},+\half,+\coeff32,{\bf1},{\bf1}) + {\rm c.c}&&
	1 && \ell=2 && 0 && (0,1,0) &\cr
\noalign{\hrule}
& untwisted &&
	({\bf1},0,0,{\bf8'},{\bf8'})&&
	1 && \ell=2 && 0 && (0,1,0) &\cr
\noalign{\hrule}
& untwisted &&
	(\ts,0,-1,{\bf1},{\bf1}) + ({\bf1},+1,0,{\bf1},{\bf1}) + {\rm c.c}&&
	1 && \ell=3 && 0 && (0,0,1) &\cr
\noalign{\hrule}
& untwisted &&
	({\bf1},0,0,{\bf8''},{\bf8''})&&
	1 && \ell=3 && 0 && (0,0,1) &\cr
\noalign{\hrule}
& $\Theta_1$ &&
	({\bf1},+\half,0,{\bf1},{\bf8''})
	+ ({\bf1},-\half,0,{\bf8''},{\bf1}) &&
	16 && \eta=(\half,\half,0) && 0 && (\half,\half,0) &\cr
\noalign{\hrule}
& $\Theta_2$ &&
	({\bf1},-\coeff14,-\coeff34,{\bf1},{\bf8'})
	+ ({\bf1},+\coeff14,+\coeff34,{\bf8'},{\bf1}) &&
	16 && \eta=(\half,0,\half) && 0 && (\half,0,\half) &\cr
\noalign{\hrule}
& $\Theta_1\Theta_2$ &&
	({\bf1},-\coeff14,+\coeff34,{\bf1},{\bf8})
	+ ({\bf1},+\coeff14,-\coeff34,{\bf8},{\bf1})&&
	16 && \eta=(0,\half,\half) && 0 && (0,\half,\half) &\cr
\noalign{\hrule}
}}\!
$$
which leads to the following values of the modular anomalies~\Alphadef:
$$
\displaylines{
\Alpha^i_{E(6)}\ =\ -6,\qquad
\Alpha^i_{SO(8)}\ =\ \Alpha^i_{SO(8)'}\ =\ +18,
\qquad\hfill\eqname\ZtwoZtwoSOeight\crr
\Alpha^1_{U(1)}\ =\ \smallmat{18}{-72}{198},\qquad
\Alpha^2_{U(1)}\ =\ \smallmat{18}{+72}{198},\qquad
\Alpha^3_{U(1)}\ =\ \smallmat{90}{0}{-18}.\hfill\cr }
$$
Comparing these results with eqs.~\threeembeddings, we see that again
eqs.~\coeffrel\ hold true for $\vgs^i=0$.
In fact, we shall momentarily see that $\vgs^i=0$ for all five $Z^2\times Z^2$
orbifolds.

For the next $Z_2\times Z_2$ orbifold, the little groups $D_i$ of the three
internal planes act similarly on the first $E_8$ but
not on the second $E_8$.
This time, the $N=1$ gauge group is $G=E_6\times U(1)^2\times SO(16)$
and the $N=2$ renormalization group coefficients $\cren^{N=2}_a(i)$
are given by eqs.~\threeembeddings\ for $i=1,2$ but by eq.~\firstZtwo\
for $i=3$.
The matter fields of this orbifold are
$$
\!\vcenter{\Tenpoint
\ialign{%
	\tablerule #&
	\hfil\enspace #\unskip \enspace\hfil &\vrule #&
	\hfil\enspace${#}$\enspace\hfil &\vrule #&
	\hfil\enspace #\unskip \enspace &\vrule #&
	\hfil\enspace${#}$\enspace\hfil &\vrule #&
	\hfil\enspace${#}$\enspace\hfil &\vrule #&
	\hfil\enspace${#}$\enspace\hfil &\vrule #\cr
\noalign{\hrule}
& sector && E_6\times U(1)^2\times SO(16) && \# &&
	\ell_I {\rm\ or\ } \eta_I^i && \rm osc. &&
	{\rm avg.}\ q_I^i &\cr
\noalign{\hrule}
& untwisted &&
	(\ts,+\half,+\half,{\bf1})
	+ ({\bf1},-\half,+\coeff32,{\bf1}) + {\rm c.c}&&
	1 && \ell=1 && 0 && (1,0,0) &\cr
\noalign{\hrule}
& untwisted &&
	(\ts,-\half,+\half,{\bf1})
	+ ({\bf1},+\half,+\coeff32,{\bf1}) + {\rm c.c}&&
	1 && \ell=2 && 0 && (0,1,0) &\cr
\noalign{\hrule}
& untwisted &&
	(\ts,0,-1,{\bf1}) + ({\bf1},+1,0,{\bf1}) + {\rm c.c}&&
	1 && \ell=3 && 0 && (0,0,1) &\cr
\noalign{\hrule}
& untwisted &&
	({\bf1},0,0,{\bf 128})&&
	1 && \ell=3 && 0 && (0,0,1) &\cr
\noalign{\hrule}
& $\Theta_1$ &&
	(\ts,0,+\half,{\bf1}) + ({\bf1},0,+\coeff32,{\bf1}) &&
	16 && \eta=(\half,\half,0) && 0 && (\half,\half,0) &\cr
\noalign{\hrule}
& $\Theta_1$ &&
	({\bf1},\pm\half,0,{\bf1}) &&
	32 && \eta=(\half,\half,0) && 1 && (\half,\half,0) &\cr
\noalign{\hrule}
& $\Theta_2$ &&
	({\bf1},+\coeff14,+\coeff34,{\bf 16}) &&
	16 && \eta=(\half,0,\half) && 0 && (\half,0,\half) &\cr
\noalign{\hrule}
& $\Theta_1\Theta_2$ &&
	({\bf1},-\coeff14,+\coeff34,{\bf 16})&&
	16 && \eta=(0,\half,\half) && 0 && (0,\half,\half) &\cr
\noalign{\hrule}
}}\!
$$
Substituting this spectrum into eqs.~\Alphadef, we obtain
$$
\eqalign{
\Alpha^{1,2}_{E(6)}\ =\ -6,\enspace\qquad
\Alpha^{1,2}_{SO(16)}\ =\ +18,\qquad &
\Alpha^{1,2}_{U(1)}\ = \smallmat{18}{\mp72}{198},\crr
\Alpha^{3}_{E(6)}\ =\ +42,\qquad
\Alpha^{3}_{SO(16)}\ =\ -30,\qquad &
\Alpha^{3}_{U(1)}\ = \smallmat{42}{0}{126} =\ +42 k_{U(1)} ,\cr
}\eqn\ZtwoZtwoSOsixteen
$$
which indeed agrees with eqs.~\coeffrel\ for $\vgs^1=\vgs^2=\vgs^3=0$.

In the remaining two $Z_2\times Z_2$ orbifolds, the little group $D_3$
of the third internal plane acts
differently from $D_1$ and $D_2$ on both $E_8$ factors:
In both models, $D_1$ and $D_2$ break the first
$E_8$ down to an $E_7\times SU(2)$ and the second $E_8$ down to an $SO(16)$
while $D_3$ breaks the second $E_8$ down to an $E_7\times SU(2)$;
the difference is whether $D_3$ leaves the first $E_8$ unbroken
or breaks it down to an $SO(16)$.
In the first case, the surviving gauge group is
$G=E_7\times SU(2)\times SU(8)\times U(1)$ and the spectrum of the
matter fields consists of
$$
\!\vcenter{\Tenpoint
\ialign{%
	\tablerule #&
	\hfil\enspace #\unskip \enspace\hfil &\vrule #&
	\hfil\enspace${#}$\enspace\hfil &\vrule #&
	\hfil\enspace #\unskip \enspace &\vrule #&
	\hfil\enspace${#}$\enspace\hfil &\vrule #&
	\hfil\enspace${#}$\enspace\hfil &\vrule #&
	\hfil\enspace${#}$\enspace\hfil &\vrule #\cr
\noalign{\hrule}
& sector && E_7\times SU(2)\times SU(8)\times U(1) && \# &&
	\ell_I {\rm\ or\ } \eta_I^i && \rm osc. &&
	{\rm avg.}\ q_I^i &\cr
\noalign{\hrule}
& untwisted &&
	({\bf1},{\bf1},{\bf28},-\half) + {\rm c.c} &&
	1 && \ell=1 && 0 && (1,0,0) &\cr
\noalign{\hrule}
& untwisted &&
	({\bf1},{\bf1},{\bf28},+\half) + {\rm c.c}&&
	1 && \ell=2 && 0 && (0,1,0) &\cr
\noalign{\hrule}
& untwisted &&
	({\bf1},{\bf1},{\bf70},0) + ({\bf1},{\bf1},{\bf1},\pm1)&&
	1 && \ell=3 && 0 && (0,0,1) &\cr
\noalign{\hrule}
& untwisted &&
	({\bf56},{\bf2},{\bf1},0)&&
	1 && \ell=3 && 0 && (0,0,1) &\cr
\noalign{\hrule}
& $\Theta_1$ &&
	({\bf1},{\bf1},{\bf28},0) &&
	16 && \eta=(\half,\half,0) && 0 && (\half,\half,0) &\cr
\noalign{\hrule}
& $\Theta_1$ &&
	({\bf1},{\bf1},{\bf1},\pm\half) &&
	32 && \eta=(\half,\half,0) && 1 && (\half,\half,0) &\cr
\noalign{\hrule}
& $\Theta_2$ &&
	({\bf1},{\bf2},{\bf\bar8},-\coeff14) &&
	16 && \eta=(\half,0,\half) && 0 && (\half,0,\half) &\cr
\noalign{\hrule}
& $\Theta_1\Theta_2$ &&
	({\bf1},{\bf2},{\bf\bar8},+\coeff14) &&
	16 && \eta=(0,\half,\half) && 0 && (0,\half,\half) &\cr
\noalign{\hrule}
}}\!
$$
where the abelian gauge charge is normalized to $k_{U(1)}=1$.
Therefore,
$$
\eqalign{
\Alpha^{1,2}_{SU(8)}\ =\ \Alpha^{1,2}_{U(1)}\ =\ +18,&
\quad \Alpha^{1,2}_{E(7)}\ =\ -6,
    \quad \Alpha^{1,2}_{SU(2)}\ =\ +90,\crr
\Alpha^{3}_{SU(8)}\ =\ \Alpha^{3}_{U(1)}\ =\ +42,&
\qquad \Alpha^{3}_{E(7)}\ =\ \Alpha^{3}_{SU(2)}\ =\ -30,\cr
}\eqn\ZtwoZtwoEseven
$$
which agrees with the fact that for this model, $\cren^{N=2}_a(i)$
are given by eqs.~\secondZtwo\ for $i=1,2$ and by eqs.~\firstZtwo\
for $i=3$.

\par\vskip 0pt plus 15pt \goodbreak
The unbroken gauge symmetry of the other model is
\vadjust{\penalty 5000}
$G=SO(12)\times SU(2)\times SU(2)\times SU(8)\times U(1)$
while its matter fields are as follows:
$$
\!\vcenter{\Tenpoint
\ialign{%
	\tablerule #&
	\hfil\enspace #\unskip \enspace\hfil &\vrule #&
	\hfil\enspace${#}$\enspace\hfil &\vrule #&
	\hfil\enspace #\unskip \enspace &\vrule #&
	\hfil\enspace${#}$\enspace\hfil &\vrule #&
	\hfil\enspace${#}$\enspace\hfil &\vrule #&
	\hfil\enspace${#}$\enspace\hfil &\vrule #\cr
\noalign{\hrule}
& sector && SO(12)\times SU(2)\times SU(2)\times SU(8)\times U(1) && \# &&
	\ell_I {\rm\ or\ } \eta_I^i && \rm osc. &&
	{\rm avg.}\ q_I^i &\cr
\noalign{\hrule}
& untwisted &&
	({\bf1},{\bf1},{\bf1},{\bf28},-\half) + {\rm c.c} &&
	1 && \ell=1 && 0 && (1,0,0) &\cr
\noalign{\hrule}
& untwisted &&
	({\bf 32},{\bf2},{\bf1},{\bf1},0) &&
	1 && \ell=1 && 0 && (1,0,0) &\cr
\noalign{\hrule}
& untwisted &&
	({\bf1},{\bf1},{\bf1},{\bf28},+\half) + {\rm c.c} &&
	1 && \ell=2 && 0 && (0,1,0) &\cr
\noalign{\hrule}
& untwisted &&
	({\bf 32'},{\bf1},{\bf2},{\bf1},0) &&
	1 && \ell=2 && 0 && (0,1,0) &\cr
\noalign{\hrule}
& untwisted &&
	({\bf1},{\bf1},{\bf1},{\bf70},0)
	+ ({\bf1},{\bf1},{\bf1},{\bf1},\pm1) &&
	1 && \ell=3 && 0 && (0,0,1) &\cr
\noalign{\hrule}
& untwisted &&
	({\bf12},{\bf2},{\bf2},{\bf1},0)&&
	1 && \ell=3 && 0 && (0,0,1) &\cr
\noalign{\hrule}
& $\Theta_1$ &&
	({\bf2},{\bf2},{\bf1},{\bf1},-\half) &&
	16 && \eta=(\half,\half,0) && 0 && (\half,\half,0) &\cr
\noalign{\hrule}
& $\Theta_1$ &&
	({\bf12},{\bf1},{\bf1},{\bf1},+\half) &&
	16 && \eta=(\half,\half,0) && 0 && (\half,\half,0) &\cr
\noalign{\hrule}
& $\Theta_2$ &&
	({\bf2},{\bf1},{\bf1},{\bf8},+\coeff14) &&
	16 && \eta=(\half,0,\half) && 0 && (\half,0,\half) &\cr
\noalign{\hrule}
& $\Theta_1\Theta_2$ &&
	({\bf1},{\bf2},{\bf1},{\bf\bar8},+\coeff14) &&
	16 && \eta=(0,\half,\half) && 0 && (0,\half,\half) &\cr
\noalign{\hrule}
}}\!
$$
Consequently,
$$
\displaylines{
\Alpha^1_{SO(12)}\ =\ \Alpha^1_{SU(2)}\ =\ \Alpha^2_{SO(12)}\
    =\ \Alpha^2_{SU(2)'}\ =\ \Alpha^3_{SU(8)}\ =\ -6,\hfil\cr
\Alpha^1_{SU(2)'}\ =\ \Alpha^2_{SU(2)}\ =\ \Alpha^3_{U(1)}\ =\ +90,\cr
\Alpha^1_{SU(8)}\ =\ \Alpha^1_{U(1)}\ =\ \Alpha^2_{SU(8)}\
    =\ \Alpha^2_{U(1)}\ =\ \Alpha^3_{SU(8)}\ =\ \Alpha^3_{SU(2)}\
    =\ \Alpha^3_{SU(2)'}\ =\ +18,\cr
\hfill\eqname\ZtwoZtwoSOtwelve\cr
}
$$
which agrees with the fact that for this model, all the $\cren_a^{N=2}(i)$
are given by eqs.~\secondZtwo, but the embedding of the $N=1$ gauge group
into the gauge group of the $N=2$\quad $D_i$ orbifold depends on $i$.
Again, in both models, eqs.~\coeffrel\ hold for $\vgs^i=0$.

\section{Other Orbifold Examples.}

After the $Z_3$ and $Z_2\times Z_2$ orbifolds we have presented thus far,
the next simplest group of orbifold examples consists of twelve $Z_4$
orbifolds whose rotation group is generated by $\Theta=(i,i,-1)$.
(There are twelve inequivalent twists of the $E_8\times E_8$ Kac-Moody
algebra that are compatible with this rotation.)
For these orbifolds, the little groups of the first two
planes are trivial while the little
group of the third plane is a $Z_2$.
Therefore, we expect
$$
\Alpha_a^{1,2}\ =\ k_a\,\vgs^{1,2}\qquad
\Alpha_a^3\ =\ k_a\,\vgs^{3}\ +\ \half\cren^{N=2}_a ,
\eqn\Zfourcoeffrel
$$
where the $N=2$ beta-function coefficients $\cren^{N=2}_a$ are given by
eqs.~\firstZtwo\ or \secondZtwo, whichever is appropriate for a particular
$Z_4$ orbifold.
Without going through the spectra of the twelve orbifolds, let us simply
state the results: For all twelve models,
eqs.~\Zfourcoeffrel\ are always satisfied and furthermore, $\vgs^3=0$.
On the other hand, $\vgs^{1,2}$ depend on a particular model but generally
do not vanish.

All of the above examples have a common feature that $\vgs^i=0$ whenever
some twisted sectors leave the $i$th plane unrotated.
However, a more general survey shows that $\vgs^i$ vanishes only when the
little group $D_i$ of the $i$th plane has index 2 (which happens to be the
case for all the non-trivial little groups of the $Z_3$, $Z_2\times Z_2$
and $Z_4$ orbifolds).
In particular, for the $Z_6$ orbifolds whose rotation group is generated
by the $\Theta=(e^{2\pi i/6},e^{2\pi i/3},-1)$,
the little groups are $D_1=1$, $D_2=Z_2$ and $D_3=Z_3$;
consequently, $\vgs^3=0$ but $\vgs^2\neq0$.

There are sixty one inequivalent $Z_6$ twists of the $E_8\times E_8$
Kac-Moody algebra that are compatible with
$\Theta=(e^{2\pi i/6},e^{2\pi i/3},-1)$.
The resulting list of sixty one models is clearly much
too long to be presented here in full detail,
so we decided to present only two of these $Z_6$ orbifolds as examples:
The left-right symmetric $(2,2)$ orbifold whose gauge group is
$\left(E_6\times U(1)^2\right)\times E_8$ and a $(0,2)$\ $Z_6$
orbifold with $G=\left(SU(6)\times SU(3)\times SU(2)\right)
\times\left(SU(8)\times U(1)\right)$;
their respective spectra of the matter fields are listed in tables on the
following pages.
(The abelian gauge charges are normalized to $k_{U(1)}={1\,0\choose 0\,3}$
for the $(2,2)$ orbifold and $K_{U(1)}=1$ for the $(0,2)$ orbifold.)

\topinsert
\def\b{{\bf1}}
\centerline{\caps Matter Fields of the (2,2) $Z_6$ Orbifold}
\smallskip
\Tenpoint
\ialign{%
	\vrule height 13.5pt depth 4.5pt width 0pt \vrule #&
	\hfil\enspace #\unskip \enspace\hfil &\vrule #&
	\hfil\enspace${#}$\enspace\hfil &\vrule #&
	\hfil\enspace #\unskip \enspace &\vrule #&
	\hfil\enspace${#}$\enspace\hfil &\vrule #&
	\hfil\enspace #\unskip \enspace\hfil &\vrule #&
	\hfil\enspace${#}$\enspace\hfil &\vrule #\cr
\noalign{\hrule}
& sector && E_6\times U(1)^2\times E_8 && \# &&
	\ell_I {\rm\ or\ } \eta_I^i && \rm osc. &&
	{\rm avg.}\ q_I^i &\cr
\noalign{\hrule}
& untwisted &&
	(\ts,+\half,+\half,\b) &&
	1 && \ell=1 && 0 && (1,0,0) &\cr
\noalign{\hrule}
& untwisted &&
	(\b,-1,0,\b) + (\b,+\half,-\coeff32,\b) &&
	1 && \ell=1 && 0 && (1,0,0) &\cr
\noalign{\hrule}
& untwisted &&
	(\ts,-\half,+\half,\b) + (\b,-\half,-\coeff32,\b) &&
	1 && \ell=2 && 0 && (0,1,0) &\cr
\noalign{\hrule}
& untwisted &&
	(\ts,0,-1,\b) + (\tsb,0,+1,\b) &&
	1 && \ell=3 && 0 && (0,0,1) &\cr
\noalign{\hrule}
& $\Theta$ &&
	(\ts,-\coeff{1}{12},-\coeff14,\b) &&
	12 && \eta=(\coeff16,\coeff13,\half) &&
	0 && (\coeff56,\coeff23,\half) &\cr
\noalign{\hrule}
& $\Theta$ &&
	(\b,-\coeff{7}{12},-\coeff34,\b) &&
	12 && \eta=(\coeff16,\coeff13,\half) &&
	1 && (\coeff{11}{6},\coeff23,\half) &\cr
\noalign{\hrule}
& $\Theta$ &&
	(\b,+\coeff{5}{12},-\coeff34,\b) &&
	24 && \eta=(\coeff16,\coeff13,\half) &&
	1 or 2 && (\coeff{11}{6},\coeff76,\half) &\cr
\noalign{\hrule}
& $\Theta$ &&
	(\b,-\coeff{1}{12},+\coeff34,\b) &&
	48 && \eta=(\coeff16,\coeff13,\half) &&
	1, 2 or 3 &&
	(\coeff{11}{6},\coeff{11}{12},\half) &\cr
\noalign{\hrule}
& $\Theta^2$ &&
	(\ts,-\coeff{1}{6},+\half,\b) + (\b,-\coeff{1}{6},-\coeff32,\b) &&
	6 && \eta=(\coeff13,\coeff23,0) &&
	0 && (\coeff23,\coeff13,0) &\cr
\noalign{\hrule}
& $\Theta^2$ &&
	(\tsb,-\coeff{1}{6},-\half,\b) + (\b,-\coeff{1}{6},+\coeff32,\b) &&
	3 && \eta=(\coeff13,\coeff23,0) &&
	0 && (\coeff23,\coeff13,0) &\cr
\noalign{\hrule}
& $\Theta^2$ &&
	(\b,-\coeff23,0,\b) &&
	9 && \eta=(\coeff13,\coeff23,0) &&
	1 && (\coeff43,0,0) &\cr
\noalign{\hrule}
& $\Theta^2$ &&
	(\b,+\coeff13,0,\b) &&
	24 && \eta=(\coeff13,\coeff23,0) &&
	1 or 2 &&
	(\coeff{7}{6},\coeff{-1}{24},0) &\cr
\noalign{\hrule}
& $\Theta^3$ &&
	(\ts,+\coeff14,-\coeff14,\b) &&
	8 && \eta=(\half,0,\half) &&
	0 && (\half,0,\half) &\cr
\noalign{\hrule}
& $\Theta^3$ &&
	(\tsb,-\coeff14,+\coeff14,\b) + (\b,\pm\coeff34,\mp\coeff34,\b) &&
	4 && \eta=(\half,0,\half) &&
	0 && (\half,0,\half) &\cr
\noalign{\hrule}
& $\Theta^3$ &&
	(\b,+\coeff14,+\coeff34,\b) &&
	24 && \eta=(\half,0,\half) &&
	1 && (\half,0,\half) &\cr
\noalign{\hrule}
& $\Theta^3$ &&
	(\b,-\coeff14,-\coeff34,\b) &&
	20 && \eta=(\half,0,\half) &&
	1 && (\coeff{7}{10},0,\half) &\cr
\noalign{\hrule}
& $\Theta^4$ &&
	(\ts,+\coeff{1}{6},+\half,\b) + (\b,+\coeff{1}{6},-\coeff32,\b) &&
	6 && \eta=(\coeff23,\coeff13,0) &&
	0 && (\coeff13,\coeff23,0) &\cr
\noalign{\hrule}
& $\Theta^4$ &&
	(\tsb,+\coeff{1}{6},-\half,\b) + (\b,+\coeff{1}{6},+\coeff32,\b) &&
	3 && \eta=(\coeff23,\coeff13,0) &&
	0 && (\coeff13,\coeff23,0) &\cr
\noalign{\hrule}
& $\Theta^4$ &&
	(\b,+\coeff23,0,\b) &&
	9 && \eta=(\coeff23,\coeff13,0) &&
	1 && (0,\coeff43,0) &\cr
\noalign{\hrule}
& $\Theta^4$ &&
	(\b,-\coeff13,0,\b) &&
	21 && \eta=(\coeff23,\coeff13,0) &&
	1 or 2 && (\coeff{1}{21},\coeff{23}{21},0) &\cr
\noalign{\hrule}
} 
\bigskip
\endinsert

\topinsert
\def\b{{\bf1}}
\centerline{\caps Matter Fields of a (0,2) $Z_6$ Orbifold}
\smallskip
\Tenpoint
\ialign{%
	\vrule height 13.5pt depth 4.5pt width 0pt \vrule #&
	\hfil\enspace #\unskip \enspace\hfil &\vrule #&
	\hfil\enspace${#}$\enspace\hfil &\vrule #&
	\hfil\enspace #\unskip \enspace &\vrule #&
	\hfil\enspace${#}$\enspace\hfil &\vrule #&
	\hfil\enspace #\unskip \enspace\hfil &\vrule #&
	\hfil\enspace${#}$\enspace\hfil &\vrule #\cr
\noalign{\hrule}
& sector && SU(6)\times SU(3)\times SU(2)\times SU(8)\times U(1) && \# &&
	\ell_I {\rm\ or\ } \eta_I^i && \rm osc. &&
	{\rm avg.}\ q_I^i &\cr
\noalign{\hrule}
& untwisted &&
	({\bf6},{\bf3},{\bf2},\b,0) &&
	1 && \ell=1 && 0 && (1,0,0) &\cr
\noalign{\hrule}
& untwisted &&
	(\b,\b,\b,{\bf28},+\half) + (\b,\b,\b,\b,-1) &&
	1 && \ell=1 && 0 && (1,0,0) &\cr
\noalign{\hrule}
& untwisted &&
	({\bf15},{\bf\bar3},\b,\b,0) + (\b,\b,\b,{\bf28},-\half) &&
	1 && \ell=2 && 0 && (0,1,0) &\cr
\noalign{\hrule}
& untwisted &&
	({\bf20},\b,{\bf2},\b,0) + (\b,\b,\b,{\bf 70},0) &&
	1 && \ell=3 && 0 && (0,0,1) &\cr
\noalign{\hrule}
& $\Theta^2$ &&
	(\overline{\bf15},\b,\b,\b,+\coeff13) &&
	6 && \eta=(\coeff13,\coeff23,0) &&
	0 && (\coeff23,\coeff13,0) &\cr
\noalign{\hrule}
& $\Theta^2$ &&
	({\bf 6},\b,{\bf2},\b,+\coeff13) + (\b,{\bf\bar3},\b,\b,-\coeff23) &&
	3 && \eta=(\coeff13,\coeff23,0) &&
	0 && (\coeff23,\coeff13,0) &\cr
\noalign{\hrule}
& $\Theta^2$ &&
	(\b,{\bf\bar3},\b,\b,+\coeff13) &&
	9 && \eta=(\coeff13,\coeff23,0) &&
	1 && (1,-\coeff13,0) &\cr
\noalign{\hrule}
& $\Theta^3$ &&
	(\b,\b,{\bf2},{\bf\bar8},+\coeff14) &&
	8 && \eta=(\half,0,\half) &&
	0 && (\half,0,\half) &\cr
\noalign{\hrule}
& $\Theta^3$ &&
	(\b,\b,{\bf2},{\bf8},\coeff14) &&
	4 && \eta=(\half,0,\half) &&
	0 && (\half,0,\half) &\cr
\noalign{\hrule}
& $\Theta^4$ &&
	({\bf15},\b,\b,\b,-\coeff13) &&
	3 && \eta=(\coeff23,\coeff13,0) &&
	0 && (\coeff13,\coeff23,0) &\cr
\noalign{\hrule}
& $\Theta^4$ &&
	({\bf\bar6},\b,{\bf2},\b,-\coeff13) + (\b,{\bf3},\b,\b,+\coeff23) &&
	6 && \eta=(\coeff23,\coeff13,0) &&
	0 && (\coeff13,\coeff23,0) &\cr
\noalign{\hrule}
& $\Theta^4$ &&
	(\b,{\bf 3},\b,\b,-\coeff13) &&
	9 && \eta=(\coeff13,\coeff23,0) &&
	1 && (-\coeff13,1,0) &\cr
\noalign{\hrule}
} 
\bigskip
\endinsert

For the left-right symmetric $Z_6$ orbifold,
the $N=2$ orbifolds produced by the little groups $D_2=Z_2$
and $D_3=Z_3$ are both left-right symmetric,
and for all such  orbifolds,
$$
\cren^{N=2}_{E(8)}\ =\ -60,\qquad
\cren^{N=2}_{E(6)}\ =\ +84,\qquad
\cren^{N=2}_{U(1)}\ =\ +84 k_{U(1)}\,.
\eqn\crenofKthree
$$
At the same time, eqs.~\Alphadef\ give us
$$
\mathsurround=2pt
\vcenter{\baselineskip=25pt \lineskip=5pt
\ialign{%
	\hfil$\displaystyle{\Alpha^{#}_{E(8)}={}}$&
		$\displaystyle{#}$,\hskip 12pt plus 1fil &
	\hfil$\displaystyle{\Alpha^{#}_{E(6)}={}}$&
		$\displaystyle{#}$,\hskip 12pt plus 1fil &
	\hfil$\displaystyle{\Alpha^{#}_{U(1)}={}}$&
		\hfil$\displaystyle{#}$\hfil &
		$\displaystyle{{}= # k_{U(1)}}\,$,\hfil\cr
1 & -30 & 1 & -30 & 1 & \smallmat{-30}{0}{-90} & -30\cr
2 & -30 & 2 & +18 & 2 & \smallmat{+18}{0}{+54} & +18\cr
3 & -30 & 3 & +42 & 3 & \smallmat{+42}{0}{+126} & +42\cr
}}\eqn\ZsixLR
$$
and we immediately see that eqs.~\coeffrel\ are satisfied for
$$
\vgs^1\ =\ -30,\qquad \vgs^2\ =\ -10,\qquad \vgs^3\ =\ 0.
\eqn\VZsixLR
$$
For the other $Z_6$ example, the $N=2$ orbifold produced by the little
group of the second plane has $G^{N=2}=E_7\times SU(2)\times SO(16)$
and its beta-function coefficients are given by eq.~\secondZtwo;
similarly, for the little group of the third plane,
$G^{N=2}=E_6\times SU(3)\times E_7\times U(1)$ and the beta-function
coefficients are
$$
\cren^{N=2}_{E(6)}\ =\ \cren^{N=2}_{SU(3)}\ =\ +48,\qquad
\cren^{N=2}_{E(7)}\ =\ -24,\qquad
\cren^{N=2}_{U(1)}\ =\ +120.
\eqn\NtwoZthree
$$
At the same time, the modular anomaly coefficients \Alphadef\ of this
$Z_6$ orbifold are
$$
\displaylines{
\Alpha^1_{SU(6)}\ =\ \Alpha^1_{SU(3)}\ =\ \Alpha^1_{SU(2)}\
=\ \Alpha^1_{SU(8)}\ =\ \Alpha^1_{U(1)}\ =\ +2,
\qquad\hfil\eqname\ZsixMaximal\hfilneg\cr
\Alpha^2_{SU(6)}\ =\ \Alpha^2_{SU(3)}\ =\ -2,\qquad
\Alpha^2_{SU(2)}\ =\ +62,\qquad
\Alpha^2_{SU(8)}\ =\ \Alpha^2_{U(1)}\ =\ +14,\qquad\cr
\Alpha^3_{SU(6)}\ =\ \Alpha^3_{SU(3)}\ =\ \Alpha^3_{SU(2)}\ =\ +24,\qquad
\Alpha^3_{SU(8)}\ =\ -12,\qquad
\Alpha^3_{U(1)}\ =\ +60,\qquad\cr }
$$
which satisfies eqs.~\coeffrel\ for
$$
\vgs^1\ =\ \vgs^2\ =\ +2,\qquad \vgs^3\ =\ 0.
\eqn\VZsixMaximal
$$

Notice that for both examples, $\vgs^2\neq0$ even though the little group of
the second plane is non-trivial; on the other hand, $\vgs^3=0$.
{}From the orbifolds we have studied so far, it appears that $\vgs^i$ vanishes
whenever the little group of the $i$th plane has index two (\eg,
$Z_2\subset Z_2^2$, $Z_2\subset Z_4$ or $Z_3\subset Z_6$).
For the orbifold in which the second $E_8$ remains unbroken,
this behavior results from the absence of any $E_8$-charged massless
matter fields
(which implies $\Alpha^i_{E(8)}=-T(E_8)=\half\cren^{N=2}_{E(8)}$),
but we have no idea why the $(0,2)$ orbifolds in which both $E_8$'s are
broken also follow the same pattern.

\refout
\bye